\documentclass[fleqn,usenatbib]{mnras}

\usepackage[T1]{fontenc}

\DeclareRobustCommand{\VAN}[3]{#2}
\let\VANthebibliography\thebibliography
\def\thebibliography{\DeclareRobustCommand{\VAN}[3]{##3}\VANthebibliography}

\usepackage{graphicx}	
\usepackage{amsmath}	
\usepackage{amssymb}	
\usepackage{multirow}   
\usepackage{bm}
\usepackage{newtxtext,newtxmath}

\graphicspath{{figs/}}

\defcitealias{endsley2021a}{E21a}  
\newcommand{\beagle}{\textsc{beagle}}  
\newcommand{\prospector}{\textsc{Prospector}}  
\newcommand{\fsps}{\textsc{fsps}}  
\newcommand{\Msun}{M$_\odot$}  

\title[SFHs of UV-luminous galaxies at $z \simeq 6.8$]{Star formation histories of UV-luminous galaxies at $z \simeq 6.8$: implications for stellar mass assembly at early cosmic times}

\author[Whitler et al.]{Lily Whitler,$^{1}$\thanks{email: \href{mailto:lwhitler@arizona.edu}{lwhitler@arizona.edu}}\thanks{NSF Graduate Research Fellow}
Daniel P. Stark,$^{1}$
Ryan Endsley,$^{1}$
Joel Leja,$^{2, 3, 4}$
St\'{e}phane Charlot,$^{5}$
\newauthor
and Jacopo Chevallard$^{6}$ \\
$^{1}$Steward Observatory, University of Arizona, 933 N Cherry Ave, Tucson, AZ 85721, USA \\
$^{2}$Department of Astronomy \& Astrophysics, The Pennsylvania State University, University Park, PA 16802, USA \\
$^{3}$Institute for Computational \& Data Sciences, The Pennsylvania State University, University Park, PA, USA \\
$^{4}$Institute for Gravitation and the Cosmos, The Pennsylvania State University, University Park, PA 16802, USA \\
$^{5}$Sorbonne Universit\'{e}, UPMC-CNRS, UMR7095, Institut d'Astrophysique de Paris, F-75014, Paris, France \\
$^{6}$Department of Physics, University of Oxford, Denys Wilkinson Building, Keble Road, Oxford OX1 3RH, UK}

\date{Accepted XXX. Received YYY; in original form ZZZ}

\pubyear{2022}

\begin{document}
\label{firstpage}
\pagerange{\pageref{firstpage}--\pageref{lastpage}}
\maketitle

\begin{abstract}
The variety of star formation histories (SFHs) of $z\gtrsim6$ galaxies provides important insights into early star formation, but has been difficult to systematically quantify. Some observations suggest that many $z\sim6-9$ galaxies are dominated by $\gtrsim200$\,Myr stellar populations, implying significant star formation at $z\gtrsim9$, while others find that most reionization era galaxies are $\lesssim10$\,Myr, consistent with little $z\gtrsim9$ star formation. Here, we quantify the distribution of ages of UV-bright ($-22.5\lesssim M_\textsc{uv}\lesssim-21$) galaxies colour-selected to lie at $z\simeq6.6-6.9$, an ideal redshift range to systematically study the SFHs of reionization era galaxies with ground-based observatories and \textit{Spitzer}. We infer galaxy properties with two SED modelling codes and compare results, finding that stellar masses are largely insensitive to the model, but the inferred ages can vary by an order of magnitude. We infer a distribution of ages assuming a simple, parametric SFH model, finding a median age of $\sim30-70$\,Myr depending on SED model. We quantify the fractions of $\leq10$\,Myr and $\geq250$\,Myr galaxies, finding that these systems comprise $\sim15-30$\,per\,cent and $\sim20-25$\,per\,cent of the population, respectively. With a flexible SFH model, the shapes of the SFHs are consistent with those implied by the simple model (e.g. young galaxies have rapidly rising SFHs). However, stellar masses can differ significantly, with those of young systems sometimes being more than an order of magnitude larger with the flexible SFH. We quantify the implications of these results for $z\gtrsim9$ stellar mass assembly and discuss improvements expected from \textit{JWST}.
\end{abstract}

\begin{keywords}
galaxies: evolution -- galaxies: high-redshift -- dark ages, reionization, first stars
\end{keywords}


\section{Introduction} \label{sec:intro}

Understanding the build up of stellar mass within the first billion years of cosmic time provides crucial insights into the formation of the first stars and galaxies. Within the last two decades, deep ground- and space-based observational campaigns \citep[e.g.][]{bouwens2007, bowler2012, mclure2013, finkelstein2015, ono2018} have pushed the redshift frontier back to only a few hundred million years after the Big Bang. These studies have yielded the first observations of galaxies at $z \gtrsim 10$ (e.g. \citealt{oesch2016, jiang2021} spectroscopically confirmed a galaxy at $z = 10.9$) and have laid the groundwork for new insights into $z \gtrsim 10$ galaxies that will shortly be enabled by \textit{JWST}; see \citet{robertson2022} for a review.

However, current efforts to measure the global properties and evolution of the star-forming galaxy population at these early times have painted conflicting pictures. While rest-frame UV luminosity functions (UV LFs) have been measured up to $z \sim 9 - 10$ \citep[$\sim 500$\,Myr after the Big Bang;][]{oesch2018, stefanon2019, bowler2020, bouwens2021, finkelstein2022, bagley2022, leethochawalit2022} and the first constraints at $z \sim 12 - 13$ are now emerging \citep{harikane2022}, current UV LFs do not strongly constrain the cosmic star formation history at $z \gtrsim 9$. Some results have argued that at $z \gtrsim 9$, the UV-luminous ($M_\textsc{uv} \lesssim -21$), star-forming galaxy population known to be present at $z \sim 6 - 9$ disappears rapidly \citep[e.g.][]{oesch2018}, while others suggest a more smooth decline \citep[e.g.][]{mcleod2016}. Thus, our understanding of the beginning of cosmic reionization and the properties of early galaxies is limited.

The stellar mass content of the Universe at slightly lower redshifts provides an alternative path towards understanding star formation at early cosmic times. Specifically, since the stellar mass of a galaxy represents the integral of its past star formation, a complete census of stellar mass at $z \sim 6 - 9$ can constrain earlier ($z \sim 9 - 15$) epochs of star formation where it is challenging to directly observe galaxies with current facilities. Over the last two decades, this census of stellar mass at $z \sim 6 - 9$ has been enabled by \textit{Spitzer}/Infrared Array Camera (IRAC) imaging at 3.6\,$\mu$m and 4.5\,$\mu$m \citep[e.g.][]{duncan2014, grazian2015, song2016, bhatawdekar2019, kikuchihara2020, stefanon2021}, which probes the rest-frame optical stellar continuum that is key to clean measurements of stellar mass. However, broadband IRAC photometry can also be contaminated by strong nebular emission lines, leading to systematic uncertainties on individual stellar masses (and consequently the total stellar mass content of the Universe) if the exact contribution of nebular lines to the photometry is unknown \citep[e.g.][]{schaerer2009, schaerer2010, gonzalez2012, gonzalez2014, stark2013, debarros2014}. Thus, it is ideal to focus on galaxies at redshifts where the strongest nebular lines, notably \ion{[O}{iii]} and \ion{H}{$\beta$} at $z \gtrsim 6.5$, transmit through only one IRAC filter, leaving the other to cleanly probe the rest-optical stellar continuum. This generally requires precisely characterized photometric redshifts (e.g. \citealt{smit2014, smit2015, endsley2021a} selected galaxies within $\Delta z \simeq 0.3$) or spectroscopic redshifts \citep[e.g.][]{zitrin2015, roberts-borsani2016, stark2017, hashimoto2018, laporte2021}.

There has been considerable progress towards developing a comprehensive understanding of the rest-frame optical properties of galaxies during reionization. Many rest-UV--selected galaxies have been found to have strong nebular lines (\ion{[O}{iii]}+\ion{H}{$\beta$} equivalent widths $\gtrsim 600$\,\AA) and weak underlying rest-optical continuum \citep{labbe2013, smit2014, smit2015, roberts-borsani2016, debarros2019, endsley2021a, stefanon2022}. These intense lines are consistent with a recent, rapid increase in star formation such that the rest-UV and optical spectral energy distribution (SED) is dominated by a young ($\lesssim 10$\,Myr) -- and correspondingly low mass -- stellar population \citep[e.g.][]{tang2019}. Such systems are found much less frequently among UV-selected galaxies at $z \sim 2$ \citep{boyett2022}, suggesting that young, intense line emitters become increasingly common into reionization. If these galaxies both (1) represent the majority of $z \sim 6 - 9$ galaxies, and (2) are truly young and low mass, this suggests that there is little star formation activity at $z \gtrsim 9$, possibly consistent with a rapid disappearance of UV-bright, star-forming galaxies. In contrast, evidence for a very different population of mature ($\gtrsim 200$\,Myr) galaxies (characterized by strong Balmer breaks in their SEDs) as early as $z \sim 9$ has been growing \citep[e.g.][]{egami2005, richard2011, zheng2012, huang2016, hoag2018, hashimoto2018, strait2020, laporte2021}. If these evolved systems are more common than young line emitters, this implies that much more stellar mass must be in place by $z \sim 9$, pointing to more vigorous star formation within the first $\sim 350$\,Myr of cosmic time.

The distribution of ages of UV-selected galaxies at $z \sim 6 - 9$ provides key insights into star formation at $z \gtrsim 9$. However, it is currently unclear how frequently either young line emitters or mature systems occur among the bright galaxy population. This ambiguity is largely caused by two sources of uncertainty. First, it is difficult to photometrically select large samples in sufficiently precise redshift ranges such that rest-optical stellar continuum and nebular lines can be unambiguously separated in IRAC photometry. With large redshift uncertainties, the IRAC observations could be reproduced by either strong rest-optical nebular lines that imply young ages, or Balmer breaks consistent with old ages \citep[and it has further been noted that nebular lines and Balmer breaks may be degenerate in IRAC photometry at $z > 7$ even with well known redshifts;][]{roberts-borsani2020}. Secondly, there are fundamental limitations in measuring ages and stellar masses of galaxies that appear young, since young stars dominate the observed rest-UV and optical SED and outshine any older stellar population that may exist \citep[e.g.][]{papovich2001}. Thus, the amount of early star formation, and therefore the age and stellar mass, inferred for an individual system can depend strongly on the SED model.

\citet{endsley2021a} (hereafter \citetalias{endsley2021a}) pioneered a method to confront the first of these shortcomings, leveraging both narrowband and overlapping broadband filters to substantially reduce photometric redshift uncertainties. Their colour selection precisely selects galaxies to lie at redshifts of $z \simeq 6.6 - 6.9$ \citep[a redshift range that was also employed by][]{smit2014, smit2015}, where \ion{[O}{iii]} and \ion{H}{$\beta$} transmit through the IRAC 3.6\,$\mu$m bandpass, leaving the 4.5\,$\mu$m filter free of strong lines. Since the 4.5\,$\mu$m filter probes only the rest-optical stellar continuum, the physical properties of the sample can then be robustly measured and statistically characterized. \citetalias{endsley2021a} focused on quantifying the $z \sim 7$ \ion{[O}{iii]}+\ion{H}{$\beta$} equivalent width distribution with this selection, and here we extend their work to assess the distribution of ages of UV-bright galaxies at $z \simeq 6.6 - 6.9$. For our analysis, we use a slightly updated selection to obtain a larger sample, then infer ages and stellar masses using star formation history (SFH) models with varying degrees of flexibility (including a highly flexible nonparametric model to investigate the potential for old stellar populations in galaxies that appear young). We explore the systematics of the stellar population synthesis, photoionization, and SFH models we adopt, and quantify the implications of our inferred stellar masses, ages, and SFHs (along with their uncertainties) for the assembly of stellar mass at very early times.

This paper is organized as follows. We describe the observations and our sample selection in Section\ \ref{sec:observations}. We present our SED modelling approach with two Bayesian galaxy SED modelling codes, the BayEsian Analysis of GaLaxy sEds \citep[\beagle;][]{chevallard2016} code and \prospector\ \citep{johnson2021}, in Section\ \ref{sec:sed_modelling}, then report the physical properties inferred from the SED models and compare the results of the two codes in Section\ \ref{sec:galaxy_properties}. We quantify the distribution of SED-inferred ages of UV-bright galaxies at $z \sim 7$ in Section\ \ref{sec:ages_sfhs}. In Section\ \ref{sec:discussion}, we discuss the implications of our findings for the build up of stellar mass in the early Universe and present the future outlook with upcoming \textit{JWST} data. Finally, we summarize and conclude in Section\ \ref{sec:summary}.

Throughout this paper, we adopt a flat $\Lambda$CDM cosmology with $h = 0.7$, $\Omega_M = 0.3$, and $\Omega_\Lambda = 0.7$. All magnitudes are given in the AB system \citep{oke1983}. We quote the marginalized 68\,per\,cent credible intervals for all errors, and all logarithms are base-10.

\section{Observations} \label{sec:observations}

In this work, we discuss a sample of Lyman break galaxies selected to lie at $z \simeq 6.6 - 6.9$ over 1.5\,deg$^2$ in COSMOS. We describe our rest-UV colour selection criteria in Section\ \ref{subsec:selection} and our \textit{Spitzer}/IRAC photometry in Section\ \ref{subsec:irac_phot}. We also provide a brief overview of the observed properties of our sample in Section\ \ref{subsec:observed_properties}. We note that though we provide a summary here, our selection technique was developed by \citetalias{endsley2021a}, where we refer the reader for most details.
\begin{figure}
    \centering
    \includegraphics[width=\columnwidth]{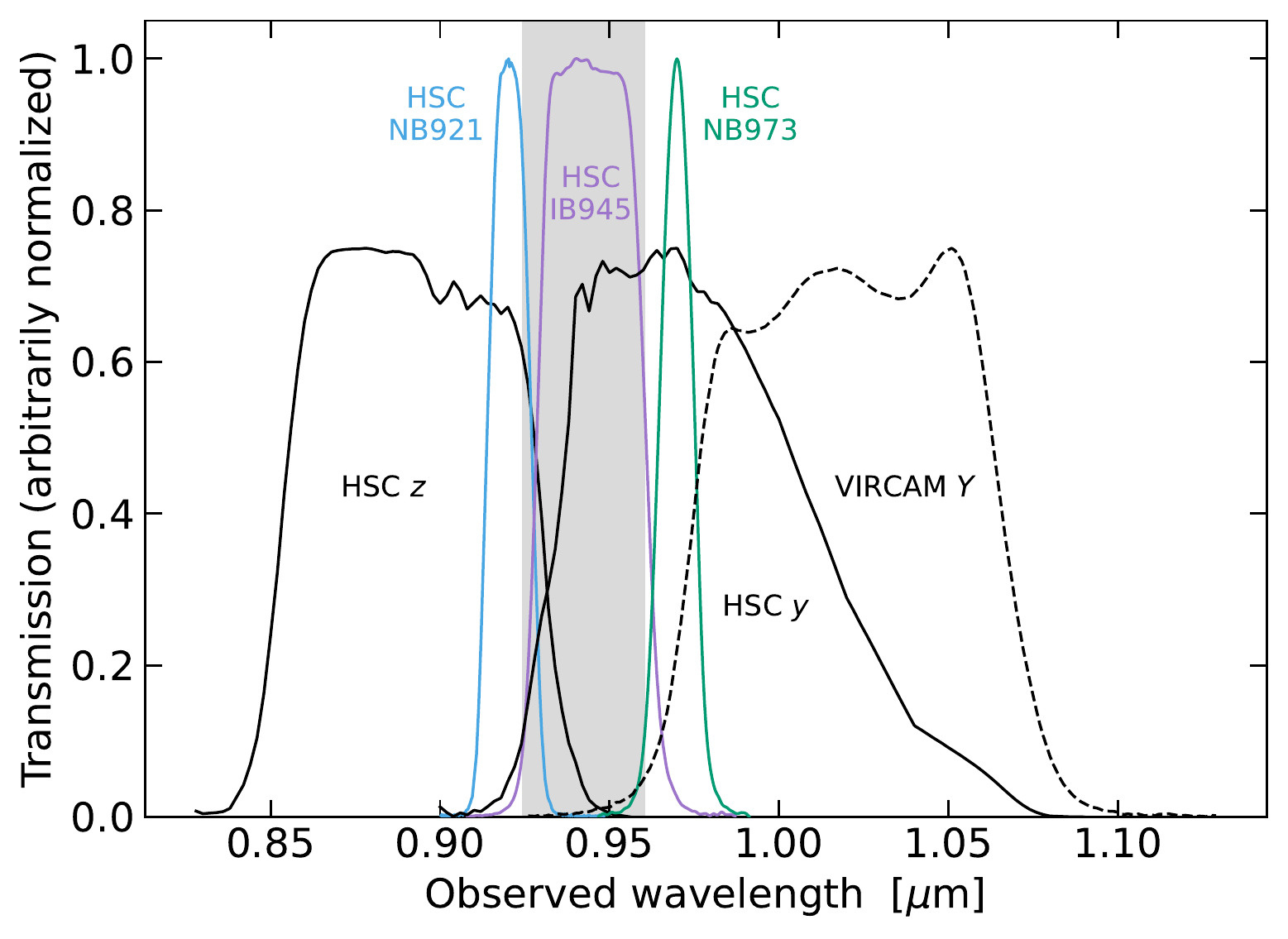}
    \caption{Transmission curves of Subaru/HSC $z$, $y$, NB921, IB945, NB973 and VISTA/VIRCAM $Y$ (arbitrarily normalized). Broadband filters are shown as black lines (solid lines denote HSC filters and dashed indicates VIRCAM) and HSC intermediate and narrowband filters are shown in colour (blue, purple, and green are HSC NB921, IB945, and NB973, respectively). The grey shaded region corresponds to the wavelength range of the \ion{Ly}{$\alpha$} break at $z = 6.6 - 6.9$. Together, these filters sample the wavelength range of the \ion{Ly}{$\alpha$} break around $z \simeq 6.8$ very finely, enabling extremely precise determinations of photometric redshifts.}
    \label{fig:filters}
\end{figure}

\subsection{Sample selection} \label{subsec:selection}

Our rest-frame UV observations come from Subaru/HSC and the Visible and Infrared Survey Telescope for Astronomy/VISTA InfraRed CAMera (VISTA/VIRCAM) imaging over COSMOS. Specifically, we utilize PDR2 of the Hyper Suprime-Cam Subaru Strategic Program \citep[HSC SSP;][]{aihara2019}, PDR1 of the Cosmic HydrOgen Reionization Unveiled with Subaru survey \citep[CHORUS;][]{inoue2020}, and DR4 of the UltraVISTA survey \citep{mccracken2012}. Together, these observations constitute a dataset spanning 14 optical and near-infrared ($\lambda < 3$\,$\mu$m) filters, and we highlight a set of filters that finely sample the wavelength range of the \ion{Ly}{$\alpha$} break around $z \simeq 6.8$: three HSC narrow and intermediate band filters, plus two overlapping $Y$-band filters from HSC and VIRCAM, shown in Figure\ \ref{fig:filters}. This unique combination of filters provides an excellent opportunity to select galaxies in a narrow redshift range with small photometric redshift uncertainties. We specifically utilize the following colour cuts to select galaxies over the small redshift interval of $z \simeq 6.6 - 6.9$:
\begin{enumerate}
    \item $z - y > 1.5$
    \item $z - Y > 1.5$
    \item $\text{NB}921 - Y > 1.0$
    \item $y - Y < 0.4$.
\end{enumerate}
We differentiate broadband HSC and VIRCAM filters by using lowercase and uppercase letters, respectively.

\begin{table*}
\renewcommand{\arraystretch}{1.5}
\centering
\caption{The observed properties of our sample of 36 UV-luminous galaxies selected to lie at $z \simeq 6.6 - 6.9$ over COSMOS. We report the $J$-band apparent magnitude, rest-UV slope measured from VIRCAM \textit{YJHK$_\text{s}$} photometry, and IRAC photometry and $[3.6] - [4.5]$ colour for each object in our sample. In the case of non-detections, we report the $2\sigma$ upper limit. We also note the spectroscopic redshift of a source when available.}
\label{tab:observed_properties}
\begin{tabular}{c|c|c|c|c|c|c|c|c} \hline
    Object ID & RA & Dec & $J$ & $\beta$ & 3.6\,$\mu$m & 4.5\,$\mu$m & $[3.6] - [4.5]$ & Notes \\ \hline\hline
    COS-83688 & 09:58:49.20 & +01:39:09.55 & $25.6_{-0.2}^{+0.2}$ & $-2.0_{-0.4}^{+0.4}$ & $24.5_{-0.2}^{+0.2}$ & $25.3_{-0.3}^{+0.5}$ & $-0.8_{-0.4}^{+0.5}$ & -- \\
    COS-87259 & 09:58:58.27 & +01:39:20.19 & $25.0_{-0.1}^{+0.2}$ & $-0.6_{-0.2}^{+0.2}$ & $22.9_{-0.1}^{+0.1}$ & $22.9_{-0.0}^{+0.0}$ & $0.0_{-0.1}^{+0.1}$ & $z_\ion{[C}{ii]} = 6.853$ \citep{endsley2022} \\
    COS-160072 & 09:58:54.96 & +01:42:56.68 & $25.6_{-0.1}^{+0.2}$ & $-3.0_{-0.4}^{+0.4}$ & $25.4_{-0.4}^{+0.6}$ & $25.7_{-0.4}^{+0.6}$ & $-0.3_{-0.5}^{+0.8}$ & -- \\
    COS-237729 & 10:00:31.41 & +01:46:51.01 & $25.7_{-0.2}^{+0.2}$ & $-1.9_{-0.4}^{+0.4}$ & $24.9_{-0.2}^{+0.2}$ & $25.4_{-0.2}^{+0.3}$ & $-0.5_{-0.3}^{+0.3}$ & -- \\
    COS-301652 & 10:00:54.83 & +01:50:05.18 & $25.7_{-0.2}^{+0.2}$ & $-2.1_{-0.4}^{+0.4}$ & $24.4_{-0.1}^{+0.1}$ & $24.7_{-0.2}^{+0.2}$ & $-0.3_{-0.2}^{+0.2}$ & -- \\
    COS-312533 & 10:00:35.52 & +01:50:38.59 & $25.6_{-0.1}^{+0.2}$ & $-1.4_{-0.3}^{+0.3}$ & $24.9_{-0.2}^{+0.2}$ & $25.1_{-0.2}^{+0.2}$ & $-0.2_{-0.3}^{+0.3}$ & -- \\
    COS-340502 & 09:59:15.36 & +01:52:00.62 & $25.7_{-0.2}^{+0.2}$ & $-2.1_{-0.5}^{+0.5}$ & $25.3_{-0.2}^{+0.3}$ & $> 25.6$ & $< -0.3$ & -- \\
    COS-369353 & 10:01:59.06 & +01:53:27.75 & $25.6_{-0.1}^{+0.2}$ & $-1.6_{-0.3}^{+0.3}$ & $24.0_{-0.1}^{+0.2}$ & $24.8_{-0.3}^{+0.4}$ & $-0.8_{-0.3}^{+0.4}$ & $z_\ion{[C}{ii]} = 6.729$ \citep{bouwens2022} \\
    COS-378785 & 09:57:22.16 & +01:53:55.50 & $25.4_{-0.1}^{+0.1}$ & $-2.2_{-0.4}^{+0.4}$ & $24.0_{-0.4}^{+0.7}$ & $24.4_{-0.3}^{+0.4}$ & $-0.4_{-0.5}^{+0.8}$ & -- \\
    COS-400019 & 09:59:17.26 & +01:55:03.08 & $25.4_{-0.1}^{+0.2}$ & $-2.3_{-0.4}^{+0.4}$ & $25.0_{-0.3}^{+0.5}$ & $> 25.6$ & $< -0.6$ & -- \\
    COS-469110 & 10:00:04.36 & +01:58:35.53 & $25.0_{-0.2}^{+0.3}$ & $-1.7_{-0.5}^{+0.5}$ & $24.3_{-0.1}^{+0.1}$ & $24.7_{-0.2}^{+0.2}$ & $-0.4_{-0.2}^{+0.2}$ & $z_\ion{[C}{ii]} = 6.644$ \citep{bouwens2022} \\
    COS-486435 & 10:01:58.71 & +01:59:31.03 & $26.0_{-0.2}^{+0.2}$ & $-0.9_{-0.4}^{+0.4}$ & $24.7_{-0.1}^{+0.2}$ & $25.2_{-0.3}^{+0.4}$ & $-0.6_{-0.3}^{+0.4}$ & -- \\
    COS-505871 & 10:00:21.35 & +02:00:30.93 & $25.5_{-0.1}^{+0.2}$ & $-2.2_{-0.5}^{+0.5}$ & $24.4_{-0.1}^{+0.1}$ & $24.6_{-0.1}^{+0.2}$ & $-0.2_{-0.2}^{+0.2}$ & -- \\
    COS-534584 & 10:00:42.13 & +02:01:56.87 & $25.0_{-0.1}^{+0.1}$ & $-1.8_{-0.2}^{+0.2}$ & $24.0_{-0.1}^{+0.1}$ & $24.3_{-0.1}^{+0.1}$ & $-0.3_{-0.2}^{+0.2}$ & $z_\ion{[C}{ii]} = 6.598$ \citep{bouwens2022} \\
    COS-559979 & 10:00:42.73 & +02:03:15.30 & $26.0_{-0.2}^{+0.2}$ & $-2.3_{-0.5}^{+0.5}$ & $25.9_{-0.4}^{+0.6}$ & $> 26.0$ & $< -0.1$ & -- \\
    COS-593796 & 10:01:53.46 & +02:04:59.62 & $25.6_{-0.2}^{+0.2}$ & $-2.7_{-0.4}^{+0.4}$ & $24.8_{-0.3}^{+0.3}$ & $> 25.5$ & $< -0.6$ & -- \\
    COS-596621 & 10:02:07.01 & +02:05:10.18 & $25.6_{-0.2}^{+0.2}$ & $-1.9_{-0.4}^{+0.4}$ & $25.4_{-0.4}^{+0.7}$ & $> 25.6$ & $< -0.2$ & -- \\
    COS-597997 & 09:57:37.00 & +02:05:11.33 & $25.6_{-0.2}^{+0.2}$ & $-1.6_{-0.3}^{+0.3}$ & $24.4_{-0.2}^{+0.2}$ & $25.3_{-0.4}^{+0.6}$ & $-0.9_{-0.4}^{+0.7}$ & $z_\ion{[C}{ii]} = 6.538$ \citep{bouwens2022} \\
    COS-627785 & 10:02:05.96 & +02:06:46.09 & $25.5_{-0.1}^{+0.2}$ & $-1.7_{-0.3}^{+0.3}$ & $24.5_{-0.1}^{+0.1}$ & $25.4_{-0.3}^{+0.4}$ & $-0.9_{-0.3}^{+0.4}$ & -- \\
    COS-637795 & 10:00:23.48 & +02:07:17.87 & $25.6_{-0.2}^{+0.2}$ & $-1.9_{-0.4}^{+0.4}$ & $24.9_{-0.2}^{+0.3}$ & $25.3_{-0.4}^{+0.6}$ & $-0.5_{-0.4}^{+0.6}$ & -- \\
    COS-703599 & 10:00:34.56 & +02:10:38.01 & $25.7_{-0.2}^{+0.2}$ & $-2.2_{-0.5}^{+0.5}$ & $24.2_{-0.1}^{+0.1}$ & $24.6_{-0.2}^{+0.2}$ & $-0.5_{-0.2}^{+0.2}$ & -- \\
    COS-705154 & 10:00:30.81 & +02:10:42.47 & $25.7_{-0.1}^{+0.2}$ & $-2.2_{-0.3}^{+0.3}$ & $25.2_{-0.3}^{+0.4}$ & $> 25.8$ & $< -0.7$ & -- \\
    COS-759861 & 10:02:06.47 & +02:13:24.06 & $24.4_{-0.1}^{+0.1}$ & $-1.9_{-0.1}^{+0.1}$ & $23.8_{-0.1}^{+0.1}$ & $24.2_{-0.1}^{+0.2}$ & $-0.4_{-0.2}^{+0.2}$ & $z_\ion{[C}{ii]} = 6.633$ \citep{bouwens2022} \\
    COS-788571 & 09:59:21.68 & +02:14:53.02 & $25.3_{-0.1}^{+0.1}$ & $-2.1_{-0.3}^{+0.3}$ & $24.4_{-0.1}^{+0.1}$ & $25.4_{-0.2}^{+0.3}$ & $-1.0_{-0.2}^{+0.3}$ & $z_\text{\ion{Ly}{$\alpha$}} = 6.883$ \citep{endsley2021b} \\
    COS-795090 & 09:57:23.92 & +02:15:13.73 & $25.5_{-0.2}^{+0.2}$ & $-2.5_{-0.5}^{+0.5}$ & $24.6_{-0.4}^{+0.8}$ & $24.6_{-0.3}^{+0.4}$ & $0.0_{-0.5}^{+0.9}$ & -- \\
    COS-810120 & 10:00:30.18 & +02:15:59.68 & $25.1_{-0.1}^{+0.1}$ & $-1.7_{-0.2}^{+0.2}$ & $23.8_{-0.1}^{+0.1}$ & $25.3_{-0.3}^{+0.4}$ & $-1.4_{-0.3}^{+0.4}$ & $z_\ion{[C}{ii]} = 6.854$ \citep{smit2018} \\
    COS-857605 & 09:59:12.35 & +02:18:28.86 & $25.8_{-0.2}^{+0.2}$ & $-1.4_{-0.3}^{+0.3}$ & $24.5_{-0.1}^{+0.1}$ & $25.0_{-0.3}^{+0.3}$ & $-0.5_{-0.3}^{+0.3}$ & -- \\
    COS-862541 & 10:03:05.25 & +02:18:42.75 & $24.5_{-0.2}^{+0.3}$ & $-1.9_{-0.3}^{+0.3}$ & $23.3_{-0.1}^{+0.1}$ & $24.7_{-0.2}^{+0.3}$ & $-1.4_{-0.3}^{+0.3}$ & $z_\ion{[C}{ii]} = 6.846$ \citep{bouwens2022} \\
    COS-955126 & 09:59:23.62 & +02:23:32.73 & $25.4_{-0.2}^{+0.2}$ & $-2.4_{-0.5}^{+0.5}$ & $24.2_{-0.2}^{+0.2}$ & $25.2_{-0.4}^{+0.6}$ & $-1.0_{-0.4}^{+0.6}$ & $z_\text{\ion{Ly}{$\alpha$}} = 6.813$ \citep{endsley2021b} \\
    COS-1099982 & 10:00:23.37 & +02:31:14.80 & $25.5_{-0.1}^{+0.1}$ & $-1.8_{-0.2}^{+0.2}$ & $24.2_{-0.1}^{+0.1}$ & $25.6_{-0.3}^{+0.4}$ & $-1.4_{-0.3}^{+0.5}$ & -- \\
    COS-1136216 & 10:01:58.50 & +02:33:08.55 & $24.9_{-0.1}^{+0.1}$ & $-2.3_{-0.3}^{+0.3}$ & $24.2_{-0.1}^{+0.1}$ & $24.5_{-0.1}^{+0.2}$ & $-0.3_{-0.2}^{+0.2}$ & -- \\
    COS-1163765 & 09:58:49.68 & +02:34:35.86 & $25.6_{-0.1}^{+0.2}$ & $-1.7_{-0.4}^{+0.4}$ & $24.8_{-0.1}^{+0.2}$ & $> 25.7$ & $< -0.9$ & -- \\
    COS-1224137 & 10:01:36.86 & +02:37:49.18 & $24.6_{-0.1}^{+0.1}$ & $-1.6_{-0.2}^{+0.2}$ & $24.0_{-0.1}^{+0.1}$ & $24.3_{-0.1}^{+0.1}$ & $-0.3_{-0.1}^{+0.2}$ & $z_\ion{[C}{ii]} = 6.685$ \citep{bouwens2022} \\
    COS-1235751 & 10:00:11.58 & +02:38:29.81 & $25.6_{-0.2}^{+0.2}$ & $-1.1_{-0.3}^{+0.3}$ & $24.3_{-0.1}^{+0.2}$ & $24.5_{-0.2}^{+0.2}$ & $-0.2_{-0.2}^{+0.3}$ & -- \\
    COS-1304254 & 10:02:54.04 & +02:42:11.94 & $24.6_{-0.2}^{+0.2}$ & $-1.9_{-0.4}^{+0.4}$ & $23.8_{-0.1}^{+0.1}$ & $23.7_{-0.1}^{+0.1}$ & $0.1_{-0.2}^{+0.2}$ & $z_\ion{[C}{ii]} = 6.577$ \citep{bouwens2022} \\
    COS-1387948 & 09:59:19.36 & +02:46:41.40 & $25.2_{-0.1}^{+0.2}$ & $-1.5_{-0.3}^{+0.3}$ & $23.4_{-0.1}^{+0.1}$ & $23.1_{-0.1}^{+0.1}$ & $0.3_{-0.1}^{+0.1}$ & -- \\ \hline
\end{tabular}
\end{table*}

\begin{figure*}
    \centering
    \includegraphics[width=\textwidth]{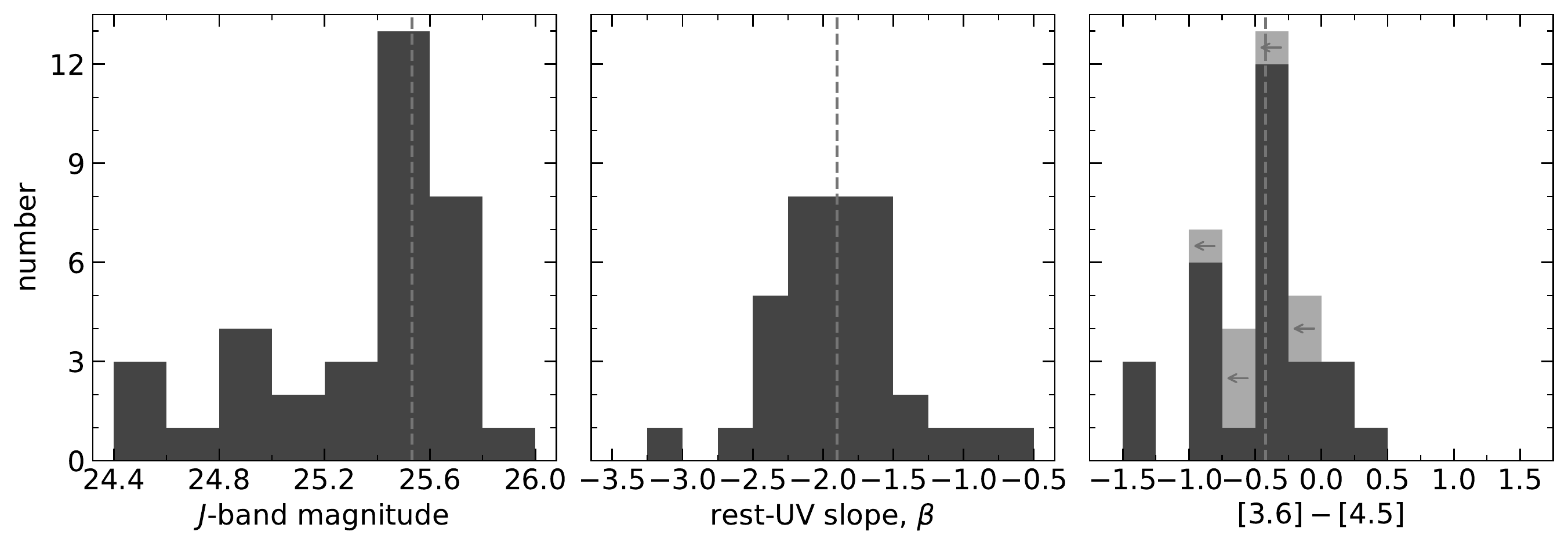}
    \caption{The distributions of selected observational properties for the 36 galaxies in our sample. Medians are shown as vertical grey dashed lines. \textit{Left:} Observed $J$-band apparent magnitudes. Our sample is relatively bright with apparent magnitudes that span $J \approx 24.5 - 26.0$ with median $J = 25.5$. \textit{Center:} Observed rest-UV slopes, $\beta$, determined by fitting $F_\nu \propto \lambda\,^{\beta + 2}$ to our observed \textit{YJHK$_\text{s}$} photometry. We observe UV slopes of $\beta$ from $-3.0$ to $-0.6$ with a median of $-1.9$, consistent with the UV slopes of moderately fainter objects at $z \sim 7$ identified in CANDELS \citep{bouwens2014} and the sample of more UV-luminous sources reported by \citet{bowler2017}. \textit{Right:} Observed IRAC $[3.6] - [4.5]$ colours, spanning $[3.6] - [4.5] \approx -1.5 - 0.3$ with median $[3.6] - [4.5] = -0.4$. Light grey histogram bars containing leftward pointing arrows denote where we have adopted a $2\sigma$ upper limit on the [4.5] photometry, leading to an upper limit on the IRAC colour. We highlight that many sources in our sample, though not all, have blue IRAC colours ($[3.6] - [4.5] < 0$), indicative of strong nebular emission in the redshift range of our selection.}
    \label{fig:properties}
\end{figure*}

As discussed by \citetalias{endsley2021a}, flux redward of the \ion{Ly}{$\alpha$} break does not contribute significantly to $z$ and NB921 at $z \gtrsim 6.6$, but does significantly contribute to $y$ until $z \simeq 6.9$. Thus, $z \simeq 6.6 - 6.9$ galaxies will drop out strongly in $z$ and NB921 but have relatively flat $y - Y$ colours. We note that the selection window of $z \simeq 6.6 - 6.9$ presented here assumes no \ion{Ly}{$\alpha$} emission, but as with any Lyman break selection using broadband photometry, the exact range depends slightly on the assumed \ion{Ly}{$\alpha$} equivalent width. We refer the reader to figure 2 of \citetalias{endsley2021a} for an exploration of the impact of \ion{Ly}{$\alpha$} on our redshift window.

We apply these colour cuts to sources identified by running \textsc{Source Extractor} on a \textit{yYJHK$_\text{s}$} $\chi^2$ detection image and measuring HSC and VIRCAM fluxes in 1.2\arcsec\ diameter apertures. Following \citetalias{endsley2021a}, we also apply additional signal-to-noise requirements to ensure that all sources are real (>3$\sigma$ detections in all of $y$, $Y$, and $J$ and a >5$\sigma$ detection in at least one of the three filters) and require that candidates are undetected ($< 2\sigma$) in HSC $g$ and $r$ (both filters are blueward of the Lyman limit at $z \geq 6.6$). We also apply additional colour cuts in the near-infrared to remove T-type brown dwarfs (either $Y - J < 0.45$, or both $J - H > 0$ and $J - K_s > 0$). Finally, in order to approximately match IRAC sensitivities, all candidates are required to have apparent magnitudes of $J < 25.7$ (or if $J < 25.7$ is not satisfied, we also allow sources with $K_s < 25.5$). As in \citetalias{endsley2021a}, our colour cuts to remove brown dwarfs and our apparent magnitude requirements are carefully chosen to retain red galaxies at $z \simeq 6.6 - 6.9$ and mitigate systematics that could be introduced to our inferred age distribution by preferentially selecting young, blue galaxies. Thus, we expect the age distribution that we infer to well sample the ages of UV-bright galaxies in the redshift range of our selection.

We emphasize that our colour criteria are designed to restrict the redshift range of our selection such that \ion{[O}{III]}+\ion{H}{$\beta$} emission falls in the IRAC [3.6] bandpass, leaving the [4.5] filter free of strong nebular lines. This reduces the degeneracy between the relative contributions of nebular emission and continuum emission from old stellar populations to our broadband rest-optical fluxes, therefore improving the reliability of our inferred physical properties.

\subsection{\textit{Spitzer}/IRAC photometry} \label{subsec:irac_phot}

To measure the rest-frame optical photometry for our sample, we use mid-infrared ($\lambda > 3$\,$\mu$m) IRAC imaging from the \textit{Spitzer} Extended Deep Survey \citep{ashby2013}, the \textit{Spitzer}-Cosmic Assembly Deep Near-infrared Extragalactic Legacy Survey \citep{ashby2015}, Star Formation at $4 < z < 6$ from the \textit{Spitzer} Large Area Survey with Hyper-Suprime-Cam \citep{steinhardt2014}, the \textit{Spitzer} Matching survey of the UltraVISTA ultra-deep Stripes \citep{ashby2018}, and Completing the Legacy of \textit{Spitzer}/IRAC over COSMOS (P.I. I. Labb\'{e}). We obtain images from the \textit{Spitzer} Legacy Archive and perform background subtraction using \textsc{Source Extractor}. We then co-add the background subtracted images using the \textsc{mopex} \citep{makovoz2005} software and astrometrically match the resulting mosaics to the \textit{Gaia} reference frame (on which the publicly released HSC SSP, CHORUS, and UltraVISTA images are provided) using the package \textsc{ccmap} from IRAF.

Since the PSF of IRAC is considerably larger than that of the ground-based imaging, we measure the IRAC photometry in 2.8\arcsec\ diameter apertures and use a deconfusion algorithm similar to previous studies \citep[e.g.][]{labbe2010, labbe2013, bouwens2015a} to remove contaminating flux from neighbouring sources. Specifically, we use the same deconfusion technique as \citet{endsley2021b}. First, we convolve the flux profile of each nearby source detected in our \textit{yYJHK$_\text{s}$} $\chi^2$ detection image with the local IRAC PSF measured empirically using unsaturated stars <3\arcsec\ away from each source, where the flux profiles are constructed using the high-resolution \textit{HST}/F814W imaging available over COSMOS \citep{scoville2007}.\footnote{We note that one object, COS-862541, is not within the F814W imaging, but it is fortunately not strongly confused. Thus, we use the deconfusion algorithm described by \citetalias{endsley2021a}, which constructs the flux profile of neighbouring sources for fitting and subtraction slightly differently. We calculate the square root of the $\chi^2$ image using the \textsc{Source Extractor} segmentation map to determine source footprints, then convolve with a 2D Gaussian with full width at half-maximum (FWHM) equal to the quadrature difference of the IRAC FWHM and the median seeing from each band in the detection image.} We then fit the convolved flux profiles to the IRAC image, leaving the total flux as a free parameter, and subtract the best-fitting profile of each neighbouring source before measuring IRAC photometry.

Our IRAC deconfusion and selection is modified slightly from that of \citetalias{endsley2021a}, resulting in a slightly different sample than their original list of COSMOS sources. Compared to the ground-based imaging originally used by \citetalias{endsley2021a} to construct flux profiles, the higher resolution F814W images allow us to obtain smoother neighbour-subtracted residual images for sources in crowded regions. Additionally, we do not place any further requirements on measured IRAC photometry or errors. We only remove five sources from our initial sample due to insufficiently smooth IRAC residuals after deconfusion, which arise due to these objects lying within the IRAC FWHM of a bright neighbouring source.

\subsection{Observed sample properties} \label{subsec:observed_properties}

Our final sample consists of 36 sources selected to lie at $z \simeq 6.6 - 6.9$, adding 16 objects to the original sample of 20 in COSMOS found by \citetalias{endsley2021a}. The selection criteria described in Section\ \ref{subsec:selection} originally identify 41 candidates, five of which are removed from the sample due to poor IRAC residuals as described in Section\ \ref{subsec:irac_phot}. We report the observed $J$-band magnitudes, rest-UV slopes, IRAC [3.6] and [4.5] magnitudes, and IRAC colours ($[3.6] - [4.5]$) for our entire sample in Table\ \ref{tab:observed_properties}, but summarize these quantities below.

Our galaxies have observed $J$-band magnitudes ranging from $J \approx 24.5 - 26.0$ with median of $J = 25.5$ (left panel of Figure\ \ref{fig:properties}). We observe rest-UV slopes, measured by fitting $F_\nu \propto \lambda\,^{\beta + 2}$ to our measured \textit{YJHK$_\text{s}$} photometry, of $\beta \approx -3.0$ to $-0.6$ with median $\beta = -1.9$ (middle panel of Figure\ \ref{fig:properties}). These UV slopes are consistent with moderately fainter objects found in CANDELS \citep{bouwens2014} and the sample of UV-luminous galaxies at $z \sim 7$ identified by \citet{bowler2017} over COSMOS. Finally, we observe relatively blue IRAC $[3.6] - [4.5]$ colours ranging from approximately $-1.5$ to $0.3$ with a median of $-0.4$ (right panel of Figure\ \ref{fig:properties}). As \ion{[O}{III]} and \ion{H}{$\beta$} are confined to the [3.6] bandpass, these blue colours are likely indicative of strong nebular emission lines \citep{smit2014, smit2015, endsley2021a}. We also note that our sample has a relatively high spectroscopic confirmation rate of 33\,per\,cent with redshifts ranging from  $z = 6.538 - 6.883$, generally consistent with our targeted colour selection redshift interval. Ten of our 36 candidates have been detected in \ion{[C}{II]} \citep{smit2018, bouwens2022, endsley2022} \citep[with two also having a \ion{Ly}{$\alpha$} detection;][]{endsley2021a, endsley2021b}, and another two sources have \ion{Ly}{$\alpha$} detections only \citep{endsley2021b}. Notably, the reddest galaxy in our sample (COS-87259) is spectroscopically confirmed via \ion{[C}{II]} and likely hosts a heavily obscured active galactic nucleus \citep{endsley2022}, leading to possible uncertainties on its age and stellar mass. However, we investigate the influence of COS-87259 on our inferred age distribution and find that the distribution is only minimally affected by its inclusion (see Section\ \ref{sec:discussion}).

\section{SED modelling} \label{sec:sed_modelling}

We infer the physical properties of the galaxies in our sample by fitting their photometry in the 16 available optical and infrared bands with two galaxy SED modelling codes: \beagle\ \citep{chevallard2016} and \prospector\ \citep{johnson2021}. Both \beagle\ and \prospector\ self-consistently compute stellar and nebular emission and infer galaxy properties in a Bayesian manner, though they use different physical models. \beagle\ is based on an updated version of the \citet{bruzual2003} stellar population synthesis models \citep[hereafter CB16, see description by][]{gutkin2016}, and \prospector\ utilizes the Flexible Stellar Population Synthesis \citep[\fsps;][]{conroy2009, conroy2010} code. We use \beagle\ and \prospector\ with the default stellar evolution models of their respective stellar population synthesis codes, which are different; the CB16 models are underpinned by isochrones computed by the PAdova and TRieste Stellar Evolution Code \citep[\textsc{parsec};][]{bressan2012, chen2015}, while \fsps\ is based on tracks calculated by the MESA Isochrones and Stellar Tracks project \citep[\textsc{mist};][]{choi2016}. However, we note that \prospector, with \fsps, can also use other stellar isochrones.

We begin by adopting a fiducial set of model parameters to make a one-to-one comparison of the results from \beagle\ and \prospector. We present the full model as implemented in \beagle\ in Section\ \ref{subsec:beagle_modelling} and describe the modifications required to set up the same model in \prospector\ in Section\ \ref{subsec:prospector_csfh}. We also test a nonparametric SFH model with \prospector, described in Section\ \ref{subsec:prospector_nonpar}, which we use to examine the systematics of the SFH model we assume.

\subsection{Constant SFH \beagle\ models} \label{subsec:beagle_modelling}

We fit our sample with \beagle\ version 0.20.4, which calculates both stellar and nebular emission using the \citet{gutkin2016} photoionization models of star-forming galaxies. These, in turn, were derived by combining updated \citet{bruzual2003} stellar population synthesis models (which are based on \textsc{parsec} isochrones) with the photoionization code \textsc{cloudy} \citep{ferland2013}. \beagle\ then uses the Bayesian inference tool \textsc{multinest} \citep{feroz2008, feroz2009, feroz2019} to calculate the posterior probability distributions of the model parameters.

Throughout our modelling process, we adopt a fiducial set of model parameters consisting of the intergalactic medium (IGM) attenuation model of \citet{inoue2014}, a \citet{chabrier2003} initial mass function with a stellar mass range of $0.1 - 300$\,\Msun, fixed dust-to-metal mass ratio of $\xi_d = 0.3$ \citep[similar to that found when assuming a total interstellar metallicity equal to solar, see][]{gutkin2016}, a restricted parameter space for metallicity and ionization parameter, and a less restrictive parameter space for $V$-band optical depth, stellar mass, and age. This allows us to primarily focus on our parameters of interest, stellar mass and age.

We place narrow log-normal priors on stellar metallicity and ionization parameter. Motivated by the properties implied by spectroscopic observations of high-ionization emission lines during reionization \citep[e.g.][]{stark2017, hutchison2019}, our priors are centered on $\mu_{\log(Z / \text{Z}_\odot)} = -0.7$ and $\mu_{\log(U)} = -2.5$ with standard deviations of $\sigma_{\log(Z / \text{Z}_\odot)} = 0.15$ and $\sigma_{\log(U)} = 0.25$. The total interstellar metallicity (dust and gas-phase) is assumed to be the same as the stellar metallicity, and \beagle\ self-consistently treats the depletion of metals onto dust grains \citep{gutkin2016, chevallard2016}.

For age, $V$-band optical depth, and stellar mass, we adopt log-uniform priors. We assume a constant SFH (CSFH) with allowed ages ranging from 1\,Myr to the age of the Universe at the redshift being considered. We adopt an SMC dust prescription \citep{pei1992}, as some studies have found that it matches the IRX--$\beta$ relation observed at $z \sim 2 - 3$ well \citep{bouwens2016, reddy2018}, though we note that alternative dust prescriptions may be favoured by other studies \citep[e.g.][]{mclure2018}. We allow the $V$-band optical depth to vary in the range of $-3.0 \leq \log(\tau_\textsc{v}) \leq 0.7$. Finally, we allow stellar mass to vary in the range of $5 \leq \log(M_* / \text{M}_\odot) \leq 12$.

We place a uniform prior on redshift from $6 \leq z \leq 8$ unless a spectroscopic redshift is available. When a systemic redshift from \ion{[C}{II]} is known, we fix redshift to $z_\ion{[C}{II]}$. If only a \ion{Ly}{$\alpha$} detection is available, we allow redshift to range uniformly from $z_\text{\ion{Ly}{$\alpha$}} - 0.013 \leq z \leq z_\text{\ion{Ly}{$\alpha$}}$ to account for possible velocity offsets up to 500\,km\,s$^{-1}$. Throughout this fitting process, we remove \ion{Ly}{$\alpha$} from the nebular templates, motivated by the small median \ion{Ly}{$\alpha$} equivalent width ($\sim 10$\,\AA) of bright galaxies at $z \sim 7$ found by \citet{endsley2021b}.

We test the validity of this assumption by modelling the ten objects in our sample with systemic redshifts measured from \ion{[C}{ii]} assuming \ion{Ly}{$\alpha$} equivalent widths (EWs) of 0\,\AA, 5\,\AA, 10\,\AA, and 20\,\AA, leaving redshift free to vary in the range of $z = 6 - 8$. In many cases, strong \ion{Ly}{$\alpha$} emission (EW$_{\text{Ly}\,{\alpha}} = 20$\,\AA) is less consistent with the data; for six objects, the EW$_{\text{Ly}\,{\alpha}} = 0$\,\AA\ model recovers the systemic redshift more accurately (within $\text{d}z \leq 0.07$) than the model with EW$_{\text{Ly}\,{\alpha}} = 20$\,\AA\ ($\text{d}z \leq 0.16$). For the remaining four objects, all four models recovered the systemic redshift equally well for one, and two others have been observed to emit \ion{Ly}{$\alpha$} \citep{endsley2021a, endsley2021b}. Moreover, the median inferred ages and stellar masses are relatively insensitive to the inclusion of \ion{Ly}{$\alpha$} in the SED models, only varying by $\sim 0.2$\,dex. We therefore model our sample with \ion{Ly}{$\alpha$} removed from the templates for the remainder of this work.

\begin{figure*}
    \centering
    \includegraphics[width=\textwidth]{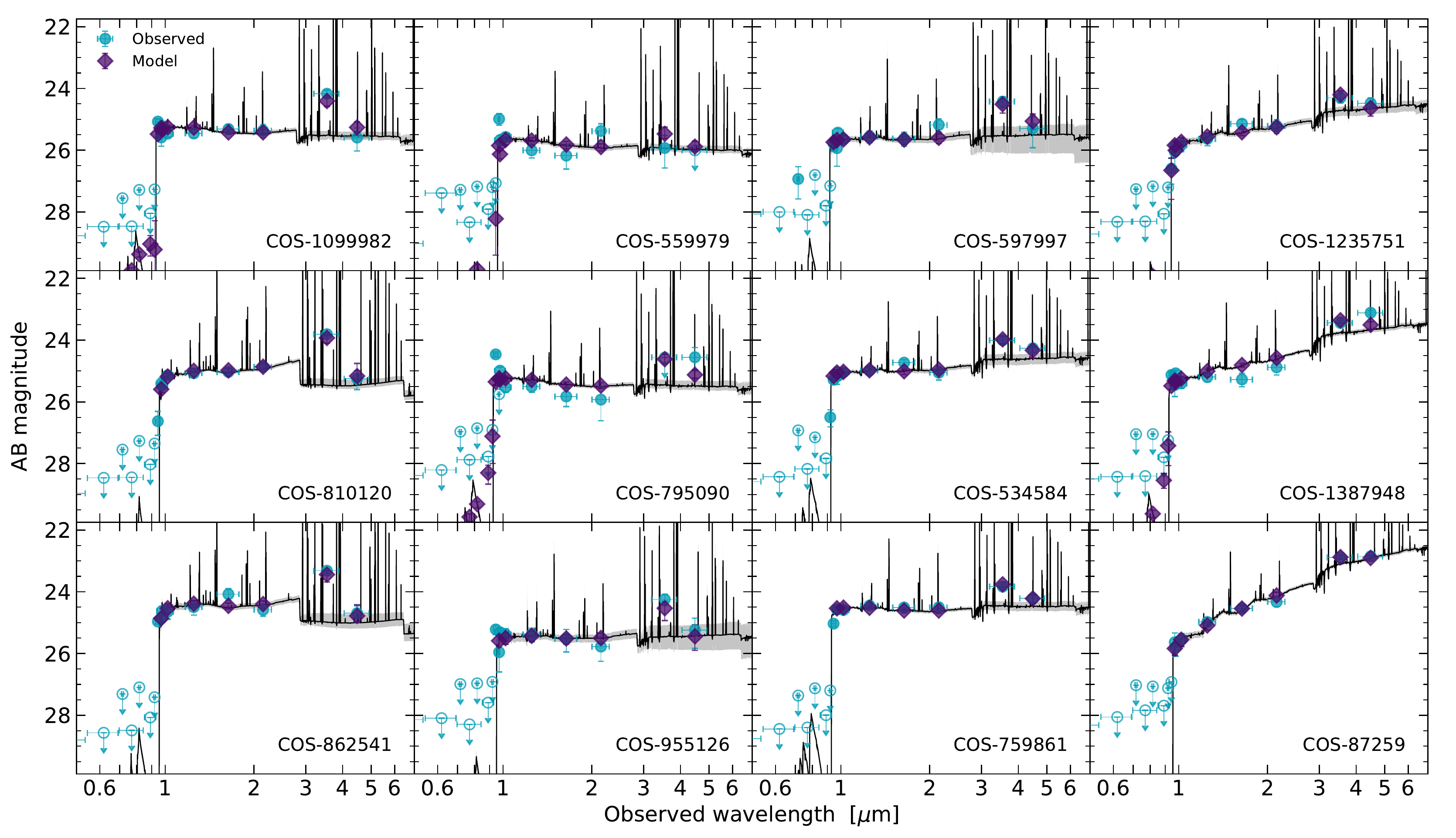}
    \includegraphics[width=\textwidth]{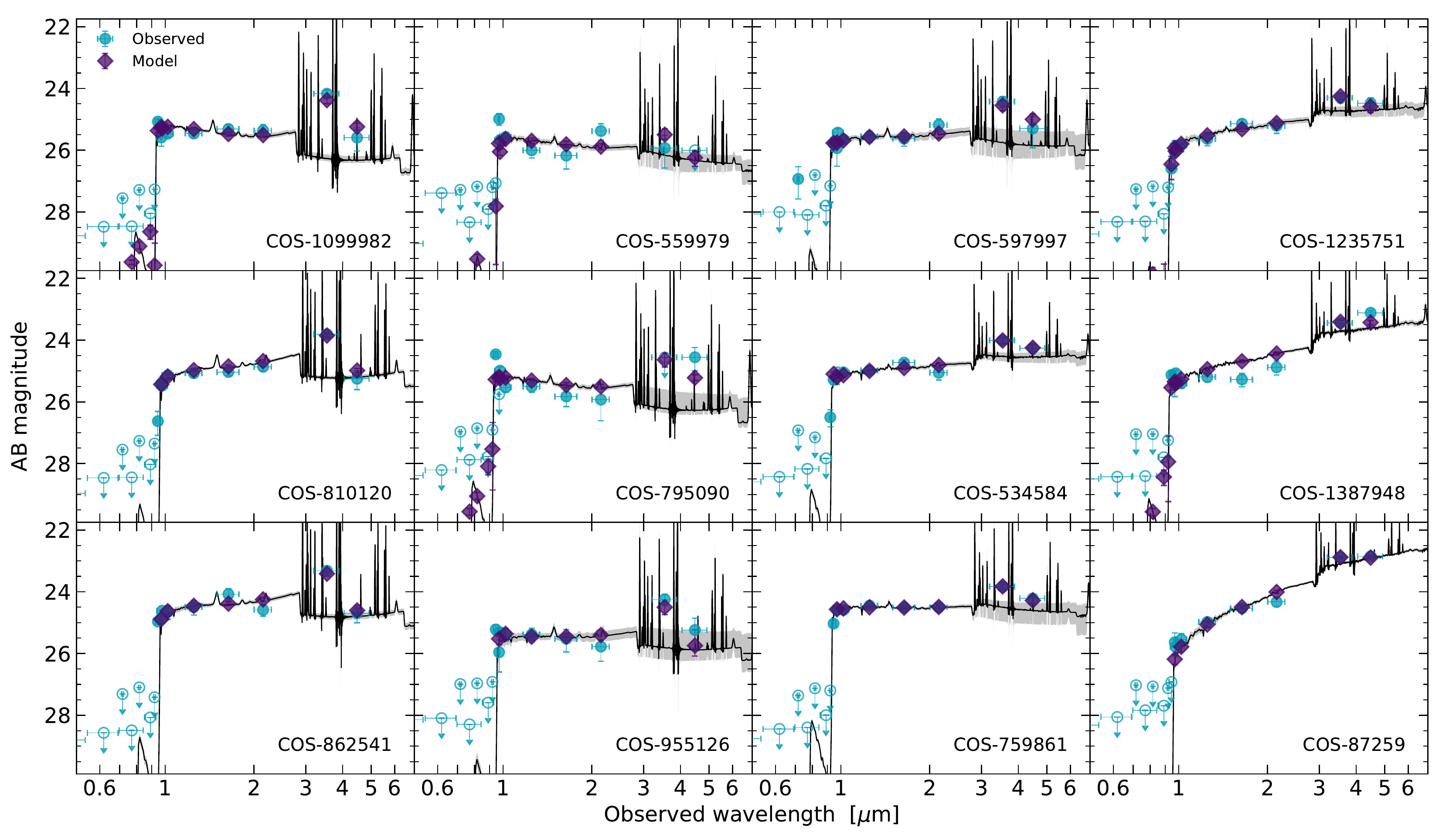}
    \caption{\beagle\ (top) and \prospector\ (bottom) SEDs for a representative subset of our sample. We show increasingly UV-luminous objects from top to bottom. The left column corresponds to objects with the bluest observed IRAC colours ($[3.6] - [4.5] \leq -1$) and the right column shows objects with the reddest $J - [4.5] \geq 1$ colours (approximately capturing the rest-UV--optical colour). We show objects with a variety of rest-UV slopes in the the middle columns. We plot the observed photometry as teal circles, with open symbols denoting $2\sigma$ upper limits for non-detections. We show the median model photometry as purple diamonds, and the median model spectrum and inner 68\,per\,cent credible interval from the posterior distribution of the SED as the black line and grey shaded region in each panel.}
    \label{fig:sample_seds}
\end{figure*}

\subsection{Constant SFH \prospector\ models} \label{subsec:prospector_csfh}

We also fit the photometry of our galaxies using \prospector\ version 1.1.0 \citep{johnson2021}. \prospector\ computes emission from stellar populations using \fsps\ and nebular emission from the \citet{byler2017} models derived by combining \textsc{cloudy} photoionization models with stellar spectra from \fsps. We emphasize that we use the default \textsc{mist} isochrones of \fsps, though other isochrone libraries are also available \citep[for a full list, see appendix A of][]{johnson2021}.

We adopt model parameters as similar as possible to those we used for our \beagle\ fits with minor modifications. To allow the upper limit on the age to vary according to the age of the Universe at the redshift under consideration, we do not fit directly for age, but rather for the fraction of the age of the Universe, $t_\text{frac}$. We adopt a log-uniform prior on $t_\text{frac}$ between $1\,\text{Myr} / \text{age}_\text{Universe}(z)$ and 1, then convert back to age from $t_\text{frac}$ and redshift after fitting is complete. Additionally, \fsps\ implements the \citet{madau1995} prescription for IGM attenuation whereas we have adopted \citet{inoue2014} for our \beagle\ models. In order to enable a more direct comparison between our results from \beagle\ and \prospector, we include an overall scaling factor of the optical depth to approximately match the \citet{inoue2014} model between \ion{Ly}{$\alpha$} and \ion{Ly}{$\beta$}. Finally, we highlight that we remove \ion{Ly}{$\alpha$} from the \prospector\ templates as we did for the \beagle\ models, which requires at least version 1.1.0 of \prospector.

\begin{table*}
\renewcommand{\arraystretch}{1.5}
\centering
\caption{The physical properties inferred for our sample. We model the observed photometry with the Bayesian SED modelling codes \beagle\ \citep{chevallard2016} and \prospector\ \citep{johnson2021}, adopting a CSFH as our fiducial model. We report the median values and marginalized 68\,per\,cent credible intervals for the inferred photometric redshift (or the spectroscopic redshift without errors, if available) and absolute UV magnitude from the \beagle\ models, as the results from \prospector\ are not significantly different. We also report the inferred stellar mass, age, and $V$-band optical depth from both models.}
\label{tab:inferred_properties}
\begin{tabular}{c|c|c|c|c|c|c|c|c} \hline
    \multirow{2}{*}{Object ID} & \multirow{2}{*}{Redshift} & \multirow{2}{*}{$M_\textsc{uv}$} & $\log\left(M_* / \text{M}_\odot\right)$ & $\log\left(M_* / \text{M}_\odot\right)$ & Age [Myr] & Age [Myr] & $\tau_\textsc{v}$ & $\tau_\textsc{v}$ \\ 
    & & & (\beagle) & (\prospector) & (\beagle) & (\prospector) & (\beagle) & (\prospector) \\ \hline\hline
    COS-83688 & $6.70_{-0.04}^{+0.04}$ & $-21.6_{-0.1}^{+0.1}$ & $9.2_{-0.7}^{+0.5}$ & $8.8_{-0.6}^{+0.8}$ & $98_{-86}^{+237}$ & $35_{-31}^{+350}$ & $0.01_{-0.01}^{+0.04}$ & $0.02_{-0.02}^{+0.04}$ \\
    COS-87259 & $6.853$ & $-21.6_{-0.1}^{+0.2}$ & $10.8_{-0.1}^{+0.1}$ & $10.9_{-0.1}^{+0.0}$ & $569_{-196}^{+133}$ & $700_{-151}^{+71}$ & $0.48_{-0.04}^{+0.05}$ & $0.47_{-0.03}^{+0.04}$ \\
    COS-160072 & $6.63_{-0.02}^{+0.02}$ & $-21.4_{-0.1}^{+0.1}$ & $8.7_{-0.2}^{+0.3}$ & $8.5_{-0.3}^{+0.4}$ & $44_{-22}^{+52}$ & $20_{-15}^{+44}$ & $0.00_{-0.00}^{+0.01}$ & $0.00_{-0.00}^{+0.01}$ \\
    COS-237729 & $6.89_{-0.09}^{+0.14}$ & $-21.1_{-0.2}^{+0.2}$ & $9.4_{-0.7}^{+0.3}$ & $9.5_{-0.5}^{+0.3}$ & $251_{-220}^{+319}$ & $270_{-215}^{+310}$ & $0.04_{-0.04}^{+0.10}$ & $0.08_{-0.07}^{+0.09}$ \\
    COS-301652 & $6.62_{-0.02}^{+0.05}$ & $-21.2_{-0.1}^{+0.1}$ & $9.6_{-0.9}^{+0.3}$ & $8.5_{-0.2}^{+1.3}$ & $282_{-263}^{+320}$ & $9_{-7}^{+387}$ & $0.12_{-0.06}^{+0.06}$ & $0.16_{-0.04}^{+0.04}$ \\
    COS-312533 & $6.73_{-0.06}^{+0.06}$ & $-21.1_{-0.1}^{+0.1}$ & $9.5_{-0.5}^{+0.3}$ & $9.4_{-0.5}^{+0.3}$ & $260_{-200}^{+309}$ & $171_{-129}^{+303}$ & $0.08_{-0.07}^{+0.06}$ & $0.14_{-0.05}^{+0.05}$ \\
    COS-340502 & $6.68_{-0.03}^{+0.04}$ & $-21.0_{-0.1}^{+0.2}$ & $8.5_{-0.4}^{+0.5}$ & $8.2_{-0.3}^{+0.7}$ & $34_{-25}^{+96}$ & $15_{-12}^{+97}$ & $0.01_{-0.01}^{+0.02}$ & $0.01_{-0.01}^{+0.03}$ \\
    COS-369353 & $6.729$ & $-21.2_{-0.2}^{+0.2}$ & $9.3_{-0.8}^{+0.6}$ & $8.7_{-0.2}^{+1.0}$ & $61_{-54}^{+307}$ & $8_{-6}^{+184}$ & $0.21_{-0.07}^{+0.06}$ & $0.23_{-0.05}^{+0.06}$ \\
    COS-378785 & $6.83_{-0.06}^{+0.07}$ & $-21.4_{-0.1}^{+0.1}$ & $9.4_{-0.9}^{+0.4}$ & $8.8_{-0.7}^{+0.8}$ & $212_{-196}^{+345}$ & $49_{-46}^{+397}$ & $0.01_{-0.01}^{+0.05}$ & $0.02_{-0.01}^{+0.05}$ \\
    COS-400019 & $6.84_{-0.04}^{+0.04}$ & $-21.4_{-0.1}^{+0.2}$ & $8.5_{-0.4}^{+0.7}$ & $8.6_{-0.4}^{+0.7}$ & $23_{-17}^{+120}$ & $23_{-19}^{+143}$ & $0.01_{-0.01}^{+0.03}$ & $0.01_{-0.01}^{+0.04}$ \\
    COS-469110 & $6.644$ & $-21.6_{-0.2}^{+0.2}$ & $9.6_{-0.8}^{+0.3}$ & $9.5_{-0.6}^{+0.4}$ & $222_{-197}^{+330}$ & $147_{-124}^{+341}$ & $0.05_{-0.04}^{+0.07}$ & $0.12_{-0.06}^{+0.06}$ \\
    COS-486435 & $6.76_{-0.06}^{+0.06}$ & $-20.8_{-0.2}^{+0.2}$ & $9.2_{-0.7}^{+0.5}$ & $9.1_{-0.6}^{+0.6}$ & $71_{-60}^{+232}$ & $47_{-38}^{+226}$ & $0.22_{-0.07}^{+0.06}$ & $0.24_{-0.05}^{+0.05}$ \\
    COS-505871 & $6.62_{-0.03}^{+0.06}$ & $-21.1_{-0.2}^{+0.2}$ & $9.8_{-0.6}^{+0.2}$ & $8.6_{-0.3}^{+1.3}$ & $390_{-323}^{+268}$ & $11_{-9}^{+517}$ & $0.16_{-0.06}^{+0.06}$ & $0.20_{-0.05}^{+0.05}$ \\
    COS-534584 & $6.598$ & $-21.9_{-0.1}^{+0.1}$ & $9.6_{-0.5}^{+0.4}$ & $9.8_{-0.4}^{+0.3}$ & $143_{-107}^{+315}$ & $219_{-158}^{+296}$ & $0.12_{-0.06}^{+0.04}$ & $0.14_{-0.04}^{+0.04}$ \\
    COS-559979 & $6.88_{-0.03}^{+0.03}$ & $-21.3_{-0.1}^{+0.2}$ & $8.6_{-0.2}^{+0.3}$ & $8.3_{-0.3}^{+0.4}$ & $33_{-16}^{+36}$ & $16_{-14}^{+34}$ & $0.00_{-0.00}^{+0.01}$ & $0.00_{-0.00}^{+0.01}$ \\
    COS-593796 & $6.76_{-0.03}^{+0.03}$ & $-21.5_{-0.1}^{+0.1}$ & $8.7_{-0.3}^{+0.4}$ & $8.2_{-0.0}^{+0.0}$ & $34_{-21}^{+55}$ & $2_{-1}^{+3}$ & $0.00_{-0.00}^{+0.01}$ & $0.00_{-0.00}^{+0.01}$ \\
    COS-596621 & $6.76_{-0.04}^{+0.04}$ & $-21.2_{-0.1}^{+0.1}$ & $8.6_{-0.4}^{+0.5}$ & $8.2_{-0.2}^{+0.5}$ & $34_{-24}^{+96}$ & $11_{-9}^{+40}$ & $0.01_{-0.00}^{+0.02}$ & $0.01_{-0.01}^{+0.02}$ \\
    COS-597997 & $6.538$ & $-21.2_{-0.1}^{+0.1}$ & $8.6_{-0.6}^{+0.8}$ & $8.4_{-0.2}^{+0.7}$ & $18_{-14}^{+135}$ & $8_{-6}^{+57}$ & $0.08_{-0.07}^{+0.06}$ & $0.13_{-0.04}^{+0.05}$ \\
    COS-627785 & $6.78_{-0.02}^{+0.02}$ & $-21.7_{-0.1}^{+0.1}$ & $8.9_{-0.4}^{+0.4}$ & $8.3_{-0.0}^{+0.0}$ & $46_{-29}^{+67}$ & $2_{-1}^{+1}$ & $0.01_{-0.00}^{+0.02}$ & $0.01_{-0.00}^{+0.01}$ \\
    COS-637795 & $6.72_{-0.03}^{+0.03}$ & $-21.3_{-0.1}^{+0.1}$ & $9.2_{-0.7}^{+0.4}$ & $9.0_{-0.7}^{+0.6}$ & $118_{-98}^{+234}$ & $74_{-65}^{+323}$ & $0.01_{-0.01}^{+0.03}$ & $0.02_{-0.02}^{+0.05}$ \\
    COS-703599 & $6.64_{-0.03}^{+0.06}$ & $-21.0_{-0.1}^{+0.1}$ & $9.7_{-1.2}^{+0.3}$ & $8.4_{-0.1}^{+0.1}$ & $232_{-224}^{+370}$ & $2_{-1}^{+3}$ & $0.22_{-0.05}^{+0.06}$ & $0.23_{-0.04}^{+0.03}$ \\
    COS-705154 & $6.65_{-0.01}^{+0.01}$ & $-21.4_{-0.1}^{+0.1}$ & $8.7_{-0.3}^{+0.4}$ & $8.4_{-0.3}^{+0.4}$ & $40_{-21}^{+64}$ & $21_{-15}^{+50}$ & $0.00_{-0.00}^{+0.01}$ & $0.01_{-0.00}^{+0.01}$ \\
    COS-759861 & $6.633$ & $-22.4_{-0.1}^{+0.1}$ & $9.5_{-0.4}^{+0.3}$ & $9.4_{-0.7}^{+0.5}$ & $81_{-57}^{+151}$ & $67_{-61}^{+214}$ & $0.06_{-0.04}^{+0.03}$ & $0.08_{-0.03}^{+0.02}$ \\
    COS-788571 & $6.883$ & $-21.5_{-0.1}^{+0.1}$ & $8.4_{-0.1}^{+0.7}$ & $8.6_{-0.1}^{+0.1}$ & $4_{-3}^{+37}$ & $3_{-1}^{+3}$ & $0.11_{-0.04}^{+0.04}$ & $0.22_{-0.03}^{+0.03}$ \\
    COS-795090 & $6.61_{-0.01}^{+0.01}$ & $-21.6_{-0.1}^{+0.1}$ & $8.9_{-0.3}^{+0.4}$ & $8.2_{-0.1}^{+0.5}$ & $52_{-29}^{+78}$ & $3_{-2}^{+30}$ & $0.00_{-0.00}^{+0.01}$ & $0.00_{-0.00}^{+0.01}$ \\
    COS-810120 & $6.854$ & $-21.9_{-0.1}^{+0.1}$ & $8.5_{-0.0}^{+0.0}$ & $8.8_{-0.0}^{+0.0}$ & $2_{-0}^{+1}$ & $2_{-1}^{+1}$ & $0.14_{-0.04}^{+0.04}$ & $0.23_{-0.02}^{+0.02}$ \\
    COS-857605 & $6.73_{-0.04}^{+0.05}$ & $-21.3_{-0.1}^{+0.1}$ & $9.4_{-0.6}^{+0.3}$ & $9.4_{-0.6}^{+0.4}$ & $146_{-116}^{+290}$ & $110_{-92}^{+310}$ & $0.14_{-0.06}^{+0.05}$ & $0.17_{-0.04}^{+0.04}$ \\
    COS-862541 & $6.846$ & $-22.5_{-0.1}^{+0.1}$ & $8.7_{-0.1}^{+0.1}$ & $8.9_{-0.1}^{+0.1}$ & $2_{-0}^{+1}$ & $2_{-1}^{+1}$ & $0.09_{-0.07}^{+0.05}$ & $0.20_{-0.03}^{+0.03}$ \\
    COS-955126 & $6.813$ & $-21.5_{-0.2}^{+0.2}$ & $8.4_{-0.3}^{+1.2}$ & $8.4_{-0.2}^{+1.0}$ & $9_{-6}^{+293}$ & $5_{-3}^{+144}$ & $0.05_{-0.04}^{+0.08}$ & $0.11_{-0.07}^{+0.06}$ \\
    COS-1099982 & $6.65_{-0.01}^{+0.01}$ & $-21.6_{-0.1}^{+0.1}$ & $8.4_{-0.3}^{+0.4}$ & $8.2_{-0.0}^{+0.1}$ & $11_{-6}^{+22}$ & $2_{-1}^{+2}$ & $0.01_{-0.01}^{+0.03}$ & $0.01_{-0.01}^{+0.03}$ \\
    COS-1136216 & $6.63_{-0.01}^{+0.02}$ & $-21.9_{-0.1}^{+0.1}$ & $9.8_{-0.5}^{+0.2}$ & $9.5_{-1.1}^{+0.5}$ & $376_{-280}^{+230}$ & $140_{-137}^{+530}$ & $0.01_{-0.01}^{+0.04}$ & $0.05_{-0.04}^{+0.04}$ \\
    COS-1163765 & $6.61_{-0.02}^{+0.03}$ & $-21.2_{-0.1}^{+0.1}$ & $8.6_{-0.5}^{+0.7}$ & $8.1_{-0.1}^{+0.3}$ & $33_{-25}^{+162}$ & $4_{-2}^{+11}$ & $0.01_{-0.01}^{+0.03}$ & $0.02_{-0.02}^{+0.04}$ \\
    COS-1224137 & $6.685$ & $-22.1_{-0.1}^{+0.1}$ & $9.7_{-0.3}^{+0.3}$ & $9.9_{-0.3}^{+0.2}$ & $131_{-84}^{+183}$ & $201_{-110}^{+205}$ & $0.12_{-0.06}^{+0.04}$ & $0.12_{-0.03}^{+0.03}$ \\
    COS-1235751 & $6.75_{-0.06}^{+0.08}$ & $-21.2_{-0.1}^{+0.1}$ & $9.7_{-0.4}^{+0.3}$ & $9.9_{-0.5}^{+0.2}$ & $199_{-145}^{+273}$ & $262_{-196}^{+324}$ & $0.24_{-0.05}^{+0.05}$ & $0.23_{-0.05}^{+0.05}$ \\
    COS-1304254 & $6.577$ & $-22.0_{-0.1}^{+0.1}$ & $10.0_{-0.5}^{+0.2}$ & $10.1_{-0.6}^{+0.2}$ & $316_{-250}^{+305}$ & $376_{-318}^{+311}$ & $0.16_{-0.05}^{+0.06}$ & $0.17_{-0.04}^{+0.05}$ \\
    COS-1387948 & $6.61_{-0.01}^{+0.01}$ & $-21.7_{-0.1}^{+0.1}$ & $10.5_{-0.1}^{+0.1}$ & $10.6_{-0.1}^{+0.0}$ & $714_{-146}^{+69}$ & $776_{-97}^{+45}$ & $0.26_{-0.03}^{+0.03}$ & $0.25_{-0.02}^{+0.02}$ \\ \hline
\end{tabular}
\end{table*}

\begin{figure*}
    \centering
    \includegraphics[width=\textwidth]{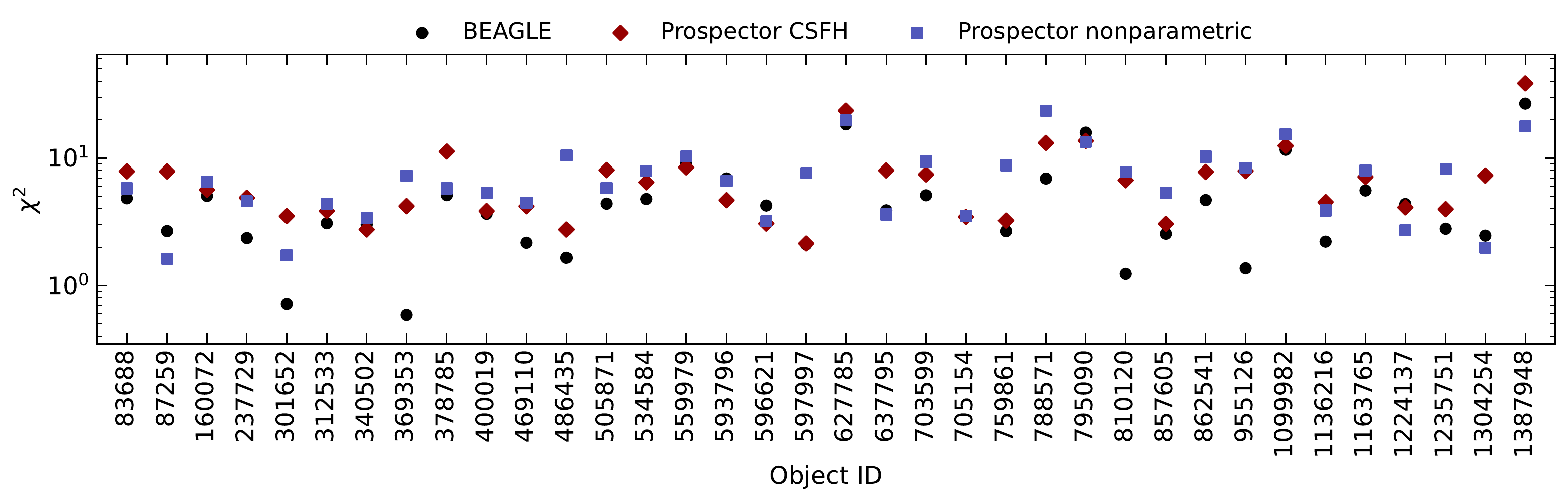}
    \caption{The $\chi^2$ values for the \beagle\ and \prospector\ CSFH models (black circles and red diamonds, respectively), and the \prospector\ nonparametric models (blue squares). Using the best-fitting model photometry, we calculate the $\chi^2$ statistic as $\chi^2 = \sum (F_\text{model} - F_\text{obs})^2 / \sigma_\text{obs}^2$, where $F_\text{model}$ and $F_\text{obs}$ are the observed and model fluxes, respectively, and $\sigma_\text{obs}$ is the observed photometric error. We use only the six broadband filters expected to be redward of the \ion{Ly}{$\alpha$} break at $z = 7$, \textit{YJHK$_\text{s}$}, IRAC [3.6], and IRAC [4.5] (i.e. we expect $\chi^2 = 6$ if the model reproduces the observed photometry exactly within uncertainties of all six filters), and find that all three models generally fit the data well though the \beagle\ models have slightly lower $\chi^2$ values on average.}
    \label{fig:chi_squared}
\end{figure*}

\subsection{Nonparametric \prospector\ models} \label{subsec:prospector_nonpar}

We also explore the systematic impact of the SFH model we assume on the physical properties we infer. For our fiducial models, we have adopted an SFH parametrization that explicitly disallows any star formation prior to the age inferred for the stellar population that dominates the SED. However, light from young, extremely luminous, blue stars can cover up emission from old, faint, red stars (the `outshining' effect), limiting our ability to extract the full SFH from the SED \citep[e.g.][]{papovich2001, pforr2012, conroy2013}. We emphasize that the problem of outshining is particularly egregious during reionization, when a significant number of galaxies have large specific star formation rates (sSFRs) and strong emission lines \citepalias{endsley2021a} that suggest they are undergoing a rapid upturn of star formation (possibly a burst, or at least a rapidly rising SFH). We acknowledge that UV selections preferentially identify UV-bright, unobscured star-forming systems, potentially leading to higher sSFRs than samples at fixed stellar mass. Nevertheless, sSFRs are observed to increase with increasing redshift at fixed mass \citep[e.g.][]{khusanova2021, stefanon2022, topping2022}, consistent with expectations from rising baryon accretion rates at earlier times. Thus, our relatively simple fiducial SFH model may be missing early epochs of star formation activity in the growing fraction of high-redshift galaxies being observed in this phase. Fortunately, more flexible `nonparametric' models can offer additional insights into the SFHs of these systems and may be able to capture complex features (e.g. rapid bursts or quenching events) that cannot be modelled by simple parametric SFHs with a functional form \citep[for detailed discussions of both parametric and nonparametric SFH models, see][]{carnall2019, leja2019a, lower2020, johnson2021, tacchella2022}. In particular, such flexible models may provide constraints on the amount of early star formation that could be physically plausible.

With this goal in mind, we fit our sample with \prospector's nonparametric SFHs \citep{leja2019a, johnson2021}, leaving all other model parameters unchanged. The nonparametric SFHs implemented in \prospector\ are step functions in time where the mass formed in each of $N$ bins of lookback time are free parameters of the model. For our models, we use $N = 8$ bins spanning the time of observation to a fixed redshift at which star formation starts; we adopt $z_\text{form} = 20$. The two most recent age bins are fixed to range from $0 - 3$\,Myr and $3 - 10$\,Myr and the remainder are spaced evenly in logarithmic time. We choose to fix the two recent age bins in order to allow the models to fit galaxies with extremely blue IRAC colours, which are unambiguously linked to very strong nebular emission associated with young stars in the redshift range of our selection.

The inferences from nonparametric SFH models can depend strongly on the prior assumed \citep{leja2019a, tacchella2022}, since the photometry is frequently only minimally informative about older, less luminous stellar populations. In this work, we adopt a prior that distributes the stellar mass formed equally over time and disfavors extremely rapid, potentially unphysical changes in the star formation rate (SFR); this is the `continuity' prior in \prospector, which fits for the logarithm of the ratios of SFRs between adjacent time bins with a Student's \textit{t}-distribution prior with dispersion $\sigma = 0.2$ \citep[for further details, see][]{leja2019a}. We also reiterate that we have fixed the redshift at which star formation starts to $z_\text{form} = 20$, which strongly restricts the amount of time available for star formation and the build up of a galaxy's stellar mass. We discuss alternatives to these priors and the impact on stellar masses in Appendix\ \ref{appendix:nonpar_priors}.

\begin{figure*}
    \centering
    \includegraphics[width=\textwidth]{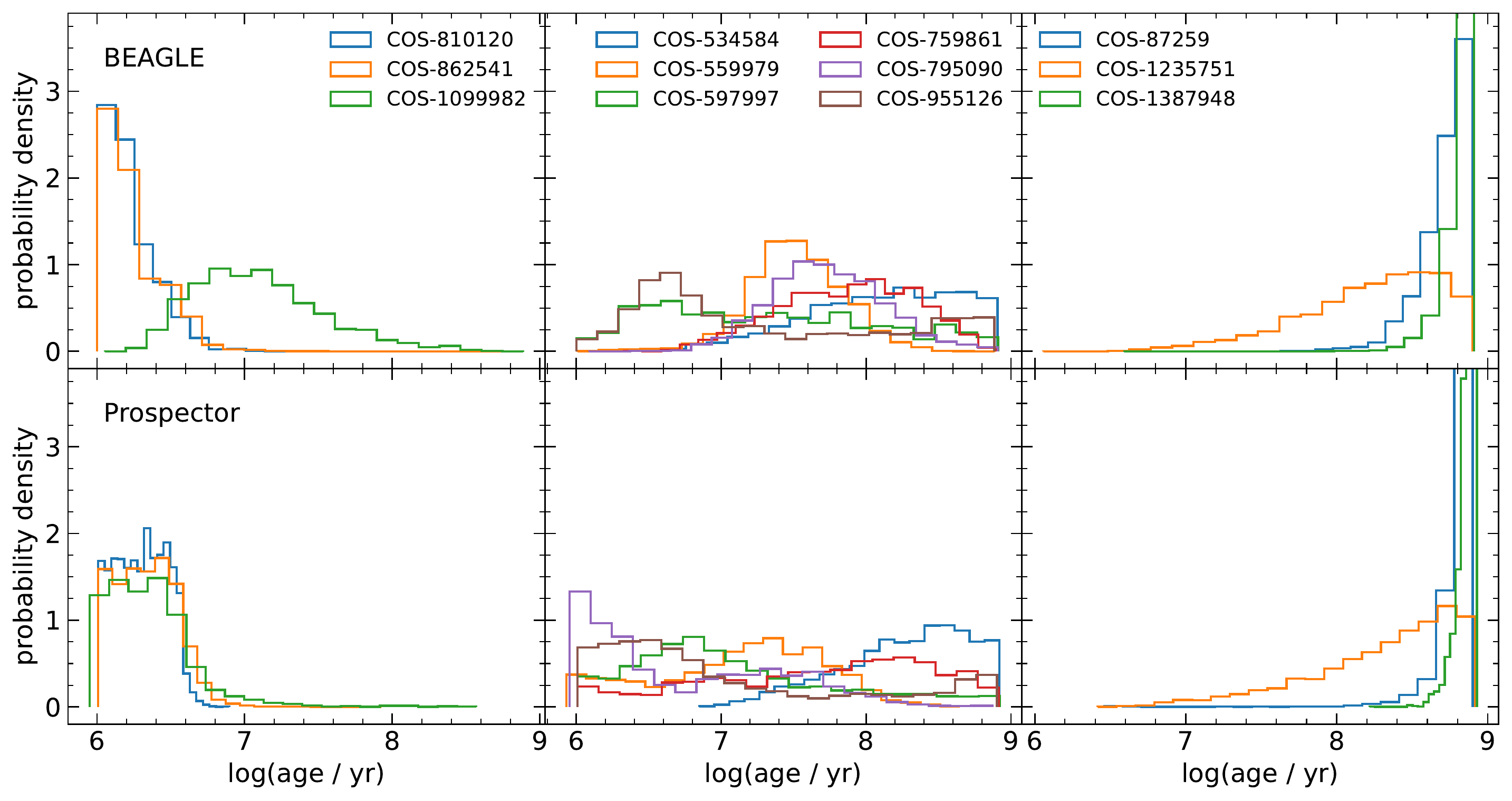}
    \caption{Posterior probability distributions for age derived from our CSFH models for a representative subset of our sample (the same as in Figure\ \ref{fig:sample_seds}). Rows correspond to our two SED modelling codes, \beagle\ in the top row and \prospector\ in the bottom row. We show sources with the most blue IRAC colours (indicative of strong \ion{[O}{III]}+\ion{H}{$\beta$} in the redshift range of our selection) in the left column, and sources with the most red $J - [4.5]$ colours (which we expect to approximately probe the rest-UV--optical colour, since the 4.5\,$\mu$m bandpass is largely free of strong nebular emission) in the right column. Finally, we show examples of more typical objects (selected to display a range of $J$-band magnitudes, observed rest-UV slopes, and IRAC colours) in the center. We generally infer young ages of $\lesssim 10$\,Myr for the objects with blue IRAC colours and older ages $\gtrsim 200$\,Myr for the galaxies with the most red $J - [4.5]$ colours. Meanwhile, the more intermediate objects have less well-constrained ages that span a wide range of the parameter space. We note that, while \beagle\ and \prospector\ generally infer similar ages, \prospector\ finds a very young age ($\sim 2$\,Myr) for COS-1099982 while \beagle\ infers an older age of $\sim 11$\,Myr. This object exemplifies a difference between \beagle\ and \prospector, where \prospector\ infers significantly younger ages than \beagle\ for a subset of objects in our sample, discussed in detail in Section\ \ref{subsec:code_compare}.}
    \label{fig:age_pdfs}
\end{figure*}

\section{Inferred galaxy properties} \label{sec:galaxy_properties}

In this section, we present the physical properties of our sample that we infer from our two Bayesian galaxy SED modelling tools, \beagle\ and \prospector. We summarize the results of the fiducial models from both codes in Section\ \ref{subsec:fiducial_models}, then compare their inferred parameters in Section\ \ref{subsec:code_compare}. We also present the results of our \prospector\ nonparametric SFH models in Section\ \ref{subsec:nonpar_results}.

\subsection{Fiducial models} \label{subsec:fiducial_models}

We show the SEDs from \beagle\ and \prospector\ for a representative subset of our sample in Figure\ \ref{fig:sample_seds}, approximately increasing in UV luminosity from top to bottom row. In the left column, we show objects that have very blue observed IRAC colours of $[3.6] - [4.5] \leq -1$, indicative of strong nebular emission lines. In the right column, we show sources with very red $J - [4.5] \geq 1$ colours, which approximately probes the rest-UV--optical colour and suggests the possible presence of a Balmer break. Finally, we show more intermediate sources with a variety of rest-UV slopes in the middle columns.

In Figure\ \ref{fig:chi_squared}, we show the $\chi^2$ statistic for both codes (and both SFH models for \prospector; see Sections\ \ref{subsec:beagle_modelling} and \ref{subsec:prospector_nonpar} for full descriptions of the SFH models). We calculate $\chi^2$ as $\sum (F_\text{model} - F_\text{obs})^2 / \sigma_\text{obs}^2$, where $F_\text{obs}$ and $\sigma_\text{obs}$ are the observed flux and error and $F_\text{model}$ is the flux from the best-fitting model. We calculate $\chi^2$ with the broadband filters expected to be redward of the \ion{Ly}{$\alpha$} break at $z = 7$ in order to (1) minimize the impact of \ion{Ly}{$\alpha$}, which will be most important in narrow and intermediate band filters, and (2) only capture differences from the stellar and nebular emission models rather than differences in the IGM attenuation model. Specifically, we calculate $\chi^2$ with six filters (\textit{YJHK$_\text{s}$}, IRAC [3.6], and IRAC [4.5]), so we expect $\chi^2 \leq 6$ if the SED models reproduce the observed photometry within uncertainties of all filters under consideration. In general, all three models from both codes perform comparably well and find acceptable fits to the majority of the sample, though the \beagle\ models have slightly lower $\chi^2$ values on average and \beagle\ and \prospector\ sometimes achieve the fits in very different ways; we discuss these differences further in Section\ \ref{subsec:code_compare}.

In Table\ \ref{tab:inferred_properties}, we report the following properties inferred by our \beagle\ and \prospector\ models: photometric redshift ($z_\text{phot}$), absolute UV magnitude ($M_\textsc{uv}$, calculated by integrating the median model spectrum over a tophat with value unity and width 100\,\AA\ centered on rest-frame 1500\,\AA), stellar mass ($M_*$), and age (defined as the time since the first star formed). With \beagle, we infer photometric redshifts ranging from $6.61 \leq z_\text{phot} \leq 6.91$, consistent with our expectations from our colour selection. Our inferred UV magnitudes vary from $-22.5 \leq M_\textsc{uv} \leq -20.8$ with median $M_\textsc{uv} = -21.4$, corresponding to $\sim 1 - 5M_\textsc{uv}^*$ with median of $\sim 2M_\textsc{uv}^*$.\footnote{We assume $M_\textsc{uv}^* = -20.6$ from the double power law UV LF at $z \sim 7$ from \citet{bowler2017}.} The inferred stellar masses span $8.4 \leq \log\left(M_* / \text{M}_\odot\right) \leq 10.8$ with median $\log\left(M_* / \text{M}_\odot\right) = 9.1$, and the inferred ages range from $\sim 2 - 670$\,Myr with median age $\sim 64$\,Myr. Our \prospector\ results are similar: we infer photometric redshifts of $6.58 \leq z_\text{phot} \leq 6.92$, absolute UV magnitudes of $-22.4 \leq M_\textsc{uv} \leq -20.9$ (median $M_\textsc{uv} = -21.4$), stellar masses of $8.1 \leq \log\left(M_* / \text{M}_\odot\right) \leq 10.9$ (median $\log\left(M_* / \text{M}_\odot\right) = 8.7$), and ages spanning $\sim 2-752$\,Myr (median age $\sim 20$\,Myr).

In Figure\ \ref{fig:age_pdfs}, we show the posterior probability distributions for age inferred from our fiducial \beagle\ and \prospector\ models for the same objects whose SEDs we show in Figure\ \ref{fig:sample_seds}. For the objects with blue IRAC colours, we infer ages $\lesssim 10$\,Myr with both \beagle\ and \prospector, and for the objects with red rest-UV--optical colours, we infer ages $\gtrsim 200$\,Myr, if not older. For the more intermediate objects, our models frequently find broad age posteriors that span the allowed parameter space of 1\,Myr to nearly 1\,Gyr.

In general, the ages inferred for the objects with the most extreme $[3.6] - [4.5]$ and $J - [4.5]$ colours are more tightly constrained than for the rest of the sample. By the design of our selection, we expect \ion{[O}{iii]} and \ion{H}{$\beta$} to be confined to IRAC [3.6]. Therefore, a very blue $[3.6] - [4.5]$ colour is a clear indicator of intense \ion{[O}{III]}+\ion{H}{$\beta$} emission produced when the UV and optical spectrum is dominated by light from by young stellar populations (left columns of Figures\ \ref{fig:sample_seds} and \ref{fig:age_pdfs}). Conversely, since IRAC [4.5] probes the rest-optical continuum while the observed near-infrared photometry corresponds to the rest-UV, a very red $J - [4.5]$ colour combined with a relatively flat (or blue) rest-UV slope suggests the presence of a prominent Balmer break created when the UV and optical spectrum is dominated by light from an old stellar population (right columns of Figures\ \ref{fig:sample_seds} and \ref{fig:age_pdfs}). However, in the absence of these extreme colours, the low signal-to-noise of the observed photometry makes the degeneracy between nebular emission lines and stellar continuum in the rest-optical more difficult to break, leading to much larger ranges of allowed ages (middle columns of Figures\ \ref{fig:sample_seds} and \ref{fig:age_pdfs}).

\begin{figure}
    \centering
    \includegraphics[width=\columnwidth]{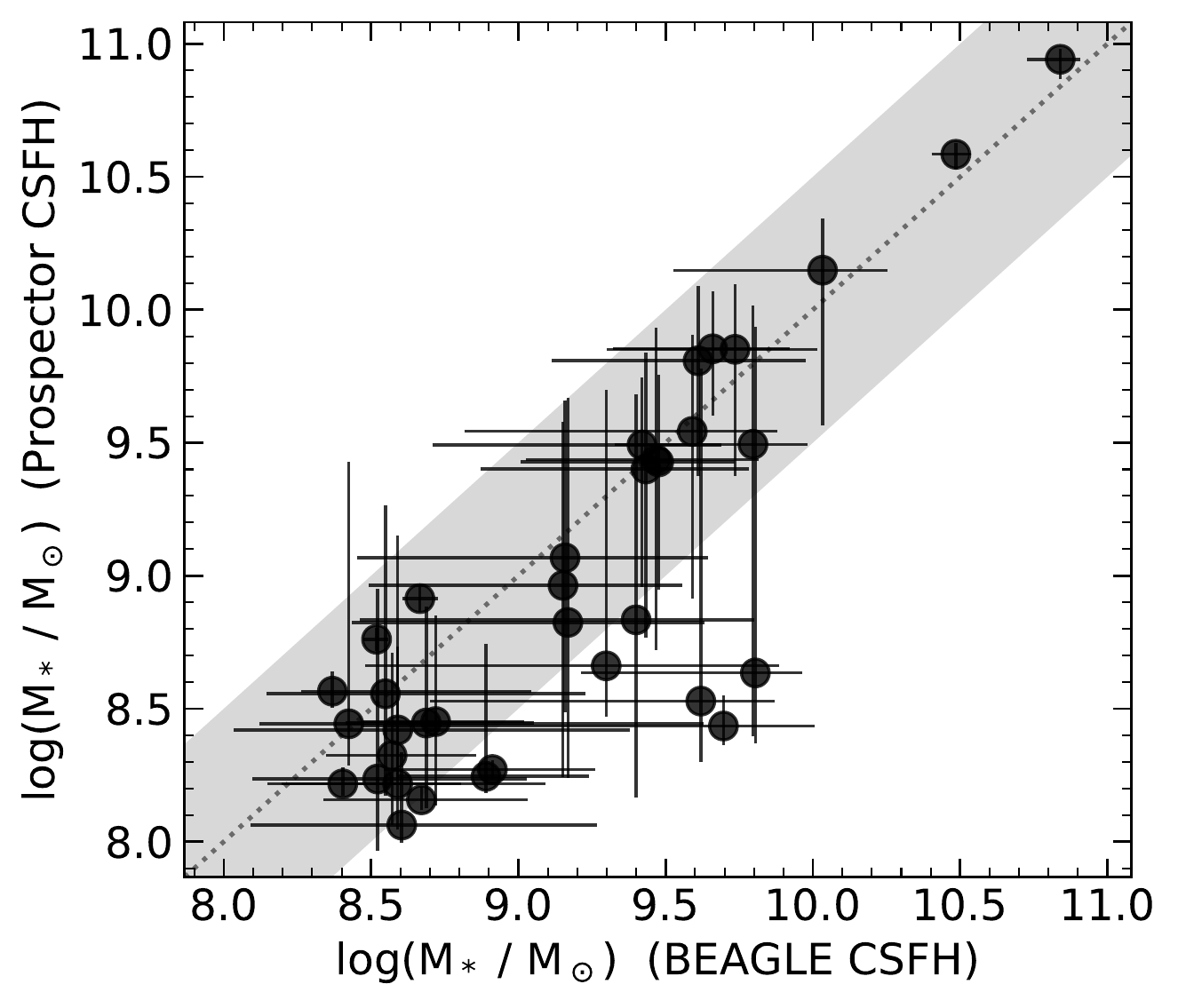}
    \includegraphics[width=\columnwidth]{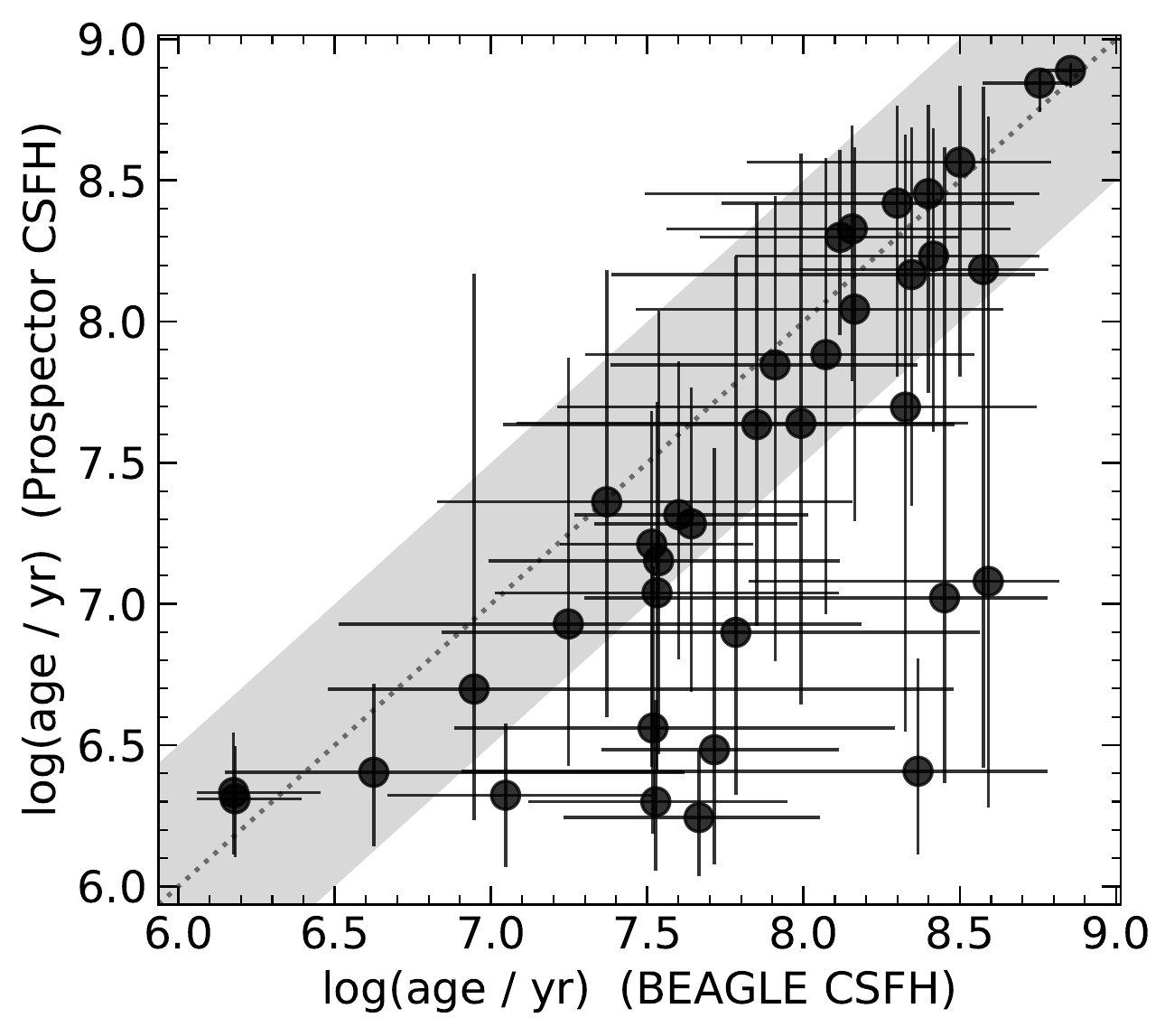}
    \caption{Comparison of the inferred stellar masses (top) and ages (bottom) from our fiducial CSFH \beagle\ and \prospector\ models. We show the 1--1 relation ($\pm 0.5$\,dex) as the dotted line (grey shaded region). Both codes broadly infer similar stellar masses with scatter on the order of $\sim 0.2$\,dex. Despite general agreement for stellar mass, \prospector, with \fsps\ using \textsc{mist} isochrones, finds slightly younger ages on average, and we specifically highlight a subset of sources for which \prospector\ infers ages up to a factor of ten younger than the ages \beagle\ infers with CB16 underpinned by \textsc{parsec}. We discuss these objects in detail in Section\ \ref{subsec:code_compare}.} 
    \label{fig:code_compare}
\end{figure}

\subsection{Comparison of \beagle\ and \prospector} \label{subsec:code_compare}

Though both codes generally fit the data well and yield similar results, they can differ significantly in some instances, which we now explore in further detail. Most strikingly, \beagle\ and \prospector\ find dramatically different ages for about 20\,per\,cent of our sample, though they still infer similar stellar masses. For these sources, \prospector\ infers ages of $1 - 10$\,Myr while \beagle\ finds solutions with ages $10 - 100$\,Myr. The SED morphologies produced by these ages are very different, especially in the rest-optical, and therefore imply markedly different interpretations of the observations. For example, an extremely young age requires that significant [3.6] flux is produced solely by very strong nebular emission lines, while an older age allows for a larger contribution from continuum flux from an old stellar population.

\begin{figure}
    \centering
    \includegraphics[width=\columnwidth]{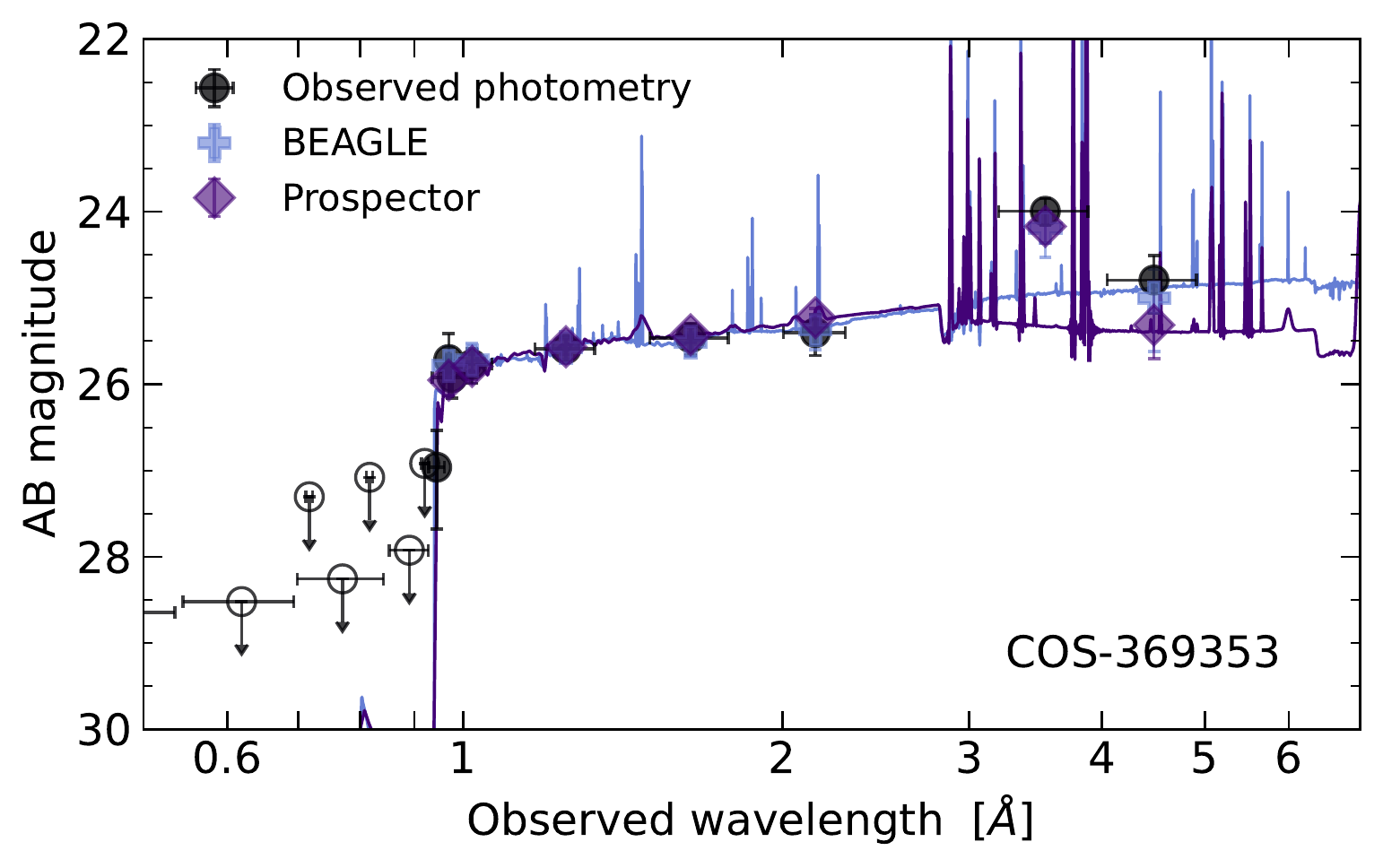}
    \caption{Comparison of the \beagle\ (blue plus signs) and \prospector\ (purple diamonds) fits of an object for which \beagle\ inferred a significantly older age than \prospector. Both codes generally have similar spectra in the rest-UV and reproduce the observed IRAC [3.6] excess well, but \prospector\ underpredicts the IRAC [4.5] flux significantly while \beagle\ matches it well. We attribute this difference to slightly fainter rest-optical stellar continuum and emission lines in the \fsps\ models used by \prospector\ compared to the CB16 models used by \beagle, which requires \prospector\ to go to younger ages with weaker rest-optical continuum to reproduce the strong emission lines.}
    \label{fig:cos369353_compare}
\end{figure}

To illustrate the similarities and differences between the codes, we compare our inferred stellar masses and ages from our fiducial \beagle\ and \prospector\ models in Figure\ \ref{fig:code_compare}. The 1--1 relation ($\pm 0.5$\,dex) is shown as the dotted line (grey shaded region). For the majority of the sample, both models tend to recover similar stellar masses with scatter on the order of $\sim 0.2$\,dex. As seen in the bottom panel of Figure\ \ref{fig:code_compare}, both models also generally recover similar ages, with the notable exception of the 20\,per\,cent of our sample for which \prospector, with \fsps\ using \textsc{mist} isochrones, infers ages up to $\sim 10$ times younger than \beagle, with CB16 using \textsc{parsec} isochrones (though we note that some of these are still consistent within the uncertainties). These objects are frequently observed to have [3.6] excesses but relatively flat rest-UV--optical colours (we show an example in Figure\ \ref{fig:cos369353_compare}). The \prospector\ models tend to match both the rest-UV photometry and the [3.6] excesses within errors but sometimes underpredict the [4.5] flux.

\begin{figure}
    \centering
    \includegraphics[width=\columnwidth]{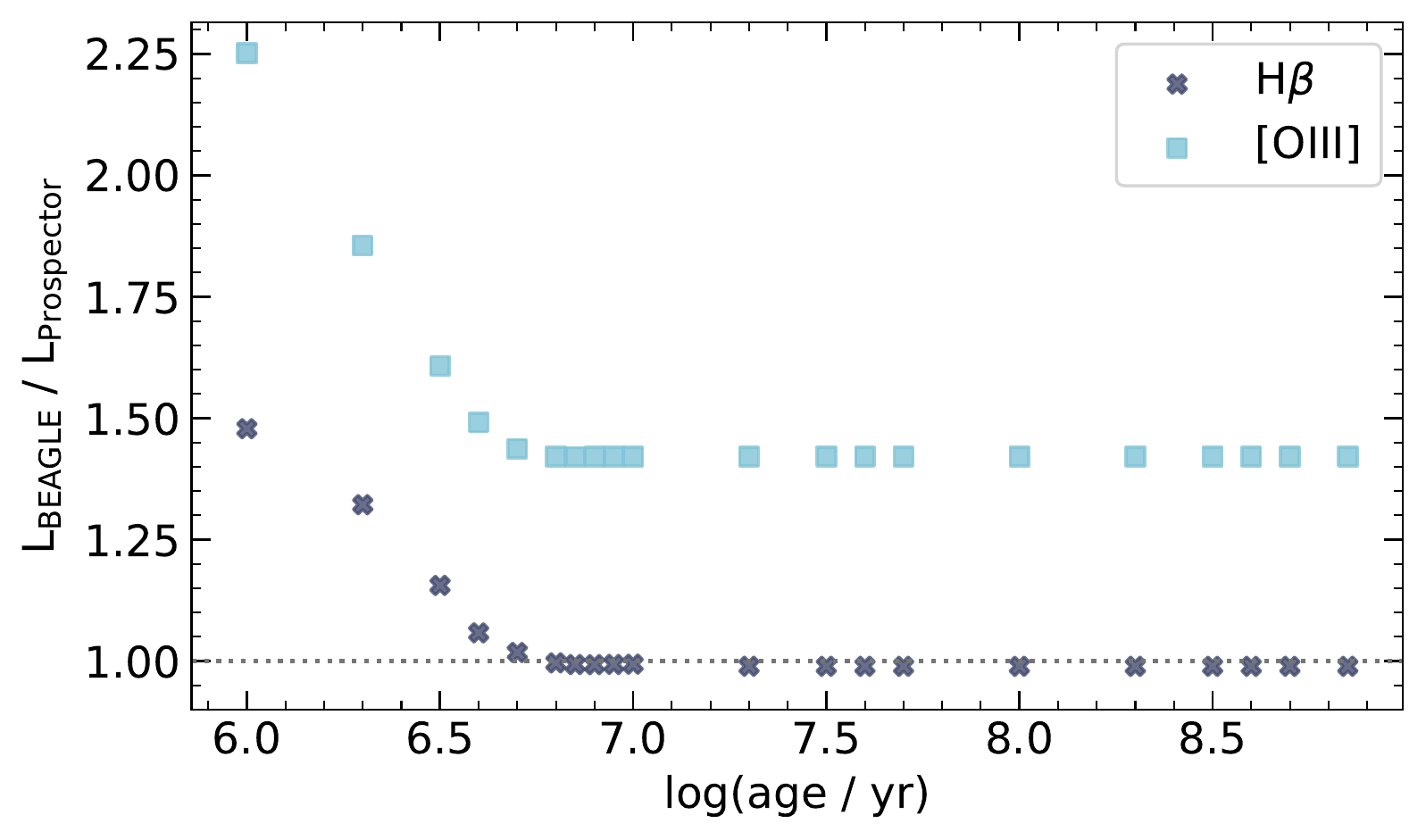}
    \includegraphics[width=\columnwidth]{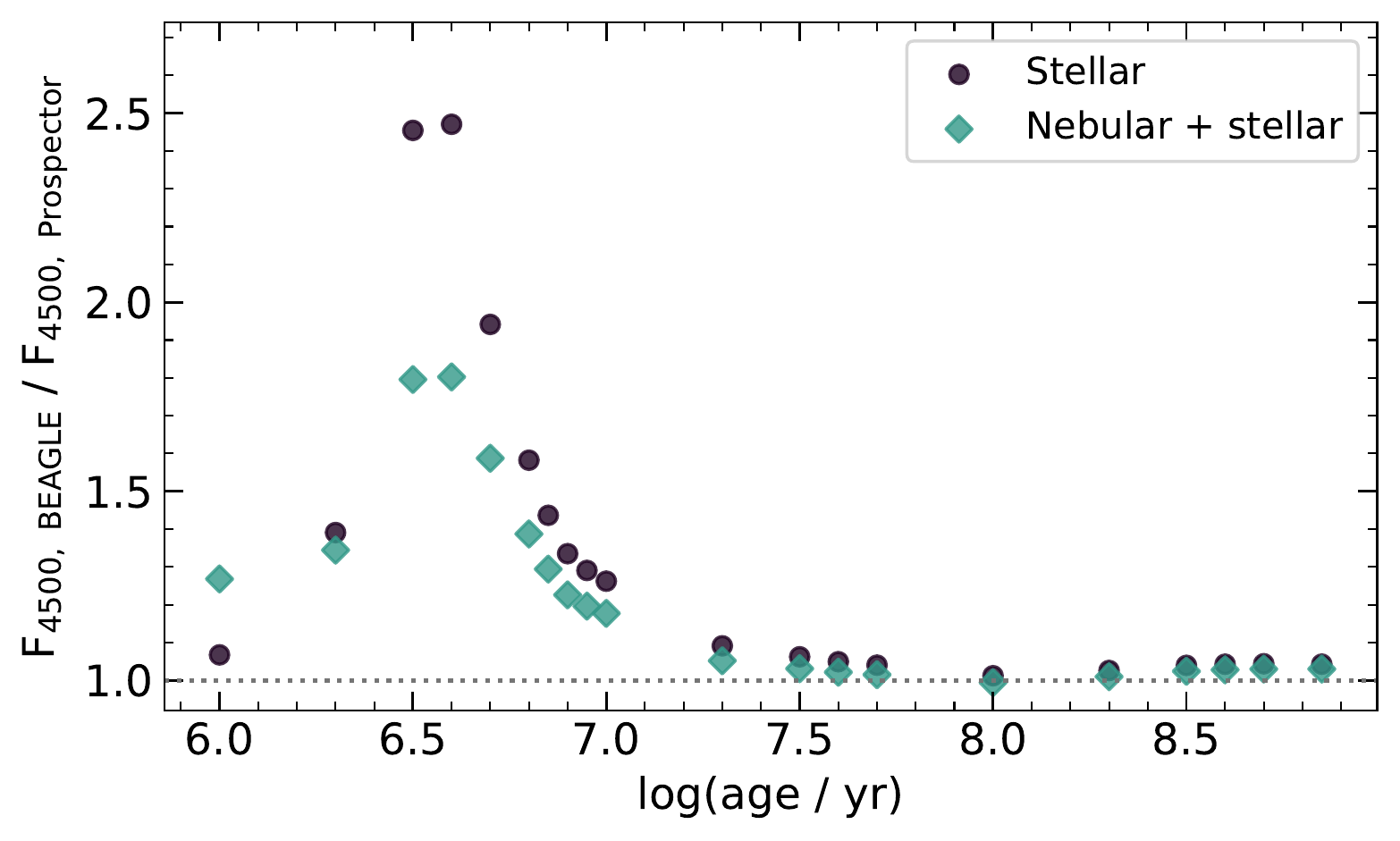}
    \caption{The ratio of line luminosities (top) and of the rest-optical ($\lambda_\text{rest} = 4500$\,\AA) continuum flux (bottom) from \beagle\ and \prospector\ CSFH models as a function of age. We adopt $\text{SFR} = 10$\,M$_\odot$\,yr$^{-1}$, $\tau_\textsc{v} = 0$, $Z = 0.2$\,Z$_\odot$, and $\log(U) = -2.5$. \beagle\ has brighter lines at all ages, both \ion{H}{$\beta$} and \ion{[O}{iii]} at young ages $\lesssim 5$\,Myr and just \ion{[O}{iii]} at older ages. \beagle\ also has brighter rest-optical continuum, particularly at ages $\lesssim 10$\,Myr. The difference between the total rest-optical continuum between the two codes as a function of age is qualitatively similar to the difference in the stellar continuum, hence we attribute the difference to variations between the stellar evolution tracks we adopt in the stellar population synthesis models for \beagle\ and \prospector\ (CB16 with \textsc{parsec}, and \fsps\ with \textsc{mist}, respectively).}
    \label{fig:model_ratios}
\end{figure}

We investigate the source of these differences by generating CSFH \beagle\ and \prospector\ models with ages from $1 - 900$\,Myr. We normalize to a star formation rate of $\text{SFR} = 10$\,M$_\odot$\,yr$^{-1}$ and fix $\tau_\textsc{v} = 0$, $Z = 0.2$\,Z$_\odot$, and $\log(U) = -2.5$. The rest-UV is very similar between the two models, but we identify two important differences in the rest-optical. Compared to \prospector, \beagle\ has (1) larger \ion{[O}{III]} and \ion{H}{$\beta$} luminosities at young ages, and larger \ion{[O}{III]} luminosities at all ages (top panel of Figure\ \ref{fig:model_ratios}), and (2) stronger rest-optical continuum at all ages, but particularly at the youngest ones (bottom panel of Figure\ \ref{fig:model_ratios}). Specifically, the rest-optical continuum flux at $\lambda_\text{rest} = 4500$\,\AA\ from \beagle\ is nearly twice that of \prospector\ at $\sim 3$\,Myr, after which the difference decreases steadily to a factor of $\sim 1.2$ at 10\,Myr and asymptotes to a factor of $\sim 1.1$ brighter than \prospector\ at ages $\gtrsim 20$\,Myr.

Deeper investigation into the models reveals that the difference in rest-optical continuum fluxes is at least partially due to the different stellar evolution tracks used by \beagle\ and \prospector. The \textsc{mist} isochrones used by default by \fsps\ \citep{choi2016, choi2017}, which we adopt in this work, produce stellar spectra that are both less luminous and have significantly bluer slopes in the rest-optical compared to the \textsc{parsec} isochrones used by the CB16 models \citep{bressan2012, chen2015}, especially from $\sim 3 - 10$\,Myr in the context of a CSFH. In this age range, the \beagle\ rest-optical continuum fluxes based on CB16 with \textsc{parsec} isochrones can be more than twice those of \prospector\ underpinned by \fsps\ with \textsc{mist}.

These differences in the rest-optical nebular emission lines and stellar continuum allow us to understand the divergent ages inferred by \prospector\ and \beagle\ for $\sim 20$\,per\,cent of our sample. As previously outlined, the \prospector\ models tend to match the near-infrared and IRAC [3.6] photometry but underpredict [4.5] for these sources, while the \beagle\ models match both [3.6] and [4.5] (Figure\ \ref{fig:cos369353_compare}). The \beagle\ models, with their stronger lines and continuum, can achieve an IRAC excess at intermediate CSFH ages ($\sim 10 - 100$\,Myr), reproducing the observed SEDs well. However, at those same intermediate ages, our \prospector\ models cannot model the [3.6] excess with their weaker lines, requiring much younger ages ($\lesssim 10$\,Myr) at the expense of modelling the continuum in [4.5]. We note that these emission line strengths are subject to systematics in the coupling of the stellar and interstellar properties in the models, and accounting for effects such as alpha enhancement may bring the solutions into closer agreement.

We emphasize that the particular combination of a strong [3.6] excess combined with flat or red rest-UV--optical colour that leads to younger \prospector-inferred ages only occurs in a subset of our sample. The differences are important to note, as these objects comprise about 20\,per\,cent of our sample, and the physical picture implied by the two model solutions are very different (extremely strong nebular emission lines coming from a very young stellar population as opposed to a more mature population producing weaker emission lines and stronger continuum emission in the rest-optical), motivating the need for deeper investigations of the differences in the physical models that lead to these diverse solutions. However, we highlight that despite the differences in age, the inferred stellar masses (and $V$-band optical depths) of these systems are nevertheless largely similar. Furthermore, barring this particular observational subset of objects, the fiducial models of both \beagle\ and \prospector\ find generally comparable ages as well as stellar masses.

\subsection{Nonparametric SFH models} \label{subsec:nonpar_results}

Thus far, we have focused on the results from our fiducial models, which assume a CSFH with a variable age. However, a known limitation of this modelling is the problem of outshining: when present, young stellar populations dominate the light in the rest-UV and optical, possibly hiding an old stellar population that is only moderately luminous. Accordingly, our CSFH models are potentially missing significant contributions from earlier epochs of star formation, \citep[the possibility and implications of which has recently been under consideration at a variety of redshifts;][]{roberts-borsani2020, topping2022, tacchella2022, tang2022}.

For sources that appear to be dominated by stars formed in a recent uptick of star formation and are subsequently inferred to have extremely young ages of $\lesssim 10$ Myr, this issue is especially troublesome. Our CSFH models for these objects are entirely insensitive to any earlier episodes of star formation, potentially leading to underpredicted stellar masses and overpredicted sSFRs. Moreover, at the early cosmic times under consideration ($z \gtrsim 7$), these objects comprise a much more significant fraction of the galaxy population than at lower redshifts \citep{endsley2021a, boyett2022}, so systematics in modelling such an important subset of the population could impact quantities such as the stellar mass function and the star-forming main sequence.

\begin{figure}
    \centering
    \includegraphics[width=\columnwidth]{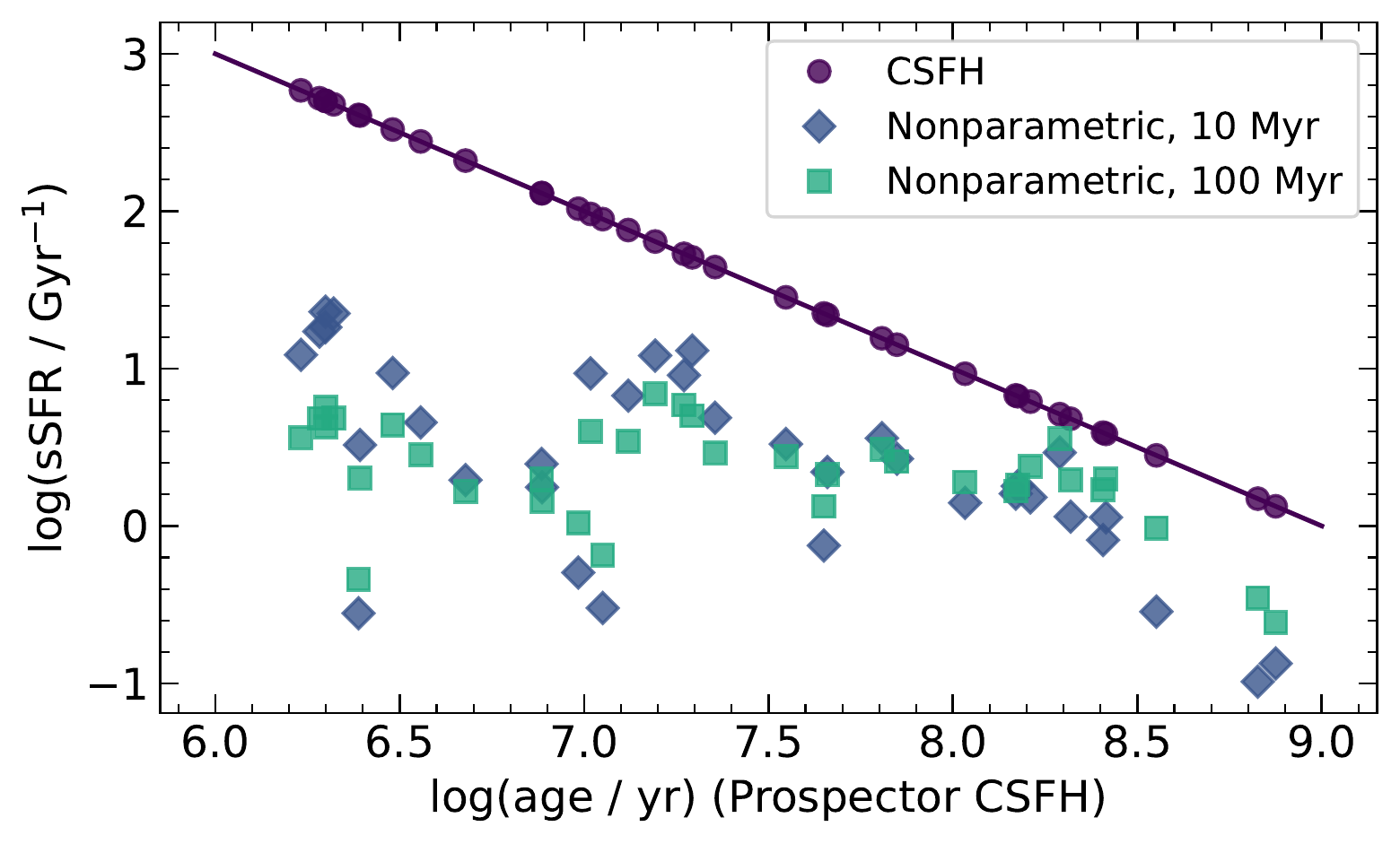}
    \caption{The sSFRs for each object inferred by the CSFH and nonparametric SFH models with continuity prior as a function of \prospector\ CSFH-inferred age. We show the CSFH sSFRs as purple circles and nonparametric sSFRs with SFR averaged over the most recent 10\,Myr and 100\,Myr as blue diamonds and green squares, respectively. The nonparametric sSFRs are systematically lower than the sSFRs inferred by the CSFH models, most significantly at the youngest ages where the nonparametric sSFR for a given object can be more than two orders of magnitude smaller than the CSFH models suggest.}
    \label{fig:ssfrs}
\end{figure}

\begin{figure}
    \centering
    \includegraphics[width=\columnwidth]{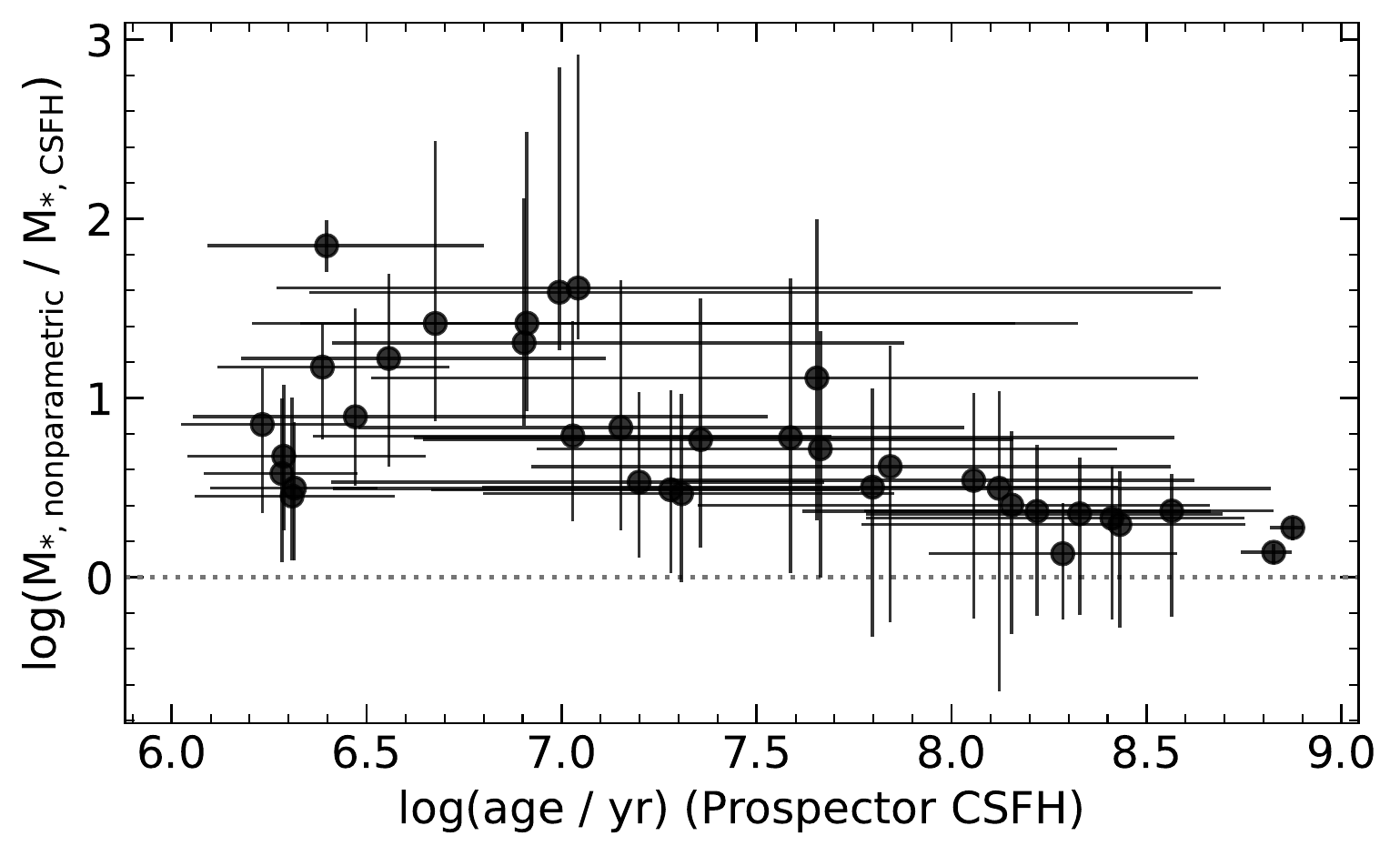}
    \caption{The offset between the stellar masses of our sample inferred by the \prospector\ constant and nonparametric SFH models with continuity prior as a function of the \prospector\ CSFH age. The horizontal dashed line shows where the CSFH- and nonparametric-inferred masses are equal. At young ages ($\lesssim 10$\,Myr), the nonparametric masses are $\sim 0.5 - 1.8$\,dex larger than the CSFH masses ($\sim 1$\,dex on average). This difference decreases as the CSFH ages increase, until the two models agree within $\sim 0.3$\,dex on average at ages $\gtrsim 250$\,Myr, comparable to the scatter between the \beagle\ and \prospector\ CSFH models.}
    \label{fig:mass_diff}
\end{figure}

Given that our fiducial CSFH models are possibly missing considerable early star formation, we now use a nonparametric SFH model to investigate the impact of the assumed SFH on the inferred physical parameters. We fit our galaxies with one of \prospector's built-in nonparametric SFH models, choosing a prior that weights against rapid, potentially unphysical, changes in the SFR over time (for further details on our nonparametric modelling technique, see Section\ \ref{subsec:prospector_nonpar}). This prior generally provides insights into the largest amount of early star formation activity and most massive stellar populations that can be accommodated by the data, since the resulting SFHs tend to evolve relatively slowly over time and have an expectation value of a constant SFH in the absence of informative data. As seen in Figure\ \ref{fig:chi_squared}, the nonparametric models generally achieve similar goodness-of-fit to the observations as the CSFH models from both \beagle\ and \prospector. However, the physical properties inferred from the nonparametric models are dramatically different from those inferred by the CSFH models. In Figure\ \ref{fig:ssfrs}, we compare the sSFRs inferred by the \prospector\ CSFH models with those inferred by the nonparametric models as a function of the CSFH-inferred age. We show sSFRs averaged over the most recent 10\,Myr and 100\,Myr for the nonparametric models. In Figure\ \ref{fig:mass_diff}, we compare the stellar masses inferred by the two SFH models.

We focus on the galaxies in our sample for which the CSFH models inferred the youngest ages. As previously outlined, these objects are likely dominated by light from stars formed in a recent burst of star formation and are therefore the most susceptible to systematic uncertainties due to outshining. It is also possible that a CSFH may infer an excessively old age for evolved systems, as it is constantly producing young, blue stars, requiring longer for a mature stellar component to build up the distinctive Balmer break. In contrast, an SFH that evolves with time allows a decrease in SFR at recent times, producing fewer young stars and allowing an evolved stellar population to become significant in less time. However, we find that our CSFH and nonparametric models generally infer similar stellar masses for these mature systems. As seen in Figures\ \ref{fig:ssfrs} and \ref{fig:mass_diff}, while the nonparametric sSFRs and stellar masses are generally systematically offset from the same parameters inferred by the CSFH models, the difference decreases with CSFH-inferred age and is most significant for the youngest sources. Notably, the sSFRs inferred by the nonparametric models are frequently more than a factor of ten, and sometimes a factor of a \textit{hundred}, lower than the sSFRs inferred by the CSFH models for sources with CSFH-inferred ages $\lesssim 10$\,Myr. Similarly, the stellar masses inferred from the nonparametric models can be a factor of ten larger than the CSFH-inferred stellar masses for the same young population.

Because our photometry cannot strongly constrain early star formation, the specifics of the mass formed at early times is highly dependent on the prior we choose. In our nonparametric models, we have required an early start to star formation ($z_\text{form} = 20$) and our fiducial nonparametric prior tends towards a relatively smooth, slow evolution of the SFH over time. In contrast, the CSFH models represent the other extreme, since they have no star formation activity before the inferred age. Thus, for galaxies that appear young, neither model is capturing other possible, potentially less extreme, behavior for the SFH before the recent, intense episode of star formation.

To investigate this possibility, we explore a variety of nonparametric priors in Appendix\ \ref{appendix:nonpar_priors}, and alternative functional parametrizations in Appendix\ \ref{appendix:parametric_sfhs}. In general, we find that some priors allow a significant, rapid decrease in SFR from recent to early times (qualitatively similar to the behavior of a CSFH that drops instantaneously to an SFR of zero at early times) and infer smaller stellar masses than our fiducial nonparametric models. Meanwhile, other priors can find extended, early periods of large SFRs and infer correspondingly large stellar masses. The alternative parametric models we explore can also introduce variations of $2 - 3$ on average, with larger stellar masses found for models with extended periods of early star formation. In general, we find that the qualitative tendency of a prior to enforce a prolonged period of early star formation can produce factor of $\sim 10$ larger stellar masses, bracketing the range of stellar masses allowed by the data, as is suggested by our findings from our fiducial CSFH and nonparametric models.

It is at least qualitatively clear that stellar masses have the potential to be enormously underpredicted if a galaxy contains a significant quantity of very young, extremely luminous stars that dominate the rest-UV and optical emission. We emphasize that these young, possibly burst-dominated systems are not an inconsequential fraction of the UV-luminous galaxy population during reionization; galaxies with CSFH ages $\lesssim 10$\,Myr constitute $\sim 10 - 30$\,per\,cent of our sample depending on the SED model used (though we acknowledge that our UV-selected sample may be additionally biased towards young, UV-bright systems with large amounts of unobscured star formation), and similar systems powering extreme emission lines are thought to be relatively ubiquitous at $z \gtrsim 7$ \citep[e.g.][]{smit2014, smit2015, roberts-borsani2016, debarros2019, endsley2021a, stefanon2022}. In other words, a large number of reionization era galaxies exist in the regime where the stellar masses may be underestimated, and sSFRs overestimated, by an order of magnitude or more. Thus, systematics in the measured SFHs could influence not only inferences of individual galaxy properties, but also have profound impacts on measurements of population properties such as the stellar mass function, the star-forming main sequence, and the cosmic SFR density. To identify the SFH model that best represents these systems, independent measurements such as dynamical masses from \textit{JWST} and the Atacama Large Millimeter/submillimeter Array (ALMA) will be important.

\begin{figure}
    \centering
    \includegraphics[width=\columnwidth]{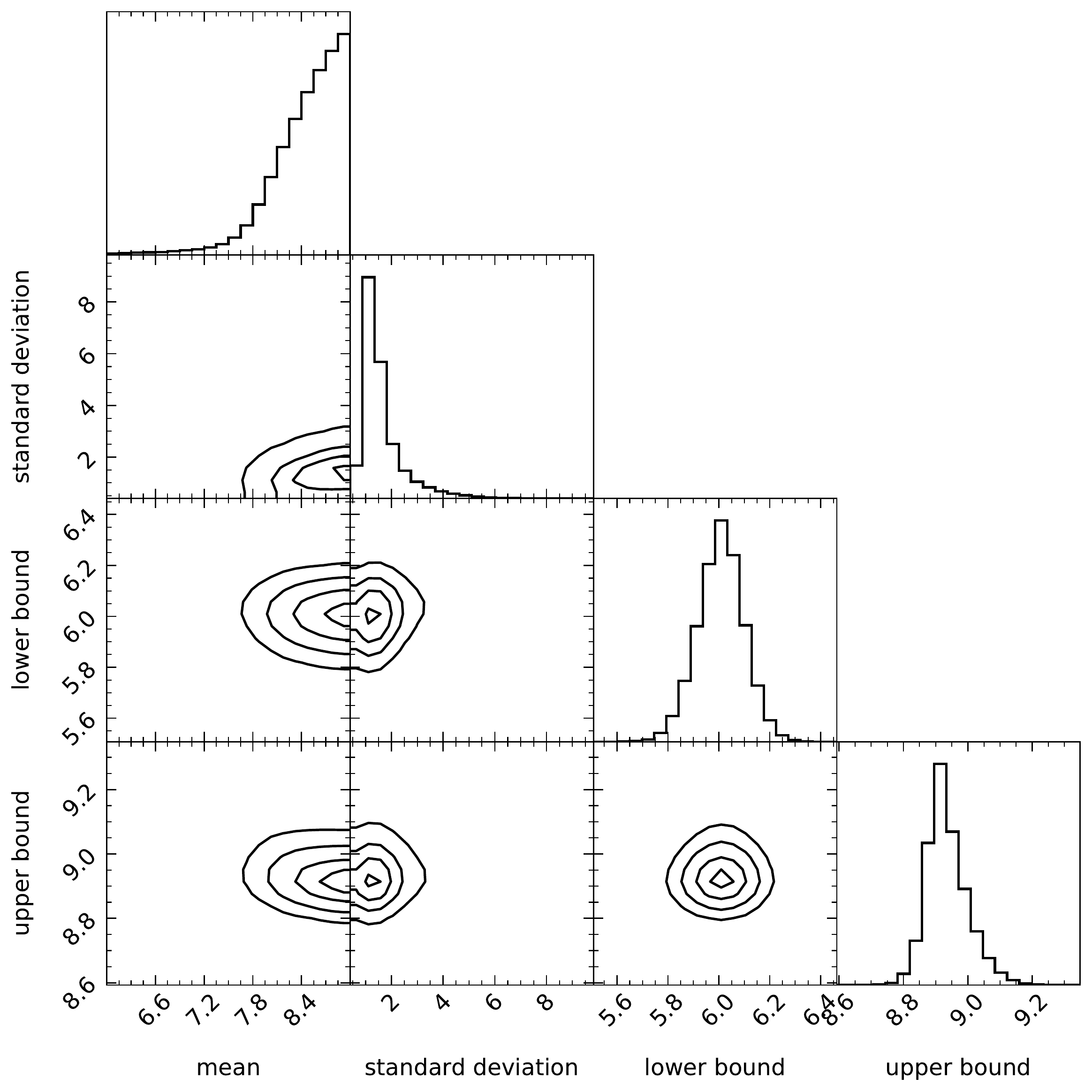}
    \caption{Constraints on the parameters of the truncated normal distribution age distribution. The panels on the diagonal show the one dimensional posteriors for each parameter marginalized over all other parameters, and the remaining panels show the two dimensional posteriors for each pair of parameters. We place a lower limit on the mean of the distribution of $\mu \geq 8.5$ and infer a standard deviation of $\sigma = 1.4_{-0.4}^{+0.8}$. The lower bound is relatively unconstrained and closely reflects its prior with value of $a = 6.0_{-0.1}^{+0.1}$, and we infer an upper bound of $b = 8.9_{-0.1}^{+0.1}$.}
    \label{fig:corner_plot}
\end{figure}

\section{Ages and star formation histories at \texorpdfstring{$\MakeLowercase{z} \simeq 6.6 - 6.9$}{z = 6.6 - 6.9}} \label{sec:ages_sfhs}

As previously introduced, galaxies at $z \sim 6 - 9$ have been found to have a wide variety of star formation histories. However, while individual examples of both young, line-emitting galaxies and more evolved systems have been observed, the relative frequencies with which they occur is not well understood. Therefore, we now undertake a systematic study of the ages and star formation histories of our sample of 36 UV-bright galaxies at $z \simeq 6.6 - 6.9$ in order to quantify the incidence rates of these two populations in the context of the full UV-luminous galaxy population. We derive the model we use to quantify the ages of the population in Section\ \ref{subsec:age_distribution_derivation} and discuss the implications of the results implied by our fiducial SED models in Section\ \ref{subsec:age_distribution_results}. We then extend the discussion to include additional insights provided by our nonparametric SFH models in Section\ \ref{subsec:sfhs_interpretation}.

\begin{figure}
    \centering
    \includegraphics[width=\columnwidth]{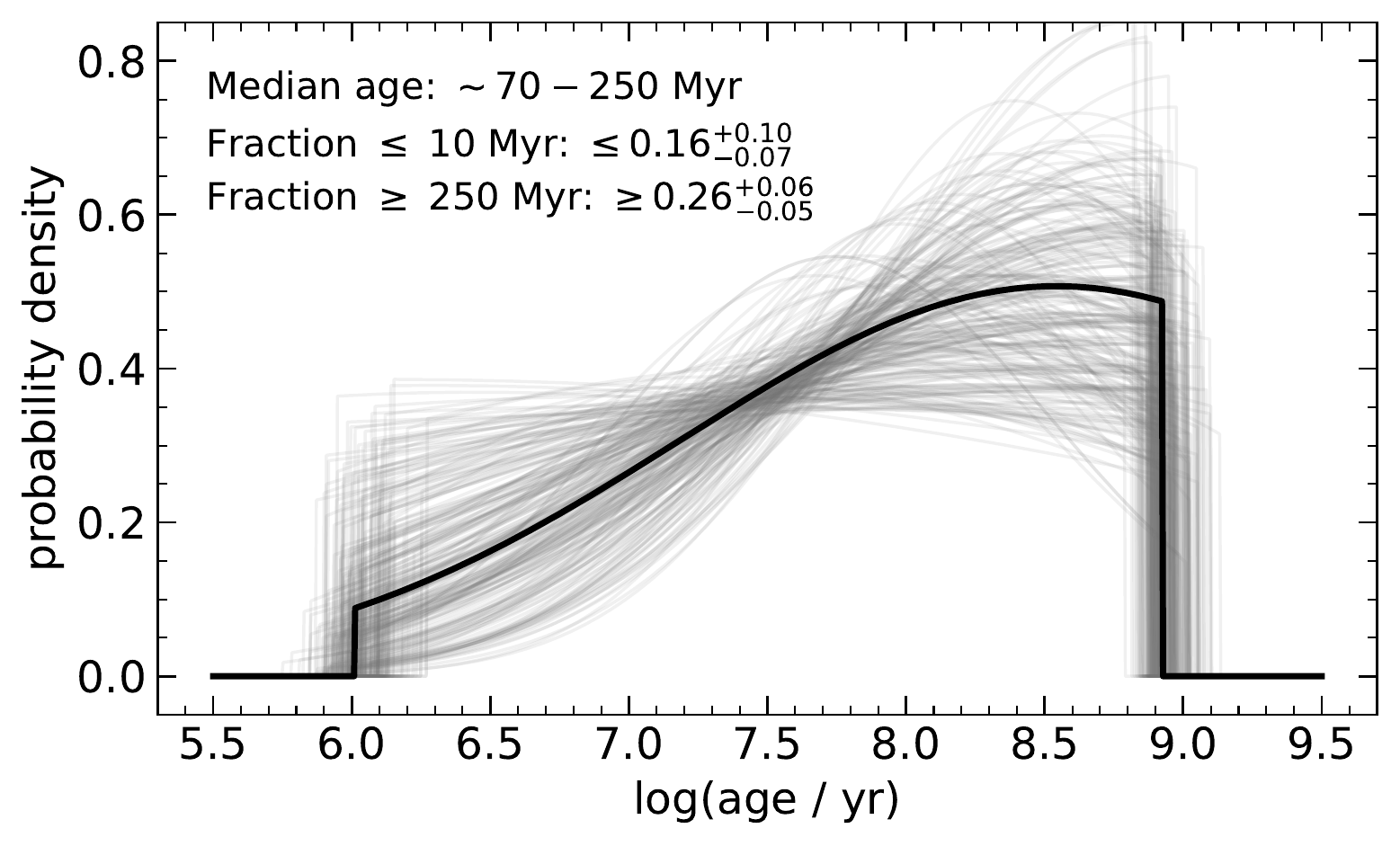}
    \caption{The distribution of ages inferred from our CSFH \beagle\ models for the UV-bright galaxy population at $z \sim 7$. We show the distribution corresponding to the median model parameters as the thick black line and distributions generated from random draws from the posterior as thin grey lines. We find a median age of at least $\sim 70$\,Myr and no more than $\sim 250$\,Myr for UV-luminous galaxies at $z \sim 7$. We also infer that $\sim 16$\,per\,cent of the population has ages $\leq 10$\,Myr, which we argue are likely young, burst-dominated sources, and that $\sim 26$\,per\,cent of the population have a significant population of old stars with ages $\geq 250$\,Myr.}
    \label{fig:age_distribution}
\end{figure}

\subsection{Ages of the population: deriving the model} \label{subsec:age_distribution_derivation}

\begin{table*}
\renewcommand{\arraystretch}{1.5}
\centering
\caption{Inferred values for the parameters of the truncated normal distribution we adopt to characterize the distribution of ages of UV-selected galaxies at $z \sim 7$. We also report the median age, fraction of young sources, and fraction of sources with mature stellar populations we derive from the age distribution. We report the values inferred from both the \beagle\ and \prospector\ SED models, and highlight that the \prospector-derived age distribution is weighted more towards younger ages, as expected from the systematically younger ages the \prospector\ models tend to infer for individual sources.}
\begin{tabular}{c|c|c|c|c|c|c|c} \hline
    \multirow{2}{*}{Model} & Mean ($\mu$) & Standard deviation ($\sigma$) & Lower bound ($a$) & Upper bound ($b$) & Median age & $\leq 10$\,Myr & $\geq 250$\,Myr \\
    & [$\log\left(\text{age / yr}\right)$] & [$\log\left(\text{age / yr}\right)$] & [$\log\left(\text{age / yr}\right)$] & [$\log\left(\text{age / yr}\right)$] & [Myr] & fraction & fraction \\ \hline\hline
    \beagle & $\geq 8.5_{-0.5}^{+0.3}$ & $1.4_{-0.4}^{+0.8}$ & $6.0_{-0.1}^{+0.1}$ & $8.9_{-0.1}^{+0.1}$ & $\sim 70 - 250^\text{a}$ & $\leq 0.16_{-0.07}^{+0.10}$ & $\geq 0.26_{-0.05}^{+0.06}$ \\
    \prospector & $7.2_{-0.8}^{+1.1}$ & $3.0_{-1.0}^{+1.3}$ & $6.0_{-0.1}^{+0.1}$ & $9.0_{-0.1}^{+0.1}$ & $29_{-7}^{+7}$ & $0.34_{-0.04}^{+0.04}$ & $0.17_{-0.03}^{+0.03}$ \\ \hline
\end{tabular}
\begin{flushleft}
$^\text{a}$Since we place a lower limit on the parameter $\mu$ of the truncated normal age distribution, we also place a lower limit on the implied median age. However, allowing $\mu$ to range to arbitrarily high, but possibly physically unreasonable, values gives an upper limit on the median age of $\sim 250$\,Myr.
\end{flushleft}
\label{tab:age_distribution_fits}
\end{table*}

We are particularly interested in quantifying the fraction of galaxies observed during an upturn in star formation and the fraction of those that have more more steady past star formation. Fortunately, the ages inferred by our fiducial CSFH models are useful metrics for this measurement. Though quantitatively interpreting the CSFH-inferred ages for individual sources is subject to complications from outshining, they are signposts of distinctive types of objects. Galaxies that are inferred to be close to the youngest limit of the ages we allow during SED modelling are likely being observed during a rapid uptick in star formation that is driving the CSFH model. Meanwhile, galaxies at the oldest extreme of the CSFH-inferred ages imply that there was early star formation and little vigorous later activity to outshine the light from the older population. Thus, we can use the fraction of the population inferred to be young as a proxy for the burst-dominated population, while the old fraction corresponds to the sources with more steady past activity.

To characterize these two extremes in age, we first quantify how the ages of the entire UV-luminous $z \simeq 6.6 - 6.9$ galaxy population are distributed, assuming no luminosity dependence in the relatively small luminosity range of our sample ($-22.5 \lesssim M_\textsc{uv} \lesssim -21$). We adopt a truncated normal distribution in logarithmic age, which has four parameters (mean, $\mu$; standard deviation, $\sigma$; lower bound, $a$; and upper bound, $b$). The probability density function is given by

\begin{equation} \label{eqn:age_distribution}
    f\left(x; \mu, \sigma, a, b\right) =
    \begin{cases} \frac{1}{\sigma}\frac{\phi\left(\frac{x - \mu}{\sigma}\right)}{\Phi\left(\frac{b - \mu}{\sigma}\right) - \Phi\left(\frac{a - \mu}{\sigma}\right)} & a \leq x \leq b \\
    0 & \text{otherwise}
    \end{cases},
\end{equation}
where $\phi\left(\xi\right) = \tfrac{1}{\sqrt{2\pi}} \exp\left(-\tfrac{1}{2}\xi^2\right)$ and $\Phi\left(\xi\right) = \frac{1}{2}\left(1 + \text{erf}\left(\xi / \sqrt{2}\right)\right)$ are the probability density function and the cumulative distribution function of the standard normal distribution, respectively, and $x = \log\left(\text{age} / \text{yr}\right)$.

We want to obtain the posterior probability distribution of our model parameters of interest given our observed data $\boldsymbol{D}$, $P\left(\mu, \sigma, a, b \mid \boldsymbol{D}\right)$. By Bayes' Theorem, we have

\begin{equation} \label{eqn:bayes_theorem}
    P\left(\mu, \sigma, a, b \mid \boldsymbol{D}\right) \propto P\left(\boldsymbol{D} \mid \mu, \sigma, a, b \right) P\left(\mu, \sigma, a, b\right),
\end{equation}
where $P\left(\boldsymbol{D} \mid \mu, \sigma, a, b\right)$ is the likelihood of observing our data given the model with parameters $\mu, \sigma, a,$ and $b$, and $P\left(\mu, \sigma, a, b\right)$ are the priors. We adopt a uniform prior on $\mu$ from $6 - 9$ (corresponding to $1\,\text{Myr} - 1\,\text{Gyr}$), and Gaussian priors on $\sigma$ (centered at $\sigma = 0.1$ with standard deviation of 2), $a$ (centered at $a = 6$ with standard deviation of 0.1), and $b$ (centered at $b = 9$ with standard deviation of 0.1).

Since we do not infer the age distribution directly from the observed data (i.e. the photometry), but instead model the SEDs of each galaxy individually to infer their age along with other physical properties, our model is better viewed as similar to a Bayesian hierarchical model \citep{kelly2012, galliano2018, leja2020, nagaraj2022, leja2022}. The likelihood for one galaxy, $i$, with observed photometry $\boldsymbol{D_i}$ is
\begin{equation} \label{eqn:individual_likelihood}
    P\left(\boldsymbol{D_i} \mid \mu, \sigma, a, b\right) = \int P\left(\boldsymbol{D_i} \mid x_i\right) P\left(x_i \mid \mu, \sigma, a, b\right) \text{d}x_i,
\end{equation}
where $x_i = \log\left(\text{age}_i / \text{yr}\right)$ as before.

Because each object in our sample was modelled independently, the full likelihood is simply the product of each individual object's likelihood:

\begin{equation} \label{eqn:full_likelihood}
    P\left(\boldsymbol{D} \mid \mu, \sigma, a, b\right) = \prod_{i = 1}^{N_\text{gals}} \int P\left(\boldsymbol{D_i} \mid x_i\right) P\left(x_i \mid \mu, \sigma, a, b\right) \text{d}x_i.
\end{equation}
The first term in the integral is the likelihood of observing the photometry of a galaxy for a given age and the second term is simply the age distribution (Equation\ \ref{eqn:age_distribution}). We approximate the value of the integral following the method described by \citet{leja2020, leja2022}. For each object $i$, we draw a set of samples from the posterior for age from the SED models marginalized over all other parameters. We then assign each sampled age, $x_{i, j}$, an importance weight, $w_{i, j}$, to mitigate the impact of the prior we placed on age in our SED models; we take the importance weight to be the inverse of the value of the original age prior we placed on the SED models evaluated at $x_{i, j}$. Then, Equation \eqref{eqn:full_likelihood} can be approximated as

\begin{align} \label{eqn:approx_likelihood}
    \prod_{i = 1}^{N_\text{gals}} \int P\left(\boldsymbol{D_i} \mid x_i\right) & P\left(x_i \mid \mu, \sigma, a, b\right) \text{d}x_i \\
    & \approx \prod_{i = 1}^{N_\text{gals}} \frac{\sum_j w_{i,j} P\left(x_{i,j} \mid \mu, \sigma, a, b\right)}{\sum_j w_{i,j}}. \nonumber
\end{align}
In practice, the weights $w_{i,j}$ are constant because we placed a log-uniform prior on age in our SED models. This approximation is the likelihood in Equation\ \eqref{eqn:bayes_theorem} to determine the posterior probability distribution of our model parameters.

We use the Markov chain Monte Carlo sampling package \textsc{emcee} \citep{foreman-mackey2013} to sample the posteriors of the model parameters $\mu, \sigma, a,$ and $b$ and show the parameters derived from the \beagle\ SED models in Figure\ \ref{fig:corner_plot}. Next, we derive the age distribution of the UV-luminous galaxy population at $z \simeq 6.6 - 6.9$, and finally use the distribution to quantify the proportions of the population that are young and old.

\begin{figure*}
    \centering
    \includegraphics[width=\textwidth]{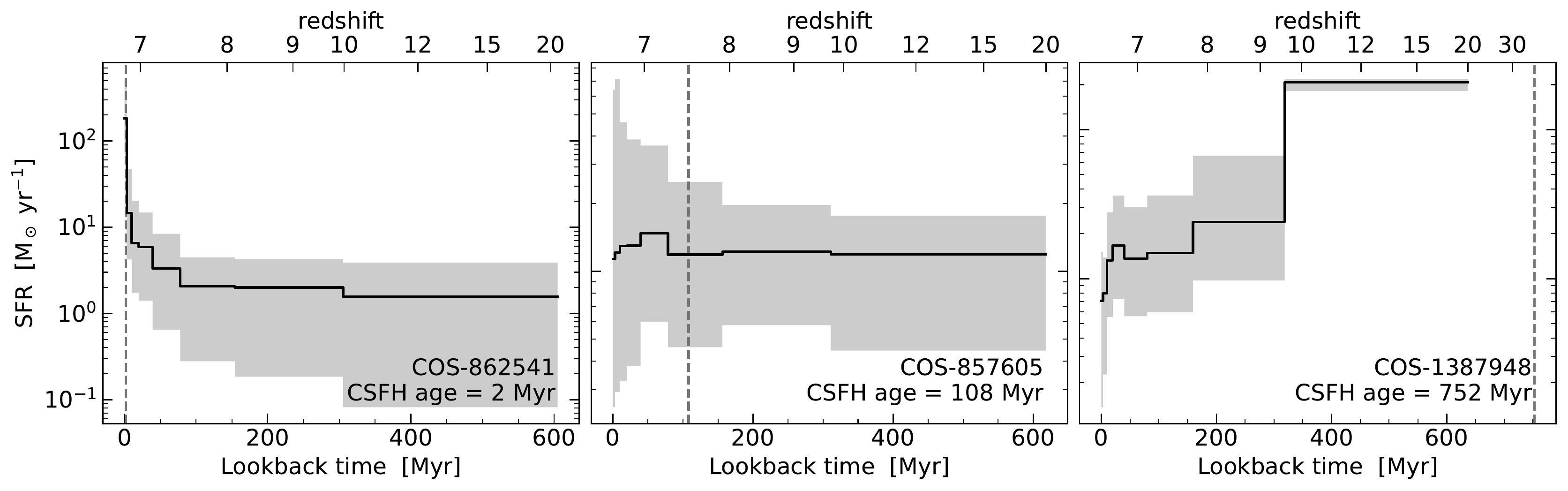}
    \caption{Examples of SFHs inferred by our nonparametric models. We show sources for which the CSFH models inferred a variety of ages (the youngest age in the left panel, an intermediate age in the middle, and the oldest age on the right). The galaxy with the youngest CSFH age has an extremely rapidly rising SFH, with SFR increasing by more than a factor of ten within a few tens of millions of years, while the galaxy with the oldest age has a declining SFH. The SFR of the more typical source changes much less over time, though we emphasize that we observe a variety of SFH morphologies for these objects with more intermediate ages.}
    \label{fig:nonparametric_sfhs}
\end{figure*}

\subsection{Ages of the population: results} \label{subsec:age_distribution_results}

Ultimately, we obtain age distributions based on our fiducial SED models from both \beagle\ and \prospector, from which we derive the fractions of the population at the two extremes in age. We quantify the subset of the population that is young (which we take to be ages $\leq 10$\,Myr) and the subset that is old (which we define as ages $\geq 250$\,Myr, corresponding to $z \gtrsim 9$ for objects at $z \simeq 6.8$). We report the inferred parameters in Table\ \ref{tab:age_distribution_fits} and show the final age distribution inferred from our fiducial \beagle\ SED models in Figure\ \ref{fig:age_distribution}. We plot the distribution corresponding to the median values for $\mu$, $\sigma$, $a$, and $b$ in black and random draws from the parameters' posteriors in grey.

As seen in Figure\ \ref{fig:age_distribution}, the age distribution inferred from the \beagle\ models is weighted significantly more towards older ages of $\gtrsim 100$\,Myr. That is, at $z \simeq 6.6 - 6.9$, there is a larger population of old galaxies with ages of a few hundred million years than young ones with ages of only a few million years. This distribution further implies that the median age of the UV-bright $z \simeq 6.6 - 6.9$ galaxy population is at least $\sim 70$\,Myr,\footnote{This median age is younger than the parameter $\mu$ we infer for the age distribution since our parametrization of the age distribution allows for skew (in this case, towards young ages).} consistent with the average ages of fainter $z \sim 8$ galaxies found by \citet{labbe2013}, though towards the older end of the ages found by \citet{stefanon2022}. We note that since we only place a lower limit on the parameter $\mu$ of the truncated normal age distribution, this median age depends somewhat on the priors we adopt for $\mu$ and $\sigma$, though further testing in which we vary these priors suggests that the median is no larger than $\sim 250$\,Myr.

As expected from the qualitative behavior of the \beagle-derived age distribution, the fraction of galaxies inferred to have ages of a few million years does not dominate the population. Taking ages $\leq 10$\,Myr, we find a young fraction of $\leq 0.16_{-0.07}^{+0.10}$, slightly more than the actual three out of 36 galaxies in our sample with \beagle\ ages $\leq 10$\,Myr inferred by our SED models. We attribute the larger fraction of young ages inferred from the age distribution to the fact that our SED models can find significant probabilities of ages $\leq 10$\,Myr even for objects that have slightly older median ages. Since we expect these young ages to roughly correlate with a recent burst of star formation, this suggests that we expect $\sim 1 - 2$ in every 10 UV-luminous galaxies observed at $z \simeq 6.6 - 6.9$ to be in a phase of extremely rapid star formation. This is broadly consistent with the $\sim 1$ in 5 galaxies found by \citetalias{endsley2021a} to have intense nebular emission (\ion{[O}{iii]}+\ion{H}{$\beta$} equivalent widths $> 1200$\,\AA), also indicative of bursts of star formation, but we emphasize that this mapping is not one-to-one.

In comparison, we find that the proportion of old objects with ages $\geq 250$\,Myr is $\geq 0.26_{-0.05}^{+0.06}$. This implies that we expect our \beagle\ SED models to infer ages $\geq 250$\,Myr for $\sim 9$ galaxies in our sample, more than the four we actually find. As for the young sources, the larger fraction of old ages inferred from the age distribution is likely due to the many objects in our sample that have significant probabilities of ages $\geq 250$\,Myr, even if their median age is slightly younger. We also note that this fraction is consistent within $2\sigma$ with the fraction of sources with evidence for strong Balmer breaks ($\simeq 36$\,per\,cent) found by \citet{laporte2021} at $z \sim 9$, though a more rigorous comparison would require a full inference of the ages of $z \sim 9$ Balmer break galaxies with our same modelling approach.

Though we have so far focused on the age distribution derived from our \beagle\ SED models, we have also derived an age distribution based on our fiducial \prospector\ SED models. We stress that we do not expect exactly the same results, since the \prospector\ models sometimes infer ages up to an order of magnitude younger than \beagle, as discussed in detail in Section\ \ref{subsec:code_compare}. Thus, we expect the \prospector-derived distribution to have a lower median age and larger probabilities of younger ages. Indeed, we find a median age of $\sim 30$\,Myr from the \prospector\ age distribution compared to the median of $\sim 70$\,Myr from the \beagle-inferred distribution. We also find that the \prospector\ age distribution is considerably more flat than the \beagle\ age distribution, leading to larger probabilities of young ages.

Together, these two differences lead to an larger young fraction and a smaller old fraction inferred from the \prospector-derived age distribution compared to the distribution based on the \beagle\ models. We find that $34_{-4}^{+4}$\,per\,cent of the \prospector\ distribution falls at ages $\leq 10$\,Myr, a factor of two larger than the fraction from \beagle, reflecting the additional objects that our \prospector\ SED models inferred to have young ages. The proportion of the \prospector-based distribution at ages $\geq 250$\,Myr is $17_{-3}^{+3}$\,per\,cent, which is also in mild tension with the old fraction from \beagle. However, this is again expected from the differences in the ages inferred from the SED models.

\begin{figure*}
    \centering
    \includegraphics[width=\textwidth]{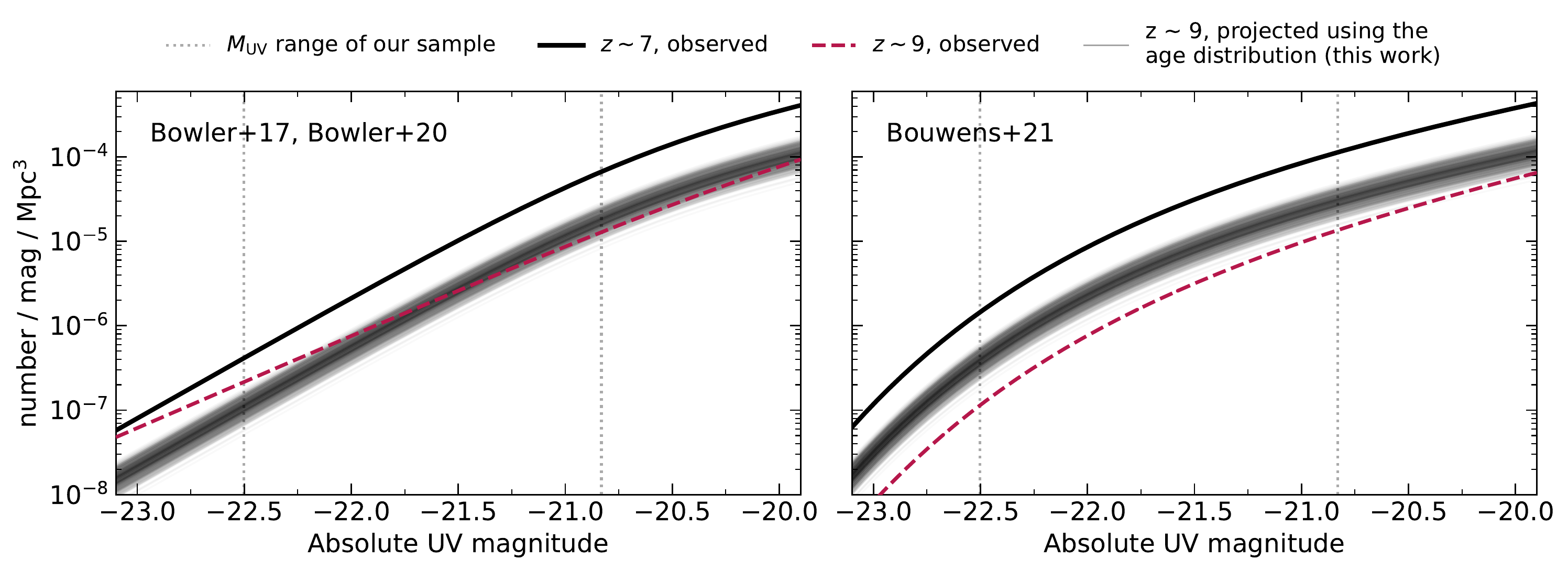}
    \caption{The evolution of the UV LF from $z \sim 7$ to $z \sim 9$ that we estimate from our distribution of the ages of galaxies at $z \sim 7$. We calculate the $z \sim 9$ UV LF by scaling the observed $z \sim 7$ luminosity function by the fraction of galaxies dominated by stellar populations with ages $\geq 250$\,Myr (and therefore expected to exist at $z \sim 9$ with the same SFR and similar UV luminosity as observed in the context of a CSFH). We show the projections based on \citet{bowler2017} in the left panel and \citet{bouwens2021} in the right panel. The thick black lines denotes the observed $z \sim 7$ UV LFs that we scale to project the $z \sim 9$ UV LF and instances of the resulting projection are shown as the grey lines. We also show the observed $z \sim 9$ luminosity functions from \citet{bowler2020} and \citet{bouwens2021} as the dashed red lines (left and right panels, respectively) and the luminosity range of our sample as the vertical grey dotted lines. Each of our predicted luminosity functions corresponds to a different realization of the projected $z \sim 9$ UV LF implied by random draws of the parameters of the age distribution (Section\ \ref{sec:ages_sfhs}) from their posteriors. We expect a decrease in the number density of UV-luminous galaxies by a factor of $2.5 - 5$, broadly consistent with the decline observed by \citet{bowler2017} and \citet{bowler2020}, though smaller than that found by \citet{bouwens2021}.}
    \label{fig:uvlf_projection}
\end{figure*}

\subsection{Insights from nonparametric star formation histories} \label{subsec:sfhs_interpretation}

In Section\ \ref{sec:galaxy_properties}, we found evidence for a large variety of SFHs in a sample of UV-bright, star-forming galaxies at $z \sim 7$ using the ages inferred from a CSFH model. We have now used the distribution of CSFH ages to infer that there is a large population of galaxies with relatively steady past star formation, as well as a smaller subset of younger, burst-like systems. However, while the ages inferred by the CSFH models are useful for identifying particular types of extreme objects, they explicitly do not allow for any star formation prior to the age of the population that dominates the SED. Instead, we turn to our nonparametric SFHs to better understand the possible past star formation activity of the population. We again stress that the quantitative details of the earliest epochs of star formation are prior-dependent. Here, we present results derived from the `continuity' prior in \prospector, which weights against large changes in SFR over short time-scales and produces a constant SFH in the absence of informative data, and we have set the onset of star formation to $z_\text{form} = 20$. However, the qualitative evolution of the SFH over time is nevertheless instructive.

In Figure\ \ref{fig:nonparametric_sfhs}, we show examples of the SFHs we infer from the nonparametric models. We include one galaxy that was inferred to have a very young age (2\,Myr) by both the \beagle\ and \prospector\ CSFH models, one inferred to have a very old age ($\sim 750$\,Myr), and one with a more intermediate age ($\sim 100$\,Myr). We expect these objects to be broadly representative of the variety of SFHs present in our sample.

We first highlight that galaxies with very young CSFH-inferred ages of only a few million years have extremely rapidly rising SFHs (see COS-862541 in the left panel of Figure\ \ref{fig:nonparametric_sfhs}). The SFRs inferred by our nonparametric models increase by an order of magnitude within as little as 10\,Myr, a factor of $\sim 30 - 40$ in 20\,Myr, and a factor of nearly 100 within 100\,Myr. Though the exact increase in SFR depends on the prior on the SFH, it is apparent that these objects undergo an extraordinarily large burst of star formation within a few tens of millions of years, increasing greatly in UV luminosity in a very short amount of time.

Secondly, we examine the SFHs of objects that were inferred to have old ages of $\geq 250$\,Myr by the CSFH models. We show COS-1387948 as an example in the right panel of Figure\ \ref{fig:nonparametric_sfhs}. These objects are best described by either constant or declining SFHs (more generally, SFHs that do not increase) in the context of the nonparametric models. For example, the two objects in our sample with the oldest CSFH ages are both modelled by declining SFHs with a decrease in SFR by factors of $\sim 30 - 40$ from the oldest age bin to the most recent 10\,Myr. Thus, in counterpoint to the objects that were inferred to be young, which were likely faint at early times and increased rapidly in luminosity right before observation at $z \sim 7$, these objects likely comprise the population of UV-luminous galaxies at high redshifts of $z \gtrsim 9$. Based on the fraction of galaxies inferred to be at ages $\geq 250$\,Myr from the age distribution, we expect the number of bright galaxies at $z \gtrsim 9$ to be approximately 30\,per\,cent of the number at $z \simeq 6.8$. We will examine this forecast in the context of the UV LF in the following section.

\section{Discussion} \label{sec:discussion}

In this work, we have systematically studied the stellar masses, ages, and SFHs of galaxies in the narrow redshift window of $z \simeq 6.6 - 6.9$, where nebular emission lines can be cleanly separated from rest-optical stellar continuum in IRAC photometry. This enables the physical properties of our sample to be robustly characterized. Now, we investigate the indirect constraints that the ages, SFHs, and stellar masses that we have measured at $z \sim 7$ can place on earlier star formation activity. We begin by exploring the implications of our results for the evolution of the UV LF up to $z \sim 9$, then turn to quantifying the evolution of the stellar mass assembly history of the Universe to gain insights into the cosmic star formation history at $z \gtrsim 9$.

\begin{figure}
    \centering
    \includegraphics[width=\columnwidth]{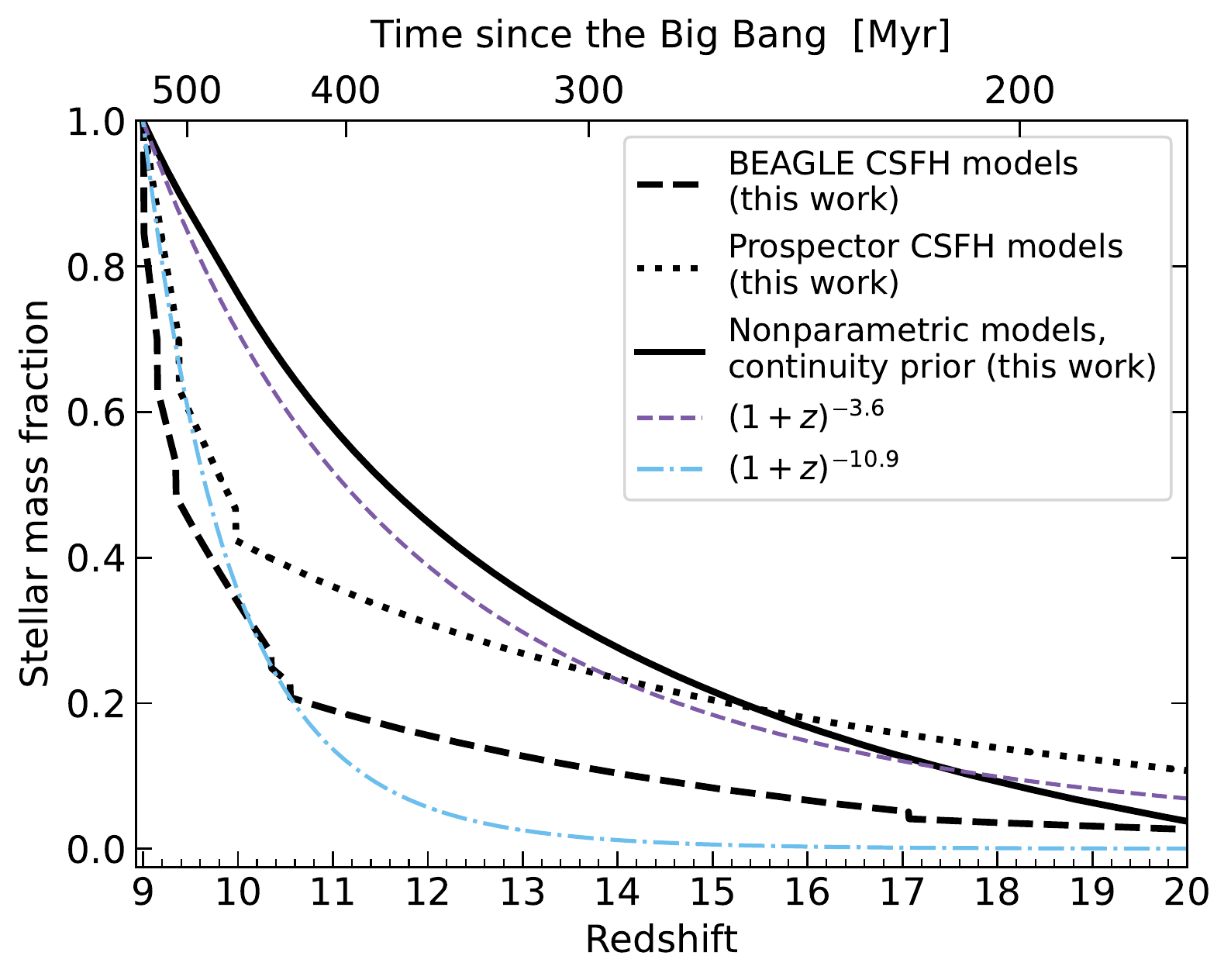}
    \caption{The redshift evolution of the stellar mass assembly history predicted by our models. We average the assembly histories of the 36 galaxies in our sample implied by the CSFH models (\beagle\ is shown as the dashed black line, \prospector\ as the dotted black line) and the \prospector\ nonparametric SFHs with continuity prior (solid black line) and normalize at $z = 9$. We compare with a smooth evolution of $\rho_\textsc{uv} \propto \left(1 + z\right)^{-3.6}$ \citep[purple dashed line;][]{mcleod2016}, a rapid evolution of $\rho_\textsc{uv} \propto \left(1 + z\right)^{-10.9}$ \citep[blue dot-dashed line;][]{oesch2018}. The rapid evolution implies that $\sim 10$\,per\,cent of the stellar mass at $z \sim 7$ is in place by $z = 9$, while the smooth evolution implies a larger fraction of $\sim 45$\,per\,cent. Both CSFH models suggest that only $\sim 5$\,per\,cent of $z \sim 7$ stellar mass has formed by $z = 9$; however, when earlier star formation is allowed with the nonparametric models, the fraction increases to $\sim 50$\,per\,cent.}
    \label{fig:stellar_mass_frac}
\end{figure}

If all $z \sim 7$ galaxies are dominated by an old stellar population, we would not expect a significant decrease in the number density of the UV-luminous, star-forming galaxy population between $z \sim 7$ and $z \sim 9$. Thus, the ages and SFHs we have inferred can provide insights into the evolution of the UV LF. In particular, since we have characterized our sample with SFH models of varying flexibility (a simple, single-parameter CSFH and a very flexible nonparametric model), we can place limits on the expected evolution. Using the distribution of CSFH-based ages we have inferred,  which encapsulates the uncertainties in the ages of individual systems, we have evaluated the fraction of UV-bright galaxies at $z \sim 7$ with ages $\geq 250$\,Myr, corresponding approximately the length of time between $z \sim 7$ and $z \sim 9$. We note that this strategy assumes that the population of bright galaxies only changes with time as new UV-bright systems are formed (i.e. that bright galaxies at $z \sim 9$ did not quench or become obscured, resulting in UV-faint systems at $z \sim 7$, and that no galaxies that are faint at $z \sim 9$ undergo increases in star formation that make them brighter with time). Fortunately, our nonparametric models place constraints on the SFRs of our sample at $z \sim 9$, providing some insights into the evolution of the UV luminosities over time of known bright galaxies at $z \sim 7$.

Our ages and SFHs imply that a large enough fraction of the $z \sim 7$ population is sufficiently young such that the number density of UV-bright galaxies decreases from $z \sim 7$ to $z \sim 9$. As an illustration of the scale of the evolution we find, we multiply the observed $z \sim 7$ UV LF of \citet{bowler2017} and \citet{bouwens2021} by the decrease we infer from the CSFH models and show the resulting $z \sim 9$ projection in Figure\ \ref{fig:uvlf_projection}. We also show examples of $z \sim 9$ UV LFs from the literature \citep{bowler2020, bouwens2021}. Our age distribution suggests that only a fraction ($26_{-5}^{+6}$\,per\,cent) of the galaxy population at $z \sim 7$ is old enough to exist at $z \gtrsim 9$. This leads to a projected decrease by a factor of $3 - 5$ in the number density from $z \sim 7$ to $z \sim 9$, though we note that since the old fraction is a lower limit (see results in Section\ \ref{subsec:age_distribution_results}), this is an upper limit on the possible UV LF evolution. Similarly, the redshift evolution of the UV luminosities predicted by our nonparametric models suggest that only 17 galaxies in our sample are brighter than $M_\textsc{uv} = -21$ at $z \sim 9$, compared to 35 at $z \sim 7$, suggesting a factor of $\sim 2$ decrease in the number density. This is broadly consistent with the factor of $2 - 5$ decrease in the luminosity range of our sample measured by \citet{bowler2017} and \citet{bowler2020}, though smaller than the factor of $10 - 15$ decrease implied by the \citet{bouwens2021} UV LFs. We note that this evolution is not driven by the sources most likely to be very evolved (i.e. the systems with the reddest UV slopes), but rather by the full age posteriors of the entire sample, which often have non-negligible probabilities of old ages. We re-derive the age distribution excluding objects with UV slopes of $\beta > -1$ (including COS-87259, which may have a slightly uncertain age due to possible contamination by an active galactic nucleus) and find a marginally larger fraction of the distribution with ages $\geq 250$\,Myr ($33_{-7}^{+8}$\,per\,cent). This implies a marginally smaller decrease in number density from $z \sim 7$ to $z \sim 9$ (a factor of $2.5 - 4$).

\begin{figure}
    \centering
    \includegraphics[width=\columnwidth]{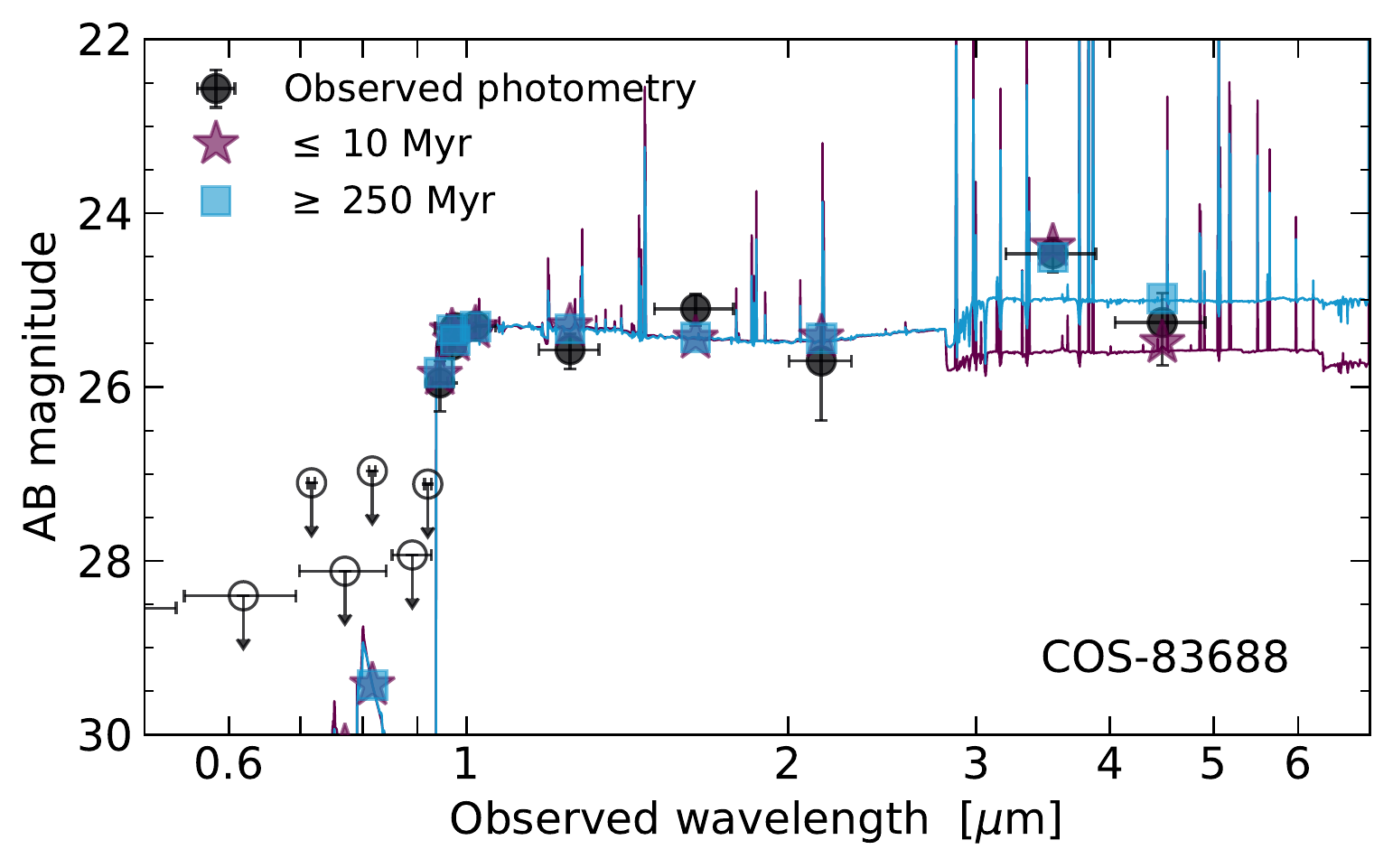}
    \caption{A comparison of the young ($\leq 10$\,Myr) and old ($\geq 250$\,Myr) solutions from the \beagle\ CSFH model for COS-83688. The observed photometry is shown as black circles, the median young solution is shown as purple stars, and the median old solution is shown as blue squares. Both ages agree well in the rest-UV, but diverge significantly in rest-optical, especially in the IRAC [4.5] filter. The young solution has a fainter rest-optical continuum and reaches the observed [3.6] flux with strong nebular emission lines (leading to a blue IRAC colour), whereas the old solution has a brighter rest-optical continuum and a redder IRAC colour. However, both are still consistent within the large uncertainties of the IRAC photometry.}
    \label{fig:young_old_sed}
\end{figure}

While the evolution of the UV LF gives insights into recent star formation in $z \sim 9$ galaxies, the full history of stellar mass assembly provides independent, integral constraints on star formation activity at very early times. In particular, because the fraction of the stellar mass that exists at $z \sim 7$ that is in place by $z = 9$ depends strongly on the evolution of the cosmic star formation history, the full SFHs of our $z \sim 7$ galaxies can provide insights into this evolution \citep{roberts-borsani2020, laporte2021, tacchella2022, tang2022}. If the decline is rapid [$\propto (1 + z)^{-10.9}$; \citealt{oesch2018}], this suggests that only $\sim 10$\,per\,cent of stellar mass in place at $z \sim 7$ is formed by $z = 9$, whereas a smooth evolution [$\propto (1 + z)^{-3.6}$; \citealt{mcleod2016}] implies that a larger fraction of $\sim 45$\,per\,cent is in place.

\begin{figure*}
    \centering
    \includegraphics[width=\textwidth]{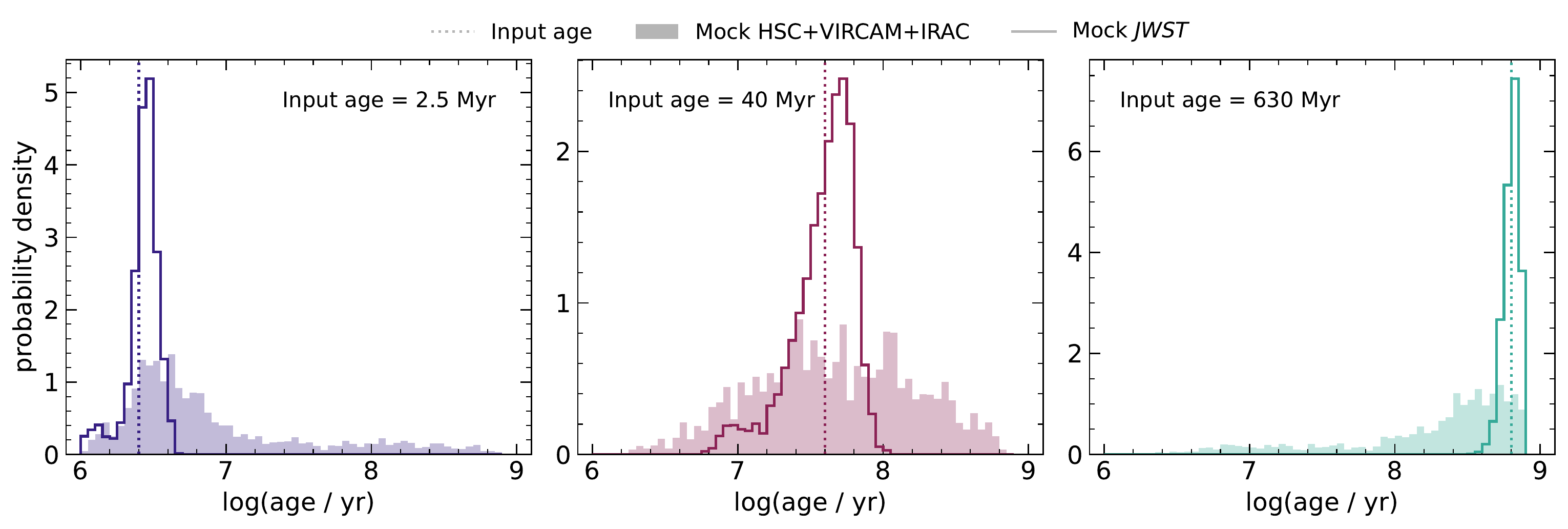}
    \caption{A demonstration of the improved constraints on age expected from \textit{JWST} observations. We compare posterior distributions for age derived from simulated photometry analogous to current data (filled histograms) and from simulated photometry similar to that expected from the medium depth of the \textit{JWST} Advanced Deep Extragalactic Survey (solid lines). We use simulated \beagle\ spectra as input to the models and show three different true ages, one in each panel. Vertical dotted lines indicate the input age. The constraints obtained by \textit{JWST}-like photometry are significantly more precise than the measurements from current data, with a reduction in error by $\sim 80$\,per\,cent.}
    \label{fig:jwst_constraints}
\end{figure*}

Averaging the SFHs implied by the CSFH models, we find that only $\sim 6$\,per\,cent of the stellar mass at $z \sim 7$ is established at $z \sim 9$. However, these models allow no star formation prior to the formation time of the population that dominates the SED, essentially providing a lower limit on early stellar mass assembly; that is, a $z \sim 7$ galaxy must be dominated by a sufficiently old ($\gtrsim 250$\,Myr) stellar population to form \textit{any} mass at $z \gtrsim 9$. Realistically, even galaxies dominated by younger stellar populations may have had earlier episodes of star formation. When earlier star formation is allowed (most importantly for systems with young CSFH-inferred ages that form no stellar mass at $z \gtrsim 9$), as in the nonparametric models with the continuity prior, a much larger fraction of stellar mass of $\sim 50$\,per\,cent is in place by $z = 9$. In Figure\ \ref{fig:stellar_mass_frac}, we show the full redshift evolution of the stellar mass assembly history we derive from both the CSFH and nonparametric models. We find that the lower limit provided by the CSFH models are more consistent with a rapid decline while the less restrictive limit provided by the nonparametric models favors a smooth decline. That is, a large range of possibilities for the evolution of the cosmic star formation history are consistent with the SFHs we infer for our galaxies with a variety of priors \citep[qualitatively consistent with the findings of][]{tacchella2022}, highlighting the growing need for independent constraints to provide a new window into burst-dominated galaxies with early SFHs that are only weakly constrained by rest-UV and optical photometry. For example, recent measurements of large dynamical masses at $z \sim 7$ are beginning to hint that massive, old stellar populations, such as the extended early SFHs inferred by our nonparametric models, can at least be accommodated \citep{topping2022}.

Despite these uncertainties, we note that a rapid decline in the cosmic star formation history requires that the true ages and SFHs of reionization era galaxies closely follows the lower limit provided by our CSFH models. That is, a rapid decline requires that these objects do not have significant star formation activity before the formation of the stellar populations that dominate the rest-UV and optical SEDs. Any amount of early star formation in these systems (e.g. if early galaxies have rising star formation histories with low, but nonzero SFR, at early times) would imply a less steep evolution in the cosmic star formation history, ultimately suggesting that a rapid, $\propto (1 + z)^{-10.9}$ evolution in the cosmic star formation history is unlikely.

Above, we have demonstrated the potential for using the ages and SFHs of galaxies at $z \sim 7$ as indirect probes of earlier epochs of star formation in the Universe. However, current conclusions are limited by (1) the low signal-to-noise of the observed SEDs in the rest-optical, which is important to constrain the age and mass of the stellar population that dominates the SED, and (2) the difficulty of constraining early epochs of star formation when a young, luminous stellar population is present. With current rest-optical photometric uncertainties, many systems can be reasonably modelled with ages spanning from a few Myr to a few \textit{hundred} Myr -- nearly two orders of magnitude. Figure\ \ref{fig:young_old_sed} shows both $\leq 10$\,Myr and $\geq 250$\,Myr model spectra from the \beagle\ CSFH model for one object in our sample, demonstrating the huge variety of ages that can provide satisfactory solutions for a single galaxy. Thus, with current data, we can only unambiguously constrain the ages of objects with the most extreme rest-optical observational signatures (either a strong [3.6] excess, indicative of intense nebular emission lines powered by young stars, or a [4.5] flux excess over the near-infrared, pointing to a strong Balmer break from an old stellar population when combined with a relatively flat or blue rest-UV slope).

In contrast, the unprecedented sensitivity of \textit{JWST} will not require such strong observational signatures in the rest-optical and allow us to place significantly better constraints on the ages of less extreme objects. To estimate the improvement that might be expected from \textit{JWST}, we use \beagle\ to simulate galaxy SEDs at $z = 6.8$ for a range of ages, normalize to $m = 26$ in \textit{JWST}/NIRCam F200W, and generate mock \textit{JWST} photometry (adopting the depth and filters of the medium depth subsurvey of the \textit{JWST} Advanced Deep Extragalactic Survey, a joint program of the NIRCam and NIRSpec Guaranteed Time Observations teams) and mock HSC+VIRCAM+IRAC observations analogous to this work. We re-fit the mock photometry with \beagle\ and show the improvement in the age posteriors expected from \textit{JWST} observations in Figure\ \ref{fig:jwst_constraints}. The mock \textit{JWST} photometry produces considerably more narrow posteriors than the mock HSC+VIRCAM+IRAC photometry, with errors reduced by $\sim 80$\,per\,cent on average (while the smallest improvement is still a $\sim 60$\,per\,cent error reduction). These improved measurements of the ages of individual systems will propagate forward to much more precise inferences of the population age distribution. Combined with independent insights into the stellar populations of reionization era galaxies, such a dynamical masses, to provide benchmarks for continuously improving the accuracy of models, much more precise measurements of the total stellar mass content of the Universe will become possible in the era of \textit{JWST}.

\section{Summary} \label{sec:summary}

We present the results of a systematic study of the star formation histories of UV-luminous galaxies at $z \sim 7$, which we use to gain insights into the assembly of stellar mass at $z \gtrsim 9$. We utilize the colour selection pioneered by \citet{endsley2021a} to select a sample of 36 UV-luminous galaxies at $z \simeq 6.6 - 6.9$ where \ion{[O}{iii]} and \ion{H}{$\beta$} transmit through the IRAC 3.6\,$\mu$m bandpass and the degeneracy between nebular and stellar emission can be broken photometrically. Thus, we can obtain robust constraints on the ages, SFHs, and stellar masses that we measure. We use this sample to characterize the distribution of SFHs expected for UV-bright galaxies at $z \sim 7$, with a particular focus on quantifying the incidence rates of two distinct subsets of the UV-bright galaxy population: sources with strong nebular emission lines powered by young stellar populations, and systems with strong Balmer breaks consistent with mature stellar populations. Our key conclusions are as follows:

\begin{enumerate}
    \item We infer the physical properties of our sample with two Bayesian galaxy SED modelling tools, \beagle\ and \prospector, and conduct an extensive comparison of their results. We adopt a single-parameter, constant SFH model and find that both codes generally infer similar stellar masses. However, \prospector\ finds systematically younger ages than \beagle\ for about $\sim 20$\,per\,cent of our sample, by up to an order of magnitude. We ascribe this difference to weaker rest-optical nebular emission lines and stellar continuum in \prospector, using \fsps\ with \textsc{mist} stellar isochrones, relative to \beagle, using CB16 with \textsc{parsec} isochrones, and emphasize the need for further investigation into the differences between the physical models that lead to these disparate solutions.
    \item We demonstrate that the physical parameters we infer changes significantly in the context of nonparametric SFH models with the built-in continuity prior of \prospector, especially for objects inferred to have the youngest ages. These nonparametric models with continuity prior find systematically larger stellar masses and lower sSFRs than the parametric models at all ages, but most notably, can infer stellar masses up to an order of magnitude larger (and therefore order of magnitude smaller sSFRs) than the CSFH models at the youngest ages of $\lesssim 10$\,Myr. At such young ages, the rest-UV and optical emission is likely dominated by light from young stars formed in a recent uptick in star formation, potentially outshining a massive, but faint, population of old stars. Our CSFH models do not probe the possible properties of this more mature population, but our more flexible nonparametric models provide some insights into these possible earlier episodes of star formation.
    \item We connect the CSFH-inferred ages to the nonparametric SFH morphologies with continuity prior. Our CSFH and nonparametric SFH models both strongly imply a subset of galaxies undergoing an upturn in star formation and a subset of galaxies with a significant population of old stars. For objects inferred to be young ($\lesssim 10$\,Myr) by our CSFH models, the nonparametric SFH models suggest a rapidly rising SFH with SFR increasing by a factor of $30 - 40$ within 20\,Myr. For the objects with the oldest CSFH-inferred ages ($\gtrsim 250$\,Myr), the nonparametric SFHs display a decline in SFR by a similar factor from the earliest times to the most recent 10\,Myr.
    \item We infer a median age of UV-selected galaxies at $z \sim 7$ of at least $\sim 70$\,Myr (and no more than $\sim 250$\,Myr) in the context of our CSFH models. We consider the full posterior probability distributions for the age of each individual object in order to quantify the distribution of ages for the entire sample, which we parametrize as a truncated log-normal distribution. We find a distribution weighted towards older ages of $\gtrsim 100$\,Myr but with a tail towards younger ages. We then use this distribution to measure the median age of $\sim 70 - 250$\,Myr.
    \item We quantify the fraction of UV-luminous galaxies at $z \simeq 6.8$ with young ages (which we interpret as being observed during a recent burst of star formation) and the fraction being dominated by an old stellar population. In the context of our \beagle\ CSFH models, we find that the rest-UV and optical SED is dominated by light from $\leq 10$\,Myr stars, which may have been produced in a recent, intense burst of star formation, in $\leq 16$\,per\,cent of the UV-bright population. Meanwhile, we expect $\geq 26$\,per\,cent of the population to be dominated by an old stellar population with age $\geq 250$\,Myr. From our \prospector\ CSFH models, we find that $\sim 30$\,per\,cent of the population has ages $\leq 10$\,Myr and $\sim 20$\,per\,cent are $\geq 250$\,Myr old.
    \item We examine the integral constraints that our measurements of stellar populations of $z \sim 7$ galaxies can place on the build up of stellar mass in the early Universe. We expect $\sim 25$\,per\,cent of UV-luminous galaxies at $z \sim 7$ to have an old stellar component, suggesting that the number density of UV-luminous galaxies at $z \sim 9$ is a factor of $3 - 5$ lower than at $z \sim 7$. This estimate is broadly consistent with some observational constraints on the high redshift UV LF evolution, though lower than others. However, we note that our analysis currently has large uncertainties due to low signal-to-noise data. We also examine the stellar mass assembly history implied by our results, and find that the rate of decline with increasing redshift is model-dependent and motivates the need for additional constraints on earlier star formation in addition to the rest-UV and optical SEDs. However, despite these uncertainties, we find that a rapid decline in the cosmic star formation history requires a close adherence to the lower limit implied by our CSFH models, suggesting that such a rapid decline is unlikely.
    \item We expect significantly improved constraints on the ages of UV-bright reionization era galaxies with \textit{JWST}, with reductions in error of $\sim 80$\,per\,cent, which will lead to more precise constraints on the distribution of ages of the population.
    \item We highlight the importance of independent mass measurements, such as dynamical masses from ALMA and shortly \textit{JWST}, to enable verification of model results, which will allow us to fully quantify the masses, ages, and SFHs of reionization era galaxies and understand their implications for early stellar mass assembly.
\end{enumerate}

\section*{Acknowledgements}

The authors thank the referee, James Dunlop, for his insightful comments that helped improve this work. We thank Ben Johnson for helpful conversations about \prospector. LW acknowledges support from the National Science Foundation Graduate Research Fellowship under Grant No. DGE-1746060. DPS acknowledges support from the National Science Foundation through the grant AST-2109066. RE acknowledges funding from NASA JWST/near-IRCam contract to the University of Arizona, NAS5-02015. JC acknowledges funding from the ``FirstGalaxies'' Advanced Grant from the European Research Council (ERC) under the European Union’s Horizon 2020 research and innovation programme (Grant agreement No. 789056).

We respectfully acknowledge the University of Arizona is on the land and territories of Indigenous peoples. Today, Arizona is home to 22 federally recognized tribes, with Tucson being home to the O’odham and the Yaqui. Committed to diversity and inclusion, the University strives to build sustainable relationships with sovereign Native Nations and Indigenous communities through education offerings, partnerships, and community service.

This material is based in part upon High Performance Computing (HPC) resources supported by the University of Arizona TRIF, UITS, and Research, Innovation, and Impact (RII) and maintained by the UArizona Research Technologies department.

This work is based in part on data products from observations made with ESO Telescopes at the La Silla Paranal Observatory under ESO programme ID 179.A-2005 and on data products produced by CALET and the Cambridge Astronomy Survey Unit on behalf of the UltraVISTA consortium.

The Hyper Suprime-Cam (HSC) collaboration includes the astronomical communities of Japan and Taiwan, and Princeton University. The HSC instrumentation and software were developed by the National Astronomical Observatory of Japan (NAOJ), the Kavli Institute for the Physics and Mathematics of the Universe (Kavli IPMU), the University of Tokyo, the High Energy Accelerator Research Organization (KEK), the Academia Sinica Institute for Astronomy and Astrophysics in Taiwan (ASIAA), and Princeton University. Funding was contributed by the FIRST program from the Japanese Cabinet Office, the Ministry of Education, Culture, Sports, Science and Technology (MEXT), the Japan Society for the Promotion of Science (JSPS), Japan Science and Technology Agency (JST), the Toray Science Foundation, NAOJ, Kavli IPMU, KEK, ASIAA, and Princeton University. 

This paper makes use of software developed for Vera C. Rubin Observatory. We thank the Rubin Observatory for making their code available as free software at \href{https://pipelines.lsst.io/}{https://pipelines.lsst.io/}.

This paper is based in part on data collected at the Subaru Telescope and retrieved from the HSC data archive system, which is operated by the Subaru Telescope and Astronomy Data Center at NAOJ. Data analysis was in part carried out with the cooperation of Center for Computational Astrophysics, NAOJ. We are honored and grateful for the opportunity of observing the Universe from Maunakea, which has the cultural, historical and natural significance in Hawaii. 

This work makes use of the following software: \textsc{numpy} \citep{harris2020}; \textsc{matplotlib} \citep{hunter2007}; \textsc{scipy} \citep{virtanen2020}; \textsc{astropy}\footnote{\url{https://www.astropy.org/}}, a community-developed core Python package for Astronomy \citep{astropy2013, astropy2018}; \textsc{Source Extractor} \citep{bertin1996} via \textsc{sep} \citep{barbary2016}; \textsc{mopex} \citep{makovoz2005}; \textsc{photutils} \citep{bradley2020}; \textsc{beagle} \citep{chevallard2016}; \textsc{Prospector} \citep{johnson2021}; \textsc{multinest} \citep{feroz2008, feroz2009, feroz2019}; \textsc{sedpy} \citep{johnson_sedpy}; \textsc{fsps} \citep{conroy2009, conroy2010} via \textsc{python}-\textsc{fsps} \citep{johnson_python_fsps}; \textsc{dynesty} \citep{speagle2020}; \textsc{emcee} \citep{foreman-mackey2013}; and \textsc{corner} \citep{foreman-mackey2016}.

\section*{Data Availability}
 
All optical and infrared images used in this work are available through their respective data archives. See \url{https://hsc-release.mtk.nao.ac.jp/} for Subaru/HSC (HSC SSP PDR2 and CHORUS PDR1), \url{http://www.eso.org/rm/publicAccess\#/dataReleases} for VISTA/VIRCAM (UltraVISTA DR4), \url{https://irsa.ipac.caltech.edu/data/COSMOS/} for \textit{HST}/F814W, and \url{https://sha.ipac.caltech.edu/applications/Spitzer/SHA/} for IRAC. The photometry and analysis code used in this work will be shared upon reasonable request to the corresponding author.



\bibliographystyle{mnras}
\bibliography{refs}



\appendix

\section{Prior-dependence of the stellar mass formed at early times} \label{appendix:nonpar_priors}

\begin{figure}
    \centering
    \includegraphics[width=\columnwidth]{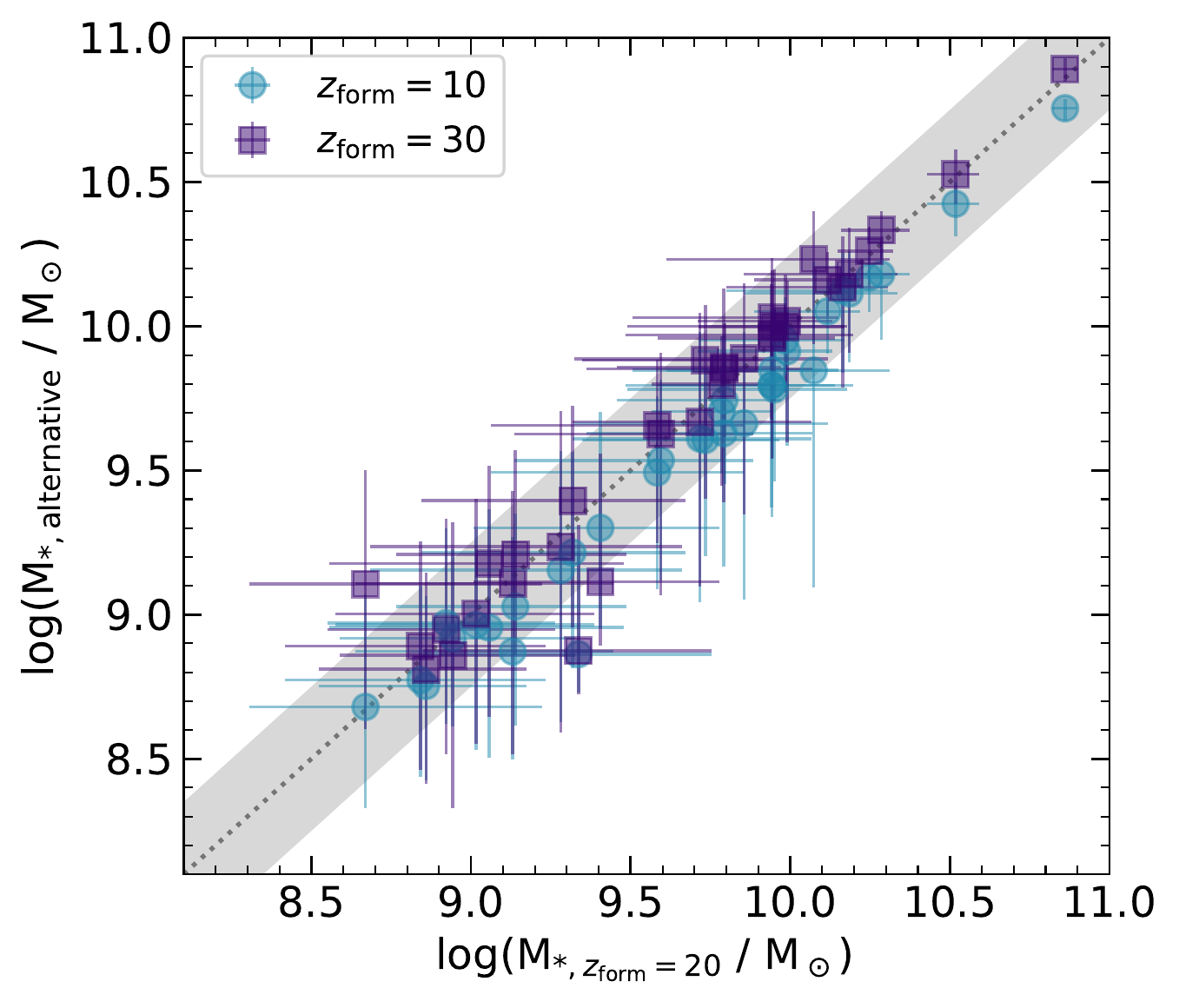}
    \caption{Comparison of the stellar masses inferred when varying the redshift at which star formation begins, $z_\text{form}$, in our nonparametric models. We show the masses inferred from models with alternative formation redshifts of $z_\text{form} = 10$ (teal circles) and $z_\text{form} = 30$ (purple squares) as a function of the stellar mass of our original nonparametric models with $z_\text{form} = 20$. Models with a later formation time ($z_\text{form} = 10$) generally find slightly smaller stellar masses by a factor of $\sim 1.25$, on average, and at most a factor of $\sim 3$, while models with an earlier start time ($z_\text{form} = 30$) are slightly larger by similar factors. We thus conclude that the exact time at which star formation begins is much less important than simply requiring an early period of star formation.}
    \label{fig:sfh_mass_comparison_zform}
\end{figure}

In Section\ \ref{subsec:nonpar_results}, we emphasized that for galaxies with SEDs dominated by stars formed during a recent, rapid increase in star formation, the data can only very weakly constrain early epochs of star formation, at best. Therefore, the stellar mass formed at early times (and so the total stellar mass) we infer from our models can be highly prior-dependent \citep[e.g.][]{leja2019a}. Understanding this prior-dependence is particularly important at the redshifts of reionization, when a significant subset of the population is being observed during a period of vigorous star formation, the light from which could be outshining light from an older stellar population.

We make two assumptions for our nonparametric SFH models that may impact the stellar masses we infer. First, we require that star formation begins at $z_\text{form} = 20$, which strongly restricts the amount of time available for star formation. Second, we adopt a prior on the SFH that favors a relatively smooth, slowly evolving SFH (i.e. weights against sharp changes in the SFH in adjacent time bins).

To investigate the impact of the assumption we make for the time at which star formation begins, we model our sample with two different formation redshifts, $z_\text{form} = 10$ and $z_\text{form} = 30$. In Figure\ \ref{fig:sfh_mass_comparison_zform}, we show the impact of these alternative values for $z_\text{form}$ on the inferred stellar masses. In general, we find that while the time at which star formation begins impacts the inferred stellar masses slightly (later formation times lead to smaller stellar masses and earlier times yield larger masses), the difference is not significant; the stellar masses differ by a factor of $\sim 3$ at most ($\sim 1.25$ on average).

\begin{figure}
    \centering
    \includegraphics[width=\columnwidth]{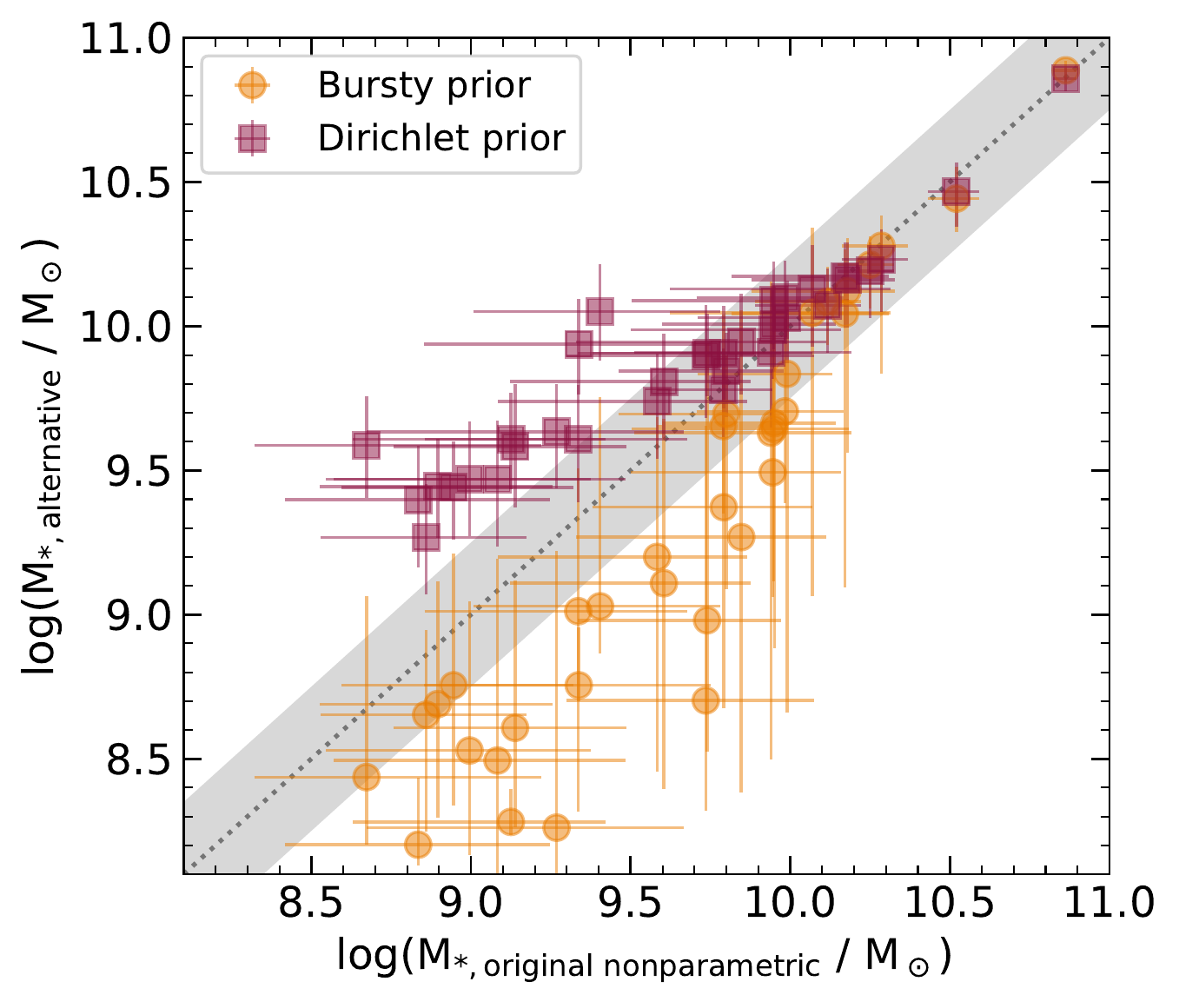}
    \caption{Comparison of the stellar masses inferred from nonparametric models with different priors and fixed $z_\text{form} = 20$. The `bursty' prior (yellow circles) infers smaller stellar masses that differ by a factor of $\sim 2$ on average, and up to a factor of $\sim 10$, from the nonparametric models described in the main text. The Dirichlet prior (red squares) can infer larger stellar masses by similar factors ($\sim 1.5$ on average, up to a factor of $\sim 8$ at low stellar masses). Thus, the general qualitative tendency of a prior to disallow a significant decrease in SFR at early times compared to the recent SFH can introduce order of magnitude differences at young ages, consistent with our findings from our fiducial CSFH and nonparametric models with standard continuity prior in the main text.}
    \label{fig:sfh_mass_comparison_prior}
\end{figure}

We also test two alternative nonparametric priors, both with $z_\text{form} = 20$: a `bursty' prior that allows for larger changes in SFR between adjacent time bins while still enforcing a relatively smooth evolution, and a prior that permits much sharper features in the SFH (e.g. rapid bursts and quenching events). For the bursty model, we adopt the same continuity prior that fits for the logarithm of the ratios between SFR in adjacent time bins with Student's $t$-distribution as the prior, and choose $\sigma = 1$ for the dispersion, as was adopted by \citet{tacchella2022}. For the prior that allows sharper features, we adopt the Dirichlet prior in \prospector, which fits for the fractional mass formed in each time bin. The Dirichlet distribution requires a concentration parameter that regulates the tendency to distribute mass equally among all bins; we choose a concentration parameter of one, which weights towards smooth SFHs. For more details about the nonparametric priors, see \citet{leja2019b}.

In Figure\ \ref{fig:sfh_mass_comparison_prior}, we compare the stellar masses inferred by the alternative priors to the masses from our original nonparametric model. Compared to the nonparametric models in the text, the bursty prior finds very similar stellar masses at large stellar masses, but up to a factor of 10 smaller masses for lower mass systems. As expected from the construction of the prior, the bursty models tend to find a larger increase in SFR from early to recent times, qualitatively more similar to a CSFH with zero early star formation than our main text nonparametric models. Meanwhile, the Dirichlet prior infers larger stellar masses than the original nonparametric models for less massive sources, up to a factor of eight larger. In general, we find that the tendency of a prior to infer large SFRs at early times (i.e. disfavor a large decrease in SFR from recent to early times) can produce stellar masses more than an order of magnitude larger than priors that allow negligible early star formation, thus bracketing the range of stellar masses that are consistent with the data.

\section{Alternative parametric SFH models} \label{appendix:parametric_sfhs}

In the main text, we adopted a simple CSFH as our fiducial model. We now test other functional parametrizations of the SFH: one that allows more complex behavior as a function of time (a delayed-tau model with a recent, ongoing burst), and one that enforces an old component of the star formation history (a constant extended component and a burst; in essence, this is a `nonparametric' model with two time bins).

\begin{figure}
    \centering
    \includegraphics[width=\columnwidth]{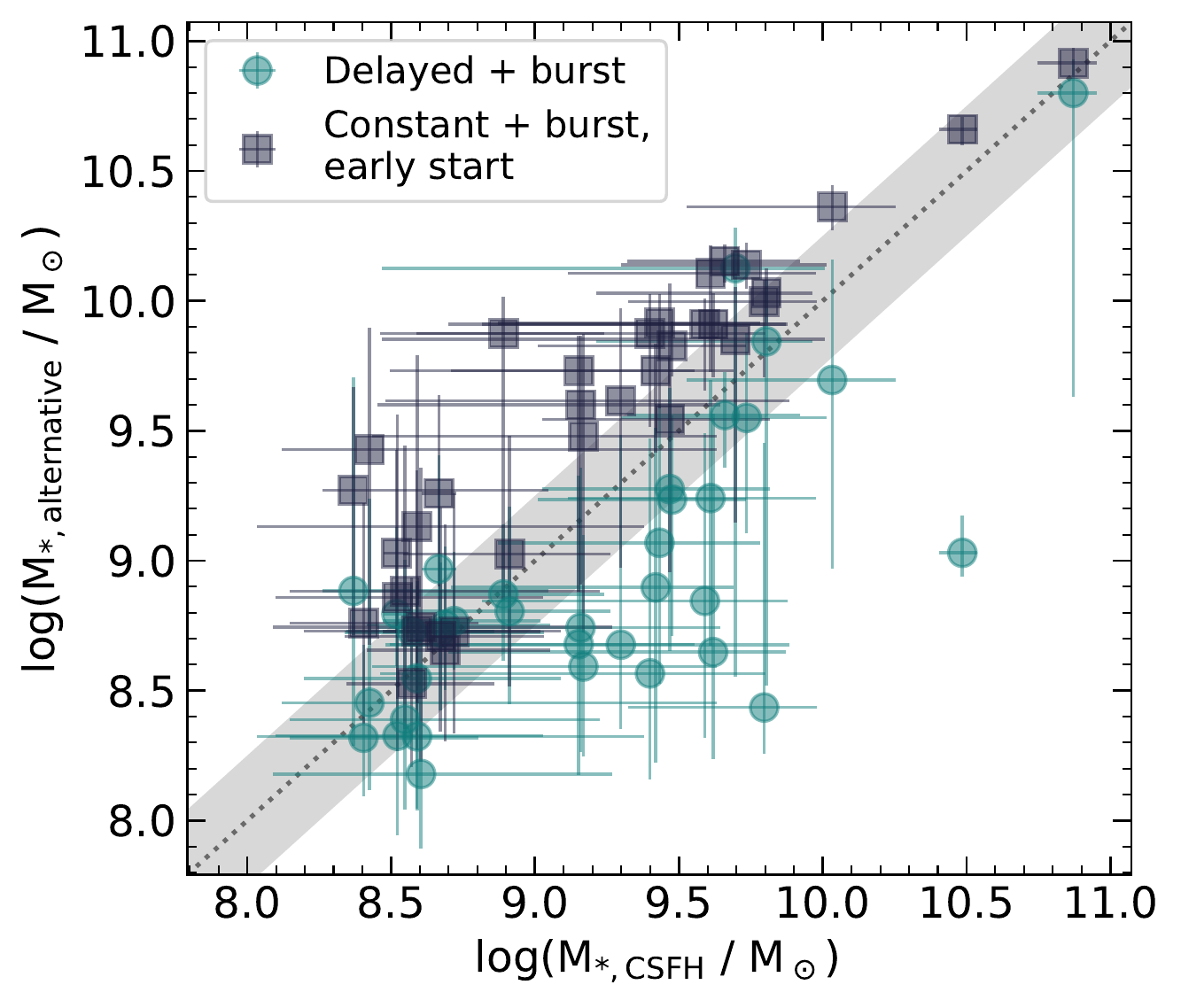}
    \caption{Comparison of the stellar masses inferred from two alternative functional SFH parametrizations. We show the delayed model with a burst as green circles, and the masses from the constant model with a burst that starts early as dark grey squares. The delayed plus burst model tends to find smaller stellar masses than the fiducial CSFH model by a factor of $\sim 3$ on average. In contrast, the constant plus burst model with an old component find systematically larger stellar masses, also by an average of a factor of $\sim 3$. We conclude that the combination of requiring an early formation time and prolonged periods of large SFRs drives the models to larger stellar masses, as they cannot rule out low-level early star formation activity.}
    \label{fig:sfh_mass_comparison_parametric}
\end{figure}

We adopt the following priors for these SFHs. For the delayed-tau ($\text{SFR}(t) \propto t\text{e}^{-t/\tau}$) SFH model with a recent ongoing burst, we fit for the time-scale for exponential decline ($\tau$) and the time when the burst begins ($\text{age}_\text{burst}$) with log-uniform priors. Specifically, we fit for $\tau$ on the interval $7 \leq \log(\tau / \text{yr}) \leq 10.5$, and the starting time of the burst from $1 \leq \text{age}_\text{burst} / \text{Myr} \leq 10$. For the constant SFH with a burst, we fit for the formation redshift with a uniform prior from $z_\text{form} = 15 - 30$. The priors on all other parameters are the same as presented in Section\ \ref{sec:sed_modelling}.

In Figure\ \ref{fig:sfh_mass_comparison_parametric}, we show the stellar masses inferred from both models, using \beagle, as a function of the CSFH stellar mass. The delayed plus burst models infer similar stellar masses at the smallest stellar masses, but smaller stellar masses for objects with CSFH masses of $M_{*,\text{CSFH}} \gtrsim 10^9$\,\Msun. On average, the delayed plus burst stellar masses are a factor of $\sim 3$ smaller, but can up to a factor of $\sim 10 - 20$ smaller at large stellar masses, since these objects are generally fit by significantly declining SFHs. In contrast, the constant plus burst models with an early start time infer systematically larger stellar masses, also by a factor of $\sim 3$ on average.

In general, we find that the enforcement of an early formation time is the primary driver of systematically larger stellar masses, even with less flexible SFH models. Since the time between $z = 20$ and $z = 7$ is $\sim 600$\,Myr, the difference in stellar mass between the constant and constant plus burst models suggests that in some cases, the models cannot rule out star formation on the order of $\text{SFR} \lesssim 5$\,M$_\odot$\,yr$^{-1}$ for a prolonged amount of time in the past, broadly consistent with the inferences from our full nonparametric models (see Figure\ \ref{fig:nonparametric_sfhs}).


\bsp	
\label{lastpage}
\end{document}